\documentclass{emulateapj}
\usepackage{apjfonts}

\newcommand{\mbh}{M_{\rm BH}}
\newcommand{\Mbh}{M_{\rm BH}}

\newcommand{\Mdot}{\dot{M}}
\newcommand{\mdot}{\dot{m}}
\newcommand{\Mdotedd}{\dot{M}_{\rm Edd}}

\newcommand{\etal}{et al.}
\newcommand{\er}{\epsilon_{r}}

\newcommand{\mdotcrit}{\mdot_{\rm crit}}

\newcommand{\EV}[1]{\langle #1 \rangle}

\newcommand{\Racc}{R_{\rm BH}} % accretion radius

\newcommand{\Raccr}{\Racc}

\newcommand{\tacc}{t_{\rm BH}} % time when hit accretion radius

\newcommand{\cs}{c_{s}} % sound speed
\newcommand{\bigfrac}[2]{{\Bigl(}\frac{#1}{#2}{\Bigr)}}
\newcommand{\Mdotbondi}{\Mdot_{\rm Bondi}}

\newcommand{\tS}{t_{S}} %Salpeter time
\newcommand{\ts}{\tS}

\newcommand{\tdyn}{t_{\rm dyn}}
\newcommand{\Mbul}{M_{\rm bul}}
\newcommand{\mbul}{\Mbul}

\newcommand{\fgas}{f_{\rm gas}}

\shorttitle{Fueling Seyfert Galaxies}
\shortauthors{Hopkins \& Hernquist}
\slugcomment{Submitted to ApJS, March 6, 2006}
\begin{document}

\title{Fueling Low-Level AGN Activity Through the Stochastic Accretion 
of Cold Gas\footnotemark[1]} \footnotetext[1]{\rm We dedicate this 
paper to John Bahcall, who inspired our interest in this problem.}
\author{Philip F. Hopkins\altaffilmark{2} \&\ 
Lars Hernquist\altaffilmark{2}
}
\altaffiltext{2}{Harvard-Smithsonian Center for Astrophysics, 
60 Garden Street, Cambridge, MA 02138, USA}

\begin{abstract}
Using a simple description of feedback from black hole growth, we
develop an analytic model for the fueling of Seyferts (low-luminosity
AGN) and their relation to their host galaxies, Eddington ratio
distributions, and cosmological evolution.  We derive a solution for
the time evolution of accretion rates in a feedback-driven blast wave,
applicable to large-scale outflows from bright quasars in galaxy
mergers, low-luminosity AGN, and black holes or neutron stars in
supernova remnants.  Under the assumption that cold gas stochastically
accretes onto a central supermassive black hole at a rate set by the
dynamics of that gas, our solution determines the evolution of Seyfert
light curves.  Using this model, we predict the Seyfert luminosity
function, duty cycles and AGN ``lifetimes,'' and the distribution of
host morphologies, Eddington ratios, and obscuration as a function of
AGN luminosity and black hole mass, and find agreement with
observations at $z=0$.  We consider the breakdown of the contribution
from this mechanism and from stellar wind and virialized hot gas
accretion and merger-driven activity.  We also make specific
predictions for the weak evolution of the Seyfert luminosity function;
i.e.\ luminosity function of quiescent, as opposed to merger-driven
activity, as a function of redshift, and for changes in both the slope
and scatter of the $M_{\rm BH}-\sigma$ relation at low-$M_{\rm BH}$.
Our modeling provides a quantitative and physical distinction between
local, low-luminosity quiescent AGN activity and violent,
merger-driven bright quasars.  In our picture, the quiescent mode of
fueling dominates over a wide range of luminosities ($-14\gtrsim
M_{B}\gtrsim-22$) at $z=0$, where most black hole growth occurs in
objects with $M_{\rm BH}\lesssim10^{7}\,M_{\sun}$, in S0 and Sa/b
galaxies.  However, quasar activity from gas-rich mergers evolves more
rapidly with redshift, and by $z=1$, quiescent fueling is important
only at luminosities an order of magnitude or more below the ``break''
in the luminosity function.  Consequently, although non-merger driven
fueling is important for black hole growth and the $M_{\rm BH}-\sigma$
relation at low $M_{\rm BH}$, it does not significantly contribute to
the black hole mass density of the Universe or to cosmological
backgrounds.
\end{abstract}


\section{Introduction}
\label{sec:intro}

While it is now generally accepted that quasars and active galactic
nuclei (AGN) are powered by the accretion of gas onto supermassive
black holes in the centers of galaxies (e.g. Salpeter 1964; Zel'dovich
\& Novikov 1964; Lynden-Bell 1969), the mechanism that fuels these
objects is uncertain.  In the local Universe, ultraluminous infrared
galaxies (ULIRGs) have bolometric luminosities similar to bright
quasars and are always in mergers \citep[see
e.g.,][]{sand86,sand88,soifer87,SM96}, and multi-wavelength studies have
shown that many contain growing, optically obscured black holes
\citep{Komossa03,ger04,max05,Alexander05a,Alexander05b,Borys05}.
Observations of low-redshift quasar hosts also reveal a connection
between galaxy mergers and quasar activity
\citep[e.g.,][]{Heckman84,stock91,HN92,bks96,Bahcall97,CS01}.  These
facts motivate a scenario where mergers of gas-rich galaxies provide
the fuel to power nuclear starbursts which evolve into bright quasars.

Numerical simulations have identified a physical mechanism for
transporting gas into the inner regions of galaxies in a merger by
showing that tidally-induced gravitational torques remove angular
momentum from the gas (e.g. Barnes \& Hernquist 1991, 1996;
Mihos \& Hernquist 1996).  Models
including supermassive black holes \citep{DSH05,SDH05b} support the
conjecture that ULIRGs evolve into quasars and suggest that feedback
from black hole growth mediates this transition by expelling obscuring
gas and dust.  This ``blowout'' results in a short-lived, bright
optical quasar \citep{H05a,H05b,H05e} and eventually terminates the
activity, leaving a remnant that quickly reddens (e.g. Springel et
al. 2005a) and satisfies observed correlations between black hole mass
and the mass \citep{Magorrian98,MD02,MH03} or velocity dispersion
(i.e. the $M_{\rm BH}$-$\sigma$ relation: Ferrarese \& Merritt 2000;
Gebhardt et al. 2000) of spheroids.

Moreover, the time evolution in these simulations reproduces many
quasar observables, including luminosity functions, host properties,
and accretion rates \citep{H05c,H05d,H05e,H05g}. The modeling also
indicates that the bright quasar population and the bulk of the cosmic
quasar luminosity density and buildup of the black hole mass density
must be dominated by merger-driven growth \citep{H05i}, to be consistent
with the evolution of the X-ray background \citep{H05e}, the red
galaxy age, metallicity, mass, and luminosity functions \citep{H05f},
and merger luminosity functions and star formation rate densities
\citep{H05h}.

However, many local, low-luminosity AGN reside in quiescent,
non-interacting galaxies
\citep[e.g.,][]{Kauffmann03,Sanchez03,Sanchez04,MD04,Hao05}.  Some of
these objects do in fact fit naturally into a merger-driven picture.
At relatively high luminosities, ellipticals with young stellar
populations and moderate Eddington ratios have observed properties
consistent with decay in the quasar light curve from a previous bright
quasar epoch in a spheroid-forming merger
\citep{Kauffmann03,Best05,H05i,H05g}.  At the lowest luminosities 
characteristic of low-luminosity AGN and LINERs, 
populations of very low accretion rate ``dead'' ellipticals dominate,
fueled via accretion of hot (virialized) spheroid gas and steady mass
loss from stars
\citep[e.g.,][]{Ciotti97,Ciotti01,Pellegrini05,Soria05b}, presumably
in radiatively inefficient accretion states \citep[e.g.,][]{NY95}.
Although the detailed properties of these objects are subject to
debate \citep[see, e.g.,][]{Narayan04}, the fuel source is reasonably
well-understood, and it is clear from comparison of black hole mass
and quasar luminosity functions
\citep[e.g.,][]{Soltan82,Salucci99,YT02,Marconi04,Shankar04},
background synthesis models
\citep[e.g.,][]{Comastri95,Gilli99,ERZ02,Ueda03,Cao05}, and
observations of the accretion rate distribution
\citep{Vestergaard04,Heckman04,MD04,H05i} that these modes do not
dominate black hole growth or cosmological backgrounds.

Nevertheless, a number of relatively high accretion rate objects are
observed at low redshift in undisturbed, late-type, star-forming
galaxies \citep[e.g.,][]{Kauffmann03} with small ($M_{\rm BH}\lesssim
10^{7}\,M_{\sun}$) black holes \citep{Heckman04}.  Indeed, the local
AGN population largely comprises black holes in non-interacting,
star-forming S0 and Sa/b hosts with no evidence of galaxy-scale
perturbations or disturbances \citep{Dong05}, spanning most of the
observed $z=0$ AGN luminosity function \citep{Hao05}.  Although these
AGN have been more thoroughly studied and are more well-understood
than bright quasars at high redshift, there is no self-consistent
model for their triggering, fueling, and evolution.  Previous
investigations of their fueling have mainly been restricted to
estimating whether or not a given fuel source could provide an
adequate mass supply
\citep[e.g.,][]{lynden69,Hills75,ShieldsWheeler78,Mathews90}, or have
examined the evolution of light curves in limited subclasses of these
objects \citep[e.g.,][]{Ciotti01}.  A critical uncertainty in previous
efforts to model AGN is the light curve evolution once a given
``trigger'' occurs.

Independent of how fuel is delivered to the supermassive black holes
in Seyferts and low-luminosity AGN, their subsequent evolution may
resemble that in bright quasars if feedback from accretion regulates
black hole growth.  In the simulations of Di Matteo et al. (2005), the
impact of this feedback resembles an explosion because the energy is
deposited on scales small compared to the host galaxy and because the
black hole grows nearly exponentially during the quasar phase with an
{\em e}-folding time that is short compared to the characteristic dynamical
time of the host potential.  Indeed, Hopkins et al. (2005g) show that
the outflow driven by this feedback can be described as a blast wave.
In what follows, we investigate whether solutions of this type can
also be used to characterize the evolution of low-luminosity AGN and
Seyferts.

There is observation evidence for outflows and winds in both AGN and
quasars \citep[for a review, see][]{Veilleux05}.  The kinematics of
gas in the narrow line regions of local, low-luminosity AGN
\citep[see, e.g.][and references therein]{Rice06} display bi-conical
or nearly isotropic (wide-angle) radial outflows at speeds
$\sim10^{2}-10^{3}\,{\rm km\,s^{-1}}$ \citep[e.g.,][]{Crenshaw00},
with wind dynamical times $\sim10^{4}-10^{6}\,$yr and entrained
molecular gas masses $\sim10^{5}-10^{7}\,M_{\sun}$
\citep[e.g.,][]{StarkCarlson84,Walter02}.  The ``warm absorber'' seen
in soft X-rays is also indicative of a significant outflow generated
local to the AGN \citep[e.g.,][and references therein]{Laor97},
accelerating radially to larger terminal velocities $\gtrsim {\rm a\
few\ }100\,{\rm km\,s^{-1}}$ \citep{Ruiz01,Kaspi02}. These absorbing
outflows appear ubiquitous 
\citep[and are likely even where not directly observed, given 
observed wind covering angles and clumping factors; see e.g.][]{Rupke05}
and are associated with clumpy,
high-ionization structures \citep[e.g.,][]{Crenshaw99}. Outflow
energetics and entrained masses appear to scale with AGN power
\citep{BaumMcCarthy00}, and intense winds with velocities
$\sim10^{4}\,{\rm km\,s^{-1}}$ are seen in bright, broad absorption
line (BAL) quasars \citep[e.g.,][]{Reeves03}. Observations of
typical, narrow-line quasars also find BAL-like outflow velocities for
some species accelerating at small radii from the central engine
\citep{Pounds03a,Pounds03b}, and outflows with large covering angles 
in central regions of ``normal'' quasars may even be detected 
in gravitational lensing signatures \citep{Green06}.

Although the relative contributions of star formation and AGN to the
energetics is unclear \citep[see e.g.,][]{Baum93,Levenson01},
high-resolution data indicate that even Seyfert II ULIRGs in which
large-scale winds may be driven by star formation have central,
AGN-driven outflows \citep{Cecil01,Rupke05}, and observations at optical,
X-ray, and radio wavelengths
\citep{Colber96a,Colber96b,Colber98,WhittleWilson04} identify a number
of mechanisms by which AGN power can couple to outflows over $\sim$kpc
scales.  The decomposition of the narrow line region components also
suggests wind ram pressure or radiation pressure from the central
source as a driver \citep{Kaiser00}, and in some cases the wind
injection zone is observed to be small relative to the scale of
nuclear star formation \citep{SmithWilson01}. Furthermore, similar
outflow structures are observed in the hot gas of some spheroidal
systems which have no rapid associated star formation
\citep[e.g.,][]{Biller04,Fabbiano04}.

The various observations fit naturally into a picture where outflows
are caused by feedback from AGN accretion.  This has motivated
theoretical studies of the properties of AGN winds
\citep[e.g.,][]{SVS85,NY94,KoniglKartje94,StoneNorman94,
Murray95,Elvis00,Proga00,Proga04}.  The models have invoked various
driving mechanisms near the black hole, including hydromagnetic disk
winds \citep[e.g.,][]{KoniglKartje94}, MHD jet outflows
\citep[e.g.,][]{BlandfordPayne82} powered by magnetic coronae
formation over the disk \citep{MillerStone00}, Comptonization in an
X-ray halo \citep[e.g.,][]{Begelman85}, and radiation pressure
coupling to dust opacity \citep[e.g.,][]{Dopita02}.  Irrespective of their 
details, these processes universally yield a multi-temperature, clumpy,
filled wind structure \citep{BegelmanMcKee83,KoniglKartje94} which
provides a good representation of the various components and spectral
properties of observed outflows
\citep[e.g.,][]{KrolikKriss01,Dopita02,Ogle03}.  The winds evolve like
a Sedov-Taylor type blast wave \citep{Begelman83}, as suggested by
observed velocity profiles \citep{ShopbellBland98}, and develop the
typical shell structure and phases of evolution of these blast waves
\citep{Schiano85}, well known from the analysis of supernova remnants
\citep[e.g.,][]{OM88}.

The theoretical works have examined the generation mechanisms of winds
near a pre-existing accretion disk, and the observable impact of these
winds on the interstellar medium (ISM) as they shock and entrain gas
\citep[e.g.,][]{Middelberg04,Machacek04,OSullivan05}.  However, the
AGN luminosity or accretion rate is generally an input parameter in
these analyses, with an undetermined macroscopic fueling mechanism.
Without a self-consistent calculation of the coupled evolution of the
wind/blast wave and AGN accretion rate, such models, while critical
for characterizing the detailed radiative properties of the ISM local
to the AGN, cannot provide a physical motivation or understanding of
the distribution of AGN luminosities, Eddington ratios, black hole
masses, and fueling mechanisms, and their evolution with redshift.

If feedback from accretion couples to the gas surrounding black holes
in Seyferts, the observations and theoretical models motivate the
following picture for AGN activity in quiescent galaxies having a
supply of cold gas.  Observations of the dynamics and distribution of
cold gas in the central regions of these AGN
\citep[e.g.,][]{KB88,Kaneko89,Heckman89,Israel90,Meixner90,
GDF97,Bock00,Schinnerer00,
Galliano03,Radomski03,Weigelt04,Jaffe04,Prieto04,Elitzur05,Mason05}
suggest that rotationally supported gas extends to the inner regions
of the galaxy.  Whether part of the galactic disk or, potentially, a
``clumpy'' torus \citep[e.g.,][]{Antonucci93} or bar-like structure,
molecular clouds or blobs of cold gas could be accreted stochastically
by the black hole.  Such gas will have some turbulent (random) motion
and corresponding probability of colliding with the central black hole
\citep[see also e.g.][]{Lauer05}.

The mass of gas required to sustain an AGN is only
$\sim10^{5}-10^{7}\,M_{\sun}$, far less than the
$\sim10^{8}-10^{9}\,M_{\sun}$ observed within the inner tens to
hundreds of pc of many late-type galaxies \citep[e.g.,][and references
therein]{Kaneko89,Heckman89,Meixner90,
GDF97,Galliano03,Mason05,Elitzur05}, and only $\sim1$ event per Hubble
time may be expected (see \S~\ref{sec:duty}).  Therefore, large-scale
gravitational torques are not required, although mechanisms such as
disk and bar instabilities
\citep[e.g.,][]{NormanSilk83,NormanScoville88, SBF89,Lin88,Lubow88},
minor mergers \citep[e.g.,][]{Roos81,Gaskell85,
Hernquist89,MH94,HM95,DeRobertis98,Tanaguchi99}, or magnetic
instabilities \citep[e.g.,][]{KrolikMeiksin90} may contribute to the
``effective'' turbulent motion.  In a ``collision,'' the black hole
will accrete at a high rate for a brief period of time until feedback
impacts the cold gas, driving a blast wave and initiating a
feedback-dominated ``blowout'' phase.  This blowout determines the
subsequent, time-averaged evolution of the Seyfert light curve,
obscured fractions, and Eddington ratio distributions, and the system
decays to lower luminosities until a potential subsequent excitation.

We use this picture to develop a model for the fueling of Seyfert
galaxies, which allows us to predict their luminosity functions and
evolution with redshift, among other quantities.  In \S~\ref{sec:deriv},
we derive a generalized Sedov-Taylor solution for feedback-driven
outflows and both a Bondi-Hoyle type and generalized perturbative
accretion solution within such a medium.  This is applicable to any
feedback regulated accretion system, including the quiescent systems
studied here, ``blowouts'' in merger-driven quasars, and accreting
black holes and neutron stars in supernova blast waves.  In
\S~\ref{sec:timescales} we apply this solution to the cases of
merger-driven activity and stochastic accretion in quiescent systems.
In \S~\ref{sec:mol.cloud.accr} we calculate the expected excitation
rates, duty cycles, and luminosity function of Seyfert galaxies as
triggered in quiescent galaxies.  We further predict and compare with
observations the distribution of host properties and morphological
types as a function of luminosity, the distributions of accretion
rates and Eddington ratios, and evolution (compared to e.g.\ evolution
in merger-driven quasar activity) with redshift.  In
\S~\ref{sec:m.sigma} we predict the effects on the $M_{\rm BH}-\sigma$
relation from this mode of AGN fueling, in particular estimating
corrections to the slope and scatter at low $M_{\rm BH}$.  In
\S~\ref{sec:torus} we determine the implications of this picture for
Seyfert obscuration and the potential buildup of the classical
molecular torus. In \S~\ref{sec:host.fx} we investigate effects on the
host galaxy from this mode of fueling. In \S~\ref{sec:winds} we
estimate and compare with other quiescent modes of AGN fueling, such
as accretion of hot spheroid gas and stellar winds from passive
evolution. Finally, in \S~\ref{sec:discuss} we summarize and discuss
predictions to test our model.

\section{Accretion in Feedback-Driven Outflows}
\label{sec:deriv}

A black hole accreting at the Eddington rate will grow
exponentially on a relatively short timescale (a few
$\times 10^{7}\,$yr). If the black hole is small, feedback is suppressed
exponentially, and surrounding gas can equilibrate with the
low rate of energy or momentum input. However, if the black hole is
already large, it will be sufficiently luminous that the energy
injected cannot be radiated by the gas in a local dynamical time,
effectively resulting in an instantaneous, point-like (relative to the
scales of the galaxy) energy injection in the center of an
(approximately spherical) bulge which dominates the local
gravitational potential. Therefore, it is appropriate to describe this
phase as a Sedov-Taylor type blast wave, which, in detail, a number of
simulations have shown is a surprisingly good approximation to a full
solution including radiative cooling, the pressure of the external
medium, magnetic and gravitational fields, and further effects, at
least on galaxy \citep{H05g} and larger \citep{FL01} scales.

Here, we determine the behavior of such a blast wave when energy is
input from black hole accretion into a medium with an initial 
density gradient.  In particular, we derive an approximation to the internal
density, velocity, and temperature structure of the blast wave as a
function of time, which becomes exact as $r\rightarrow0$.  Using this,
we can calculate the evolution of the Bondi-Hoyle accretion rate for a
central black hole at $r=0$.  However, because the Bondi rate is not
calculated self-consistently in a medium with external density and
velocity gradients, we use our solution for the internal blast wave
structure to derive a solution for a perturbative inflow driven by the
black hole gravitational potential at small radii.

Many authors have considered the evolution of blast waves and the
detailed clumping and radiative properties of their interiors
\citep[e.g.,][and references
therein]{Begelman83,BegelmanMcKee83,OM88}.  (For a discussion
of the different Sedov-Taylor radial shells or phases see e.g.\
\citet{Schiano85}.)  However, these works have focused on understanding
the radiative properties of such blast waves, and have not calculated
an accretion rate solution within them or examined the interior
properties as they affect the small radii relevant for accretion processes.

For simplicity, we assume a power-law scaling for
the external (pre-shock) density,
\begin{equation}
\rho_{0}(r)\propto{R}^{-k_{\rho}} \, ,
\end{equation}
appropriate for a spheroid or black hole-dominated potential or
molecular cloud.  Note that this need hold only over some range in $r$
for each ``stage'' of the blast wave.  We are, for now, interested in
early times during the wave and (by definition of the blowout
``triggering'' condition) blast waves with energy greater than the
binding energy of the material, so at least in this phase we can
consider gravity to be a second-order effect which we calculate in
detail below.

Under these circumstances, the system will evolve 
as a similarity solution, 
in which the shock radius $R_{s}$ expands as
\begin{equation}
R_{s}\propto t^{\eta}. 
\end{equation}
These similarity solutions are well-understood, and 
we refer to \citet{OM88} for details.
Once the system enters the blast wave-dominated 
phase, the central accretion rate will decline with time, in power-law fashion 
when the surrounding medium evolves according to a similarity solution. We 
derive this below; for now note that we expect
\begin{equation}
L\propto\dot{M}\propto t^{-\eta_{L}}
\end{equation}
with $\eta_{L}\geq0$. 
The total energy of the blast wave evolves as
\begin{equation}
E_{s}\propto R_{s}^{-k_{E}}\propto t^{\eta_{E}}
\end{equation}
with $\eta_{E}=-\eta\,k_{E}$.
Radiative losses
from the blast wave front are small, at least on the scales 
of interest here \citep{FL01}. 
However, if feedback energy continues to couple to the 
surrounding medium as the accretion rate declines, there are 
two possible cases. 
First, if 
\begin{equation}
\eta_{L}\geq1,\ \eta_{E}\approx 0,
\end{equation}
then the blast wave is dominated by the initial energy input, and 
the evolution is effectively energy conserving with $k_{E}=0$. Second, if 
\begin{equation}
0\leq\eta_{L}<1,\ \eta_{E}={1-\eta_{L}},
\end{equation}
the energy input increases with time and radius 
in a self-similar manner. 
In either case, we obtain a similarity solution for the shock radius 
of the form
\begin{equation}
\eta=\frac{2}{5-k_{\rho}+k_{E}}=\frac{2+\eta_{E}}{5-k_{\rho}}.
\end{equation}

\subsection{Internal Blast Wave Structure at Small Radii}
\label{sec:internal.structure}

We determine the internal structure of the blast wave 
using the two-power approximation (TPA) for the 
internal velocity and density structure.  Unlike e.g.\ a 
one-power approximation,
this ensures that gradients behave physically at small radii.
In general, a 
quantity $x$ will have the scaling
\begin{equation}
x(r,\,t)=x_{1}(R_{s})\,\tilde{x}(\lambda)
\end{equation}
internal to the blast wave, where
\begin{equation}
\lambda=r/R_{s}
\end{equation}
and $x_{1}$ is the post-shock value of $x$ at the shock front.
From \citet{OM88}, this gives
\begin{equation}
\tilde{\rho}=a_{\rho}\lambda^{\ell_{\rho 1}}+(1-a_{\rho})\,\lambda^{\ell_{\rho 2}}
\end{equation}
where
\begin{eqnarray} 
\label{eqn:ells}
\nonumber & & {\ell_{\rho 1}=\frac{3-\gamma\,k_{\rho}}{\gamma-1}=\frac{1}{2}(9-5 k_{\rho})}\\
\nonumber & & {a_{\rho}=\frac{\gamma (k_{\rho,\,{\rm crit}}-k_{\rho})}{10-\gamma-(\gamma+2)k_{\rho}}=\frac{5 (2-k_{\rho})}{25+11\,k_{\rho}}}\\
\nonumber & & {\ell_{\rho 2}={\rm MIN}(\frac{6-(g+1)(2\,k_{\rho}-k_{\rho,\,{\rm crit}})}{\gamma-1},\ell_{\rho 1},1)=17-8\,k_{\rho}}\\
& & {k_{\rho,\,{\rm crit}}=\frac{7-\gamma}{\gamma+1}=2}
\end{eqnarray}
(the second equalities are for $\gamma=5/3$). These are determined by requiring the 
gradients to match exact solutions at the center and shock front and 
imposing mass conservation.
The corresponding velocity approximation is 
\begin{equation}
\tilde{v}=\frac{(\gamma+1)\lambda+(\gamma-1)\lambda^{\ell_{v 2}}}{2\gamma}
=\frac{4}{5}\lambda+\frac{1}{5}\lambda^{\ell_{v 2}}
\end{equation}
with 
\begin{equation}
\ell_{v 2}=1+\frac{\gamma}{\gamma-1}(k_{\rho,\,{\rm crit}}-k_{\rho})=6-\frac{5}{2}k_{\rho}.
\end{equation}

The expressions reduce to an exact solution for $\gamma=5/3$,
$k_{\rho}=k_{\rho,\,{\rm crit}}=2$.  For other cases, they provide a
good approximation for our purposes, accurate to $\sim17\%$ even for
$k_{\rho}=0$. The accuracy improves at small radii, and these
solutions capture the correct power-law dependence as $r\rightarrow0$
where, for example, Kahn's approximation \citep{Kahn76} for the
density and velocity structure, which is accurate to $\sim4\%$
\citep{CoxFranco81} reduces to the expressions above.  These also
provide a good description of observed blast wave velocity and
temperature structure \citep[e.g.,][]{ShopbellBland98,Crenshaw00,
KrolikKriss01,SmithWilson01,Kaspi02,Veilleux05,Rice06}.

Given the density and velocity field, it is straightforward to determine the interior pressure.  For an adiabatic blast wave, 
the entropy per unit mass is conserved in comoving coordinates, 
giving 
\begin{eqnarray}
\nonumber & & s\equiv P/\rho^{\gamma} = s_{1} \tilde{M}^{\ell_{s}}\\
\nonumber & & M(r)=M\,\tilde{M}(\lambda)\\
& & \ell_{s}=\frac{\gamma k_{\rho}-3-k_{E}}{3-k_{\rho}} \, .
\end{eqnarray}
This yields 
\begin{equation}
P\propto \lambda^{\frac{\gamma}{\gamma-1}\frac{5-k_{\rho}}{2+\eta_{E}}\eta_{E}}
\end{equation}
for the pressure as $P\rightarrow0$. For a blast wave with no energy
injection, then, we recover the well-known condition that the pressure
is constant at the origin.

In detail, 
we can use the equation of motion to solve for the pressure terms. 
Inserting the density and velocity fields into the equation of motion and solving yields a 
six-term power-law solution for the pressure
\begin{eqnarray}
\frac{P}{\rho_{1}v_{1}^{2}}&=&
a_{P0}+
a_{P1}\,\lambda^{\ell_{\rho 1}+2}+
a_{P2}\,\lambda^{\ell_{\rho 2}+2}\nonumber\\
& & \mbox{}+a_{P3}\,\lambda^{\ell_{\rho 1}+\ell_{v}+1}+
a_{P4}\,\lambda^{\ell_{\rho 2}+\ell_{v}+1}\nonumber\\
& & \mbox{}+a_{P5}\,\lambda^{\ell_{\rho 1}+2\,\ell_{v}}+
a_{P6}\,\lambda^{\ell_{\rho 2}+2\,\ell_{v}}
\end{eqnarray}
with 
\begin{eqnarray}
\nonumber & & a_{P1}=\frac{a_{\rho}a_{v}(1-a_{v}\eta\nu_{1})}{(2+\ell_{\rho 1})\eta\nu_{1}}\\
\nonumber & & a_{P2}=\frac{(a_{\rho}-1)a_{v}(1-a_{v}\eta\nu_{1})}{(2+\ell_{\rho 2})\eta\nu_{1}}\\
\nonumber & & a_{P3}=\frac{a_{\rho}(a_{v}-1)(-1+\eta-\ell_{v}\eta+a_{v} (1+\ell_{v}) \eta \nu_{1})}
{(1+\ell_{\rho 1}+\ell_{v})\eta\nu_{1}}\\
\nonumber & & a_{P4}=\frac{(1-a_{\rho})(a_{v}-1)(-1+\eta-\ell_{v}\eta+a_{v} (1+\ell_{v}) \eta \nu_{1})}
{(1+\ell_{\rho 2}+\ell_{v})\eta\nu_{1}}\\
\nonumber & & a_{P5}=\frac{-a_{\rho} (1-a_{v})^{2}\ell_{v}}{\ell_{\rho 1}+2\ell_{v}}\\
& & a_{P6}=\frac{(a_{\rho}-1) (1-a_{v})^{2}\ell_{v}}{\ell_{\rho 2}+2\ell_{v}} \, ,
\end{eqnarray}
where $a_{v}=(\gamma+1)/(2\gamma)$ is the coefficient of the $\lambda$ term 
in $\tilde{v}$ and $\nu_{1}\equiv v_{1}/v_{s}=2/(\gamma+1)$. 

Because $P=P_{1}$ at $\lambda=1$, we obtain a solution for $a_{P0}$. 
In terms of $\tilde{P}$
\begin{equation}
\tilde{P}=\tilde{P}(0)+a_{\tilde{P}}\lambda^{\ell_{\rho 1}+2}+...
\end{equation}
this yields an expression for $\tilde{P}(0)$ which is quite complicated, 
reducing for e.g.\ $k_{E}=0$, $\gamma=5/3$ to 
\begin{eqnarray}
\tilde{P}(0)&=&\frac{1}{175}
{\Bigl(}
82+\frac{210}{23-10\,k_{\rho}}+\frac{266}{19-8\,k_{\rho}}\nonumber\\
& &+\frac{7}{11-5\,k_{\rho}}-\frac{560}{13-5\,k_{\rho}}-\frac{108}{16-7\,k_{\rho}} 
{\Bigr)} \, ,
\end{eqnarray}
which interpolates from $\tilde{P}(0)\approx0.32$ for $k_{\rho}=0$ to 
$\tilde{P}(0)=1/25$ for $k_{\rho}=2$.
However, the full expression for $\tilde{P}(0)$ from the six-power expansion is 
reasonably well approximated (to $\sim10\%$ for a 
variety of $\gamma,\ k_{\rho},\ k_{E}$) by the simpler 
form
\begin{eqnarray}
\tilde{P}(0)&=&\frac{(\gamma+1)^{2}(k_{\rho,\,{\rm crit}}-k_{\rho}-k_{E})}
{3\gamma^{2}+20\gamma+1-(\gamma+1)[(3\gamma+1)\,k_{\rho}-3(\gamma-1)\,k_{E}]}\nonumber\\
&=&\frac{4\,(3-k_{\rho}-k_{E})}{24-9\,k_{\rho}+6\,k_{E}} \, ,
\end{eqnarray}
which follows from the pressure-gradient approximation \citep[see][]{OM88}, 
essentially a two-power approximation in the pressure. 

The next-order term, $a_{\tilde{P}}\lambda^{\ell_{\rho 1}+2}$, 
which is proportional to the local $\rho v^{2}$, is given by 
\begin{eqnarray}
a_{\tilde{P}}&=&\frac{(\gamma^{2}-1) (k_{\rho,\,{\rm crit}}-k_{\rho}) 
((5-k_{\rho})\gamma-(2+\eta_{E}))}
{2\gamma (1+(2-k_{\rho})\gamma)\,(10-\gamma-(2+\gamma)\,k_{\rho})\,(2+\eta_{E})}\nonumber\\
&=&\frac{8 (2-k_{\rho}) (19-5 k_{\rho}-3 \eta_{E})}{5 (13-5 k_{\rho}) (25-11 k_{\rho}) (2+\eta_{E})}.
\end{eqnarray}
This goes to zero for $\gamma=5/3$, $k_{\rho}\rightarrow2$, but 
this is because in this 
case $a_{\rho}\rightarrow0$, $\ell_{\rho 1}=\ell_{\rho 2}=1$, and 
a proper solution yields 
\begin{equation}
a_{\tilde{P}}=\frac{2-\eta_{E}}{2+\eta_{E}} \, .
\end{equation}
It is straightforward to derive the remaining terms, but 
we are interested only in the behavior
as $r\rightarrow0$, so we do not need
higher-order terms in $\lambda$. 

\subsection{Evolution of the Bondi Approximation}
\label{sec:Bondi}

We can now determine the Bondi accretion rate as a function of radius, 
again for $r\rightarrow0$, given by
\begin{equation}
\Mdot=\frac{4\pi\alpha G^{2}\,\mbh^{2}\,\rho}{(v^{2}+\cs^{2})^{3/2}}, 
\label{eqn:define.mbondi}
\end{equation}
where $\alpha$ is a constant of order unity dependent on the 
gas equation of state, $\rho$ and $c_{s}^{2}\equiv\gamma P/\rho$ 
are the local gas density and 
sound speed, respectively and $v$ is the bulk motion of the gas relative to the black hole. 
Near the origin, $\tilde{P}\rightarrow\tilde{P}(0)=\,$constant, 
thus
\begin{eqnarray}
\nonumber & & P(0)=\tilde{P}(0) P_{1}=\tilde{P}(0)\theta_{0} \rho_{0}(R_{s}) v_{s}^{2}\\
\nonumber & & \theta_{0}=\frac{2}{\gamma+1}\\
\nonumber & & \rho\rightarrow a_{\rho}\lambda^{\ell_{\rho 1}} \rho_{1}
=a_{\rho}\chi_{1} \lambda^{\ell_{\rho 1}} \rho_{0}(R_{s})\\
\nonumber & & \chi_{1}=\frac{\gamma+1}{\gamma-1}\\
\nonumber & & c_{s}^{2}=\frac{\gamma \tilde{P}(0) \theta_{0}}{a_{\rho} \chi_{1}}\,\lambda^{-\ell_{\rho 1}}\,v_{s}^{2}\\
\nonumber & & \frac{\rho}{c_{s}^{3}}=\chi_{1}{\Bigl(}\frac{\gamma \tilde{P}(0) \theta_{0}}{a_{\rho} \chi_{1}}{\Bigr)}^{-3/2}\,
\bigfrac{r}{R_{s}}^{\frac{5}{2}\ell_{\rho 1}} \frac{\rho_{0}(R_{s})}{v_{s}(R_{s})^{3}}\\
& & \ \ \ \ \propto r^{\frac{5}{2}\ell_{\rho 1}} t^{3\,(1-\eta)-\eta(k_{\rho}+\frac{5}{2}\ell_{\rho 1})}.
\label{eqn:rho.cs3}
\end{eqnarray}

So, at some point, an inward-moving Bondi-like flow can grow and 
give a residual accretion rate, and if the shock wave continues to 
blow mass out of the central regions, the scaling should be similar 
to the time scaling above. The effective ``accretion radius'' at which this 
will occur is, however, ambiguous. A natural choice
is the black hole radius of influence, 
\begin{equation}
R_{\rm BH}=\frac{G\,M_{\rm BH}}{\sigma^{2}}
\end{equation}
where $\sigma$ is the bulge velocity dispersion. This 
follows from equating the black hole potential to the 
external potential, and determines where the black hole
dominates the dynamics. 
We could instead consider the trans-sonic radius, 
\begin{equation}
R_{\rm ts}=\frac{G\,M_{\rm BH}}{c_{s}(R_{\rm ts})^{2}} \, ;
\end{equation}
however, once $c_{s}(R_{\rm BH})<\sigma$, the trans-sonic radius 
will be larger than the black hole radius of influence, 
regardless of how $c_{s}$ scales with $R$, and the black hole 
will not affect the dynamics. For the inner regions 
(or intermediate regions for $k_{\rho}=2,\ \gamma=5/3$), 
we have $c_{s}\sim v\sim r/t$, so $R_{\rm ts}\gtrsim R_{\rm BH}$ 
at $t\sim R_{\rm BH}/\sigma \sim \mu\,t_{\rm dyn}$, where 
\begin{equation}
\mu\equiv\frac{M_{\rm BH}}{M_{\ast}}\approx0.001
\end{equation} 
is the ratio of the 
black hole to the (total) spheroid stellar mass \citep[e.g.,][]{MH03} and 
\begin{equation}
t_{\rm dyn}\equiv \frac{a}{\sigma}\approx 
3.6\times10^{7}\,{\rm yr}\ {\Bigl(}\frac{\sigma}{100\,{\rm km\,s^{-1}}}{\Bigr)}
\end{equation}
is defined as the spheroid dynamical time (with $a$ being the bulge scale length). Thus the 
timescale for $R_{\rm ts}\gg R_{\rm BH}$ is $\sim10^{4}$\,yr, much
less than the timescales of interest, and 
so it is inappropriate to consider accretion through $R_{\rm ts}$ as it is 
outside the region where the black hole will 
have a significant effect on the 
kinematics. Even if some of the post-shock gas avoids cooling adiabatically and 
slows/compresses to pressure equilibrium, this will necessarily
give $c_{s}\sim\sigma$, so $R_{\rm ts}\sim R_{\rm BH}$ and 
again $R_{\rm BH}$ is the appropriate radius to consider. 

Since we expect the accretion rate to decline rapidly with time, the 
black hole mass is approximately constant over this period (see \S~\ref{sec:mass.blowout} 
for a detailed calculation), so for a fixed accretion radius $\sim R_{\rm BH}$
this then gives
\begin{equation}
\dot{M}\propto\frac{\rho}{c_{s}^{3}}\propto 
t^{3\,(1-\eta)-\eta(k_{\rho}+\frac{5}{2}\ell_{\rho 1})}
\propto t^{-\eta_{L}}.
\end{equation}
This equation for $\eta_{L}$ expands to 
\begin{equation}
\eta_{L}=\frac{24-9\gamma-5\,k_{\rho}}{(\gamma-1)\,(5-k_{\rho})}=\frac{3 (9-5 k_{\rho})}{2 (5-k_{\rho})}
\end{equation}
for a rapidly declining accretion rate with $\eta_{E}=0$ (i.e.\ valid 
for $\eta_{L}>1$, $k_{\rho}<17/13$ for $\gamma=5/3$). 
For a less rapidly declining accretion rate with $\eta_{E}\ne0$ (i.e. $0<\eta_{L}<1$) we 
obtain a self-consistent solution with 
\begin{equation}
\eta_{L}=\frac{3\,(19-4\gamma-(4+\gamma)\,k_{\rho})}{(16-5\,k_{\rho})\gamma-1}
=\frac{111-51\,k_{\rho}}{77-25\,k_{\rho}} \, ,
\end{equation}
which is valid for all $k_{\rho}$ where $\eta_{L}<1$ ($k_{\rho}>17/13$ for $\gamma=5/3$). 
Together, these form a continuous class of self-consistent solutions
for all $0\leq k_{\rho}\leq 37/17$. 

\subsection{Perturbative Calculation of the Accretion Rate}
\label{sec:perturbative}

Although the above derivation
gives a reasonable solution for
$\dot{M}$ and its scaling with time, in agreement with simulations 
\citep{H05g}, it is not strictly appropriate to adopt the Bondi-Hoyle 
rate at some radius, as the full blast wave solution represents an
outflow.  However, in the Sedov-Taylor solution, a given 
mass shell interior to the blast wave decelerates 
with time, and eventually
the gravitational potential will dominate the dynamics. Shells with 
smaller initial $r/R_{s}$ begin with smaller relative velocities and 
decelerate 
more quickly, and will successively ``turn around'' and fall back establishing
a Bondi-Hoyle-like 
steady accretion flow. Therefore, we expect a
mode of inflow in the inner regions which grows 
relative to the decaying blast wave solution, with the inflow smaller at early times 
when the blast wave is propagating rapidly (although the inflow may decay with 
time in an absolute sense). The gravitational 
term in the equation of motion is a perturbation, 
which will induce corresponding perturbations in the velocity 
and density.  We can, therefore,
describe this inflow as a perturbation about the exact 
blast wave solution. 
We consider the density and velocity perturbations 
$\delta_{\rho}\ll \rho$ and $\delta_{v}\ll v$, and demand that the 
first order perturbation to the mass inflow rate 
\begin{equation}
\delta{\dot{M}}=\delta{\dot{M}}(t)=4\,\pi\,r^{2}\,(\delta_{\rho}v+\rho\,\delta_{v})
\end{equation}
be constant in radius (although not necessarily in time). Our method and 
this assumption are identical to that used to derive the Bondi solution, but 
considered in a medium with a 
pre-existing large density and velocity field (and consequently 
first-order $|\dot{M}|=4\pi r^{2}\rho |v|\gg |\delta{\dot{M}}|$).

The first-order continuity equation is 
\begin{equation}
\frac{\partial\delta_{\rho}}{\partial t}+\frac{1}{r^{2}}\frac{\partial}
{\partial r} \delta{\dot{M}(t)}=0, 
\end{equation}
so ${\partial}\delta_{\rho}/{\partial t}=0$ and 
$\delta_{\rho}=\delta_{\rho}(r)$ because 
$\delta{\dot{M}}$ is constant in $r$.
The equation of motion 
\begin{equation}
\rho\frac{{\rm d}v}{{\rm d}t}=\rho {\Bigl(}\frac{\partial v}{\partial t}+
v\,\frac{\partial v}{\partial r}{\Bigr)}=
-\frac{\partial P}{\partial r}-\rho \frac{\partial \phi}{\partial r}
\end{equation}
(where $\phi(r)$ is the gravitational potential and represents the 
driving first-order correction) is then completely determined, if we assume the 
perturbation is adiabatic; i.e.\ if
\begin{eqnarray}
\nonumber & & \frac{\delta{P}}{P}=\gamma\,\frac{\delta{\rho}}{\rho}\\
& & \frac{P}{\rho\,v^{2}}\rightarrow \frac{a_{\tilde{P}}\,\theta_{0}}{a_{\rho} a_{v} \chi_{1} \nu_{1}^{2}}\ \ (r\rightarrow0) \, .
\end{eqnarray}
The full equation of motion then reduces to the equation of motion for the perturbation 
\begin{eqnarray}
\nonumber & &  A_{t}\,\frac{\delta{\dot{M}}}{4\pi}\,t+A_{r}\,\delta_{\rho}\,r^{3}=-\rho\,r^{2}\,t^{2}\,\frac{\partial\phi}{\partial r}\\
\nonumber & &  A_{t}=\frac{{\rm d}\ln{\delta\dot{M}}}{{\rm d}\ln{t}}+\eta\,(\ell_{\rho 1}+k_{\rho})-\nu_{0}\,(\ell_{\rho 1}+1)\\
\nonumber & &  A_{r}=\nu_{0}^{2}\,{\Bigl[}
\frac{{\rm d}\ln{\delta_{\rho}}}{{\rm d}\ln{r}}\,{\Bigl(}\frac{\gamma\,a_{\tilde{P}}\,\theta_{0}}{a_{\rho} a_{v} \chi_{1} \nu_{1}^{2}}-1{\Bigr)}
-\frac{\eta\,(\ell_{\rho 1}+k_{\rho})+1}{\nu_{0}}\\
\nonumber & & \ \ \ \ \ \ \ \ \ \ \ +(\gamma-1)
-\ell_{\rho 1}\,{\Bigl(}\frac{\gamma\,a_{\tilde{P}}\,\theta_{0}}{a_{\rho} a_{v} \chi_{1} \nu_{1}^{2}}-1{\Bigr)}
{\Bigr]}\\
& &  \nu_{0}\equiv a_{v}\,\nu_{1}\,\eta=\frac{\eta}{\gamma}.
\end{eqnarray}

\subsubsection{Exact Isothermal Sphere or Wind Solution}
\label{sec:perturbative.exact}

First, consider
$\gamma=5/3$, $\eta_{E}=0$, $k_{\rho}=k_{\rho,\,{\rm crit}}=2$, for which 
the power-law approximations we have adopted and formulae above 
are an exact solution for the internal structure of the blast wave. 
Note that for this case (as 
indicated above) our $a_{v}$ is inappropriate and the 
equations above should take
\begin{equation}
\nu_{0}=\frac{2\,\eta}{\gamma+1}=\frac{1}{2}.
\end{equation}
In this case, $\ell_{\rho}=\ell_{\rho 1}=1$, $\eta=2/3$, and the 
time dependence of $\rho r^{2} t^{2} \partial\phi/\partial r$ 
cancels completely, and since $\delta\dot{M}$ is a function only of 
time and $\delta_{\rho}$ is a function only of $r$, 
it must be true that $A_{t}\,\delta\dot{M}\,t$ goes to zero 
or a constant. Then, both reduce to the same solution, 
namely 
\begin{eqnarray}
& & \frac{{\rm d}\ln\delta\dot{M}}{{\rm d}\ln t}=-1\\
\nonumber & & \frac{1}{9}{\Bigl(}\frac{{\rm d}\ln\delta_{\rho}}{{\rm d}\ln r}+11{\Bigr)}\,\delta_{\rho}=\frac{\rho\,t^{2}}{r}\,\frac{\partial \phi}{\partial r}
=\bigfrac{4\,\rho_{0}(r_{0})\,t(r_{0})^{2}}{r_{0}}\,\frac{\partial\phi}{\partial r} \, ,
\end{eqnarray}
where $r_{0}$ is an arbitrary normalization radius. 
The equation for $\delta_{\rho}$ has a general solution with $\delta_{\rho}\propto r^{-11}$, but 
this is not of interest as the central mass diverges strongly. This leaves the 
specific solution for $\delta_{\rho}$. Let us define 
\begin{equation}
\frac{\partial \phi}{\partial r}=\tilde{\phi}\, \frac{\sigma^{2}}{a} \bigfrac{r}{a}^{-2\,k_{\phi}} \, ,
\end{equation}
where e.g.\ $k_{\phi}=1$ for the potential at short distances (dominated by the black hole)
or large distances (when the galaxy potential is effectively a point-mass) 
and $k_{\phi}\approx0$ for intermediate distances in 
e.g. a Hernquist (1990) profile. 
This gives
\begin{eqnarray}
\delta_{\rho}&=&\frac{36}{11-2\,k_{\phi}}\tilde{\phi}\,\bigfrac{\sigma^{2}\,t(r_{0})^{2}}{r_{0}\,a}
\rho_{0}(r_{0})\,\bigfrac{r}{a}^{-2\,k_{\phi}}\nonumber\\
&=&\frac{36}{11-2\,k_{\phi}}\tilde{\phi}\,\bigfrac{t(a)}{t_{\rm dyn}}^{2}\,\rho_{0}(a)\,\bigfrac{r}{a}^{-2\,k_{\phi}} \, ,
\end{eqnarray}
where the second equality comes from setting the arbitrary $r_{0}=a$ and using $t_{\rm dyn}\equiv a/\sigma$. 
We show in \S~\ref{sec:timescales} that the characteristic blowout 
timescales $t(a)$ are $\ll t_{\rm dyn}$ as defined here,  
indeed giving $\delta_{\rho}\ll\rho$ at least for early times when the 
blast wave speed exceeds the escape velocity (such a condition guarantees 
$t(a)\ll t_{\rm dyn}$). 

This solution determines a self-consistent 
power-law slope for the decay of the accretion rate $\delta{\dot{M}}$. 
The perturbation $\delta{\dot{M}}$ is independent of radius, whereas 
the blast wave outflow $\dot{M}\propto r^{2}\rho v\propto r^{4}\, t^{-3}$
falls at small $r$, meaning that the perturbation will eventually dominate 
at small $r$, as expected. The perturbations $\delta_{\rho}$ and $\delta_{v}$ 
both decline with radius, meaning that each is individually also important 
only at small radii, and vanish at large radii $r\sim R_{s}$ where the 
blast wave structure is dominant. The decline of $\delta{\dot{M}}$ with 
time is less rapid than the decline of the blast wave $\dot{M}$, reflecting the 
fact that as the blast wave expands, slows down, and cools, the gravitational 
potential and accretion solution will 
increasingly dominate (albeit still with a declining accretion rate as 
gas continues to flow out of the central regions and be shocked to 
virial temperatures).

This perturbative approach to 
calculating the accretion rate does not, however, determine the 
absolute normalization of that accretion rate, essentially an arbitrary parameter 
in the above equations, but within the radius where the steady accretion 
flow dominates, a Bondi-like spherical flow should set in and detailed 
simulations suggest that the Bondi-estimated normalization is a reasonable 
approximation \citep{H05g,FL01}. 

\subsubsection{Extension to General Mass Profiles}
\label{sec:perturbative.other}

We now extend this solution to cases where 
$k_{\rho}\ne k_{\rho,\,{\rm crit}}$ (i.e.\ $\gamma\ne5/3$ or 
$k_{\rho}\ne 2$). 
Because the power-law approximations to the density and 
velocity profiles are then not exact, the 
cancellation of the time dependence of 
$\rho r^{2} t^{2} \partial\phi/\partial r$ is 
imperfect, but we can show that the remaining factor
is small. 
If we neglect this small remainder, then a solution is 
obtained for $A_{t}=0$ or 
$A_{t}\,\delta{\dot{M}}\,t=\,$constant. 
The latter case always yields the 
\begin{equation}
\frac{{\rm d}\ln\delta\dot{M}}{{\rm d}\ln t}=-\eta_{L}=-1
\end{equation}
solution. The former ($A_{t}=0$) yields one of 
two classes of solutions. 
If the accretion feedback decouples from the 
surrounding medium after the initial injection (i.e.\ 
$\eta_{E}=0$, independent of $\eta_{L}$), then 
the general solution 
\begin{equation}
\eta_{L}=\frac{2}{\gamma}\eta=\frac{4}{\gamma\,(5-k_{\rho})}=\frac{12}{5\,(5-k_{\rho})}
\label{eqn:eta.L.decoupled}
\end{equation}
is obtained. If the accretion feedback remains coupled 
to the blast wave, i.e.\ $\eta_{E}\ne0$ for $\eta_{L}<1$, then a 
self-consistent solution is obtained with 
\begin{equation}
\eta_{L}=\frac{6}{2+\gamma\,(5-k_{\rho})}=\frac{18}{31-5\,k_{\rho}} \, .
\label{eqn:eta.L.coupled}
\end{equation}
These are similar, with $\eta_{L}$ within $\sim0.1$ for a 
given $k_{\rho}$, regardless of whether the 
accretion feedback continues to couple to the surrounding blast wave.
This implies that while such coupling may significantly affect the 
growth of the blast wave front, it does not dramatically change the 
time structure of the $r\rightarrow0$ accretion rate once the initial blast wave has grown. 

For each, a corresponding solution for $\delta_{\rho}$ is obtained, 
similar to that for $\gamma=5/3$, $\eta_{E}=0$, $k_{\rho}=2$.
In all cases, the general solution for $\delta_{\rho}$ has such a steep 
$r$ dependence that it is generally not of interest, but the specific solution 
is given by
\begin{equation}
\delta_{\rho}\propto r^{\ell_{\rho 1}-1-2\,k_{\phi}}
\end{equation}
which may increase with $r$, but always does so slower 
than the internal blast wave density profile $\propto r^{\ell_{\rho 1}}$ 
and thus satisfies the necessary perturbative conditions.
When $A_{t}\,\delta{\dot{M}}\,t=\,$constant, 
i.e.\ $\eta_{L}=1$, a second specific 
solution also exists with 
\begin{equation}
\delta_{\rho}\propto r^{-3}
\end{equation}
which may be of interest near the origin, 
and although the enclosed mass formally diverges, in reality 
the density profile will flatten once the 
perturbation dominates and a standard Bondi-type solution sets in, 
giving a Coulomb logarithm. 

We noted above that the time-dependent terms in $\rho r^{2} t^{2} \partial\phi/\partial r$
do not entirely cancel in these approximations, and thus the 
solutions above are not exact. However, the residuals
are small. If we apply the solutions above, 
then considering 
$\rho r^{2} t^{2} \partial\phi/\partial r\propto t^{\delta_{t}}$, 
we find that indeed $\delta_{t}$ is small, meaning this is not a 
bad approximation over any limited time range. 
The worst case is the $\eta_{E}\ne0$, $k_{\rho}=1$ solution, for 
which $\delta_{t}=3/8$. However, generally the cancellation is 
more complete; for example for $k_{\rho}=0$ 
we obtain $\delta_{t}\approx1/10,\ 1/20$ for the 
$\eta_{E}=0$ and $\eta_{E}\ne0$ solutions, respectively. 

\subsubsection{Late-Time Behavior of the Solution}
\label{sec:perturbative.late}

At late times, the internal blast wave velocity 
falls and the gas will not continue to cool indefinitely. 
The criterion for this 
can be determined exactly, but is essentially 
\begin{equation}
t\gg t_{\rm dyn}(r)
\end{equation}
where $t_{\rm dyn}(r)$ is the {\em local} dynamical time 
(distinct from the global dynamical time $a/\sigma$ which 
we generally refer to). The entropy of the post-shock gas is large,
and the sound speed is $\gtrsim\sigma$, so 
as the gas cools and slows, a quasi-hydrostatic equilibrium is 
established, with 
\begin{eqnarray}
\nonumber & & \frac{\partial P}{\partial r}=\frac{1}{\gamma}
\frac{\partial{\rho c_{s}^{2}}}{\partial r}\approx \rho \frac{\partial \phi}{\partial r}\\
& & c_{s}^{2}\approx-(\gamma-1)\,\phi .
\end{eqnarray}
If $\rho$ can be separated into space and time dependent parts (as 
for a power-law dependence of $\rho$ on time), the 
time dependence of $\rho$ factors out of these equations, and 
$c_{s}$ remains nearly constant at fixed $r$ while 
$\rho(r)$ slowly declines as the residual outflow
carries away mass. This approximation is valid (i.e.\ 
$c_{s}$ remains roughly constant given by the equations above 
while $\rho$ decreases) so long as $v\ll v_{\rm esc}=\sqrt{-2\phi}$. 

Then, we can approximate the residual velocity and 
decay of $\rho$ with the original blast wave solution 
extrapolated to these late times. Strictly speaking, this is 
not appropriate, as at these times the 
potential becomes important and the solutions above 
are not necessarily valid. However, we can either: (1) consider 
special cases in which they remain a valid solution, or 
(2) perform a perturbative analysis again, but this time 
the blast wave solution is the perturbation to the 
quasi-static solution on large scales, determining a net decline 
in the density of the central regions. The perturbative analysis 
demonstrates that the blast wave solution is a self-consistent 
perturbation solution for $r\ll R_{s}$, implying that 
the late-time behavior 
\begin{equation}
\dot{M}\propto \frac{\rho}{c_{s}^{3}}\propto\rho\propto t^{-\eta\,(\ell_{\rho 1}+k_{\rho})}
\end{equation}
is indeed appropriate. 
This gives 
\begin{equation}
\eta_{L}=\frac{2}{\gamma-1}\bigfrac{3-k_{\rho}}{5-k_{\rho}}=3\bigfrac{3-k_{\rho}}{5-k_{\rho}}
\end{equation}
with $\eta_{L}>1$; i.e.\ $\eta_{E}=0$ for all self-consistent
solutions, as should be true for the late stages of the accretion
rate evolution when the blast wave is weak and expanding far from the
origin.  Note that for the special case $\gamma=5/3$,
$k_{\rho}=k_{\rho,\,{\rm crit}}=2$, the proper solution becomes
$\eta_{L}=2$ (instead of $\eta_{L}=1$ as implied above, this is
because the $\ell_{\rho 1}$ is not valid in this specific case as
noted above, and the exact $\ell_{\rho}=1$ should instead be used). Appropriately, 
this solution reduces to the scaling derived by 
\citet{Shu77} for the self-similar collapse of a singular isothermal 
sphere ($k_{\rho}=2$) with no initial velocity field, as the bulk outflow 
velocity is increasingly small relative to the escape velocity. 
This solution is important for late-time galactic-scale outflows, at
low $\dot{M}$, but is less relevant for e.g.\ accretion of cold
clouds or supernova remnants, as the dense medium through which the
blast wave propagates is sub-galactic.

\section{Timescale for Accretion Rate Decay}
\label{sec:timescales}

\subsection{Merger-Driven Black Hole Growth}
\label{sec:merger.timescale}

Now, we wish to determine the appropriate timescale for the
accretion rate evolution calculated
in \S~\ref{sec:deriv}.
It is convenient to employ the dimensionless accretion rate 
$\dot{m}=\dot{M}/\dot{M}_{\rm Edd}$, with 
\begin{equation}
\Mdotedd=\frac{L_{\rm Edd}}{\epsilon_{r}\,c^{2}}=\frac{\Mbh}{\ts},
\label{eqn:define.medd}
\end{equation}
where $\epsilon_{r}\sim0.1$ is the radiative efficiency and 
$\ts=4.2\times10^{7}\,(\epsilon_{r}/0.1)\,$yr is the Salpeter 
(1964) time. 

We first consider the case of gas inflows produced during a
galaxy merger. 
The quasar light curves resulting from this process have been
discussed by
Hopkins et al.\ (2005a-e; 2006a-e); here we are specifically interested 
in the ``blowout'' phase. 
As the black hole mass increases exponentially in the 
late stages of a merger, 
a threshold is eventually crossed where feedback superheats
the surrounding gas, unbinding the 
gas before it can cool. In \citet{H05g}, we study this phase during
galaxy mergers, and find that a similarity solution with 
\begin{equation}
\mdot\propto t^{-\eta_{L}},
\label{eqn:t.powerlaw}
\end{equation}
with a typical $\eta_{L}\sim2$ approximates well the accretion 
interior to the blast wave. \citet{H05g} also find a weak dependence of $\eta_{L}$ on the 
final black hole mass, expected from the differences in external profiles as 
derived in \S~\ref{sec:deriv}. 

There are three regimes relevant to our calculation of black hole
growth.  First, there is a period of rapid (high-$\mdot$) accretion,
where accretion rate is high enough to be Eddington-limited.  This
proceeds efficiently as the black hole grows at low masses, radiating
at low enough luminosity that the nearby gas can still cool and be
accreted.

The second regime is the beginning of the blowout -- when the black
hole becomes large enough that e.g.\ the feedback energy input into
the surrounding medium in a local cooling or dynamical time is
sufficient to unbind the gas.  Effectively, the gas is ``superheated''
in the potential of the galaxy and the medium enters the Sedov-Taylor
type solution of Equation~(\ref{eqn:t.powerlaw}). Of course, since the
implied Bondi-Hoyle accretion rate is much larger than the Eddington
rate, it will take some time $t_{\dot{m}}$ for the Bondi rate to fall
to $\Mdotbondi\leq\Mdotedd$.

Beyond this point, the third regime is entered and the actual 
accretion rate begins to decline.
If $t=0$ at the beginning of the ``blowout'' (second phase) and  
$t_{\dot{m}}$ is defined as the time for the Bondi rate to fall to $\mdot_{\rm Bondi}=\mdot=1$, 
then the subsequent accretion rate is given by 
\begin{equation}
\mdot=\bigfrac{t}{t_{\dot{m}}}^{-\eta_{L}}.
\label{eqn:mdot.pwrlaw}
\end{equation}
We did not attempt to calculate the timescale $t_{\dot{m}}$ from 
our analytical model
in \citet{H05g}, as it does not affect our analysis of
the faint-end slope of the quasar luminosity function.
However, this quantity determines 
the relative contributions of e.g.\ stochastic accretion and relaxation from bright, 
high-$\mdot$ phases in mergers, so we calculate it here. 

During the blowout, the accretion rate 
drops to that given by the internal structure of the 
blast wave in Equation~(\ref{eqn:rho.cs3})
\begin{eqnarray}
\dot{m}&=&\frac{4\pi\alpha G^{2} M_{\rm BH} \rho\,\tS}{(v^{2}+c_{s}^{2})^{3/2}}\\
&=& 4\pi\alpha\chi_{1}{\Bigl(}\frac{\gamma \tilde{P}(0) \theta_{0}}{a_{\rho} \chi_{1}}{\Bigr)}^{-3/2}\,
\bigfrac{r}{R_{s}}^{\frac{5}{2}\ell_{\rho 1}} 
\frac{G^{2}M_{\rm BH}\rho_{0}(R_{s})\,\tS}{v_{s}(R_{s})^{3}}.\nonumber
\end{eqnarray}
The relevant radius for calculating the accretion rate is, 
as discussed in \S~\ref{sec:deriv}, the black hole radius of 
influence $R_{\rm BH}=G\,M_{\rm BH}\,\sigma^{-2}$. 
For convenience, we normalize the accretion rate to $R_{s}=R_{\rm BH}$ 
and define $t_{\rm BH}=t(R_{s}=R_{\rm BH})$.

The accretion rate for $t>\tacc$ is then 
$\mdot=\mdot_{0}\,(t/\tacc)^{-\eta_{L}}$, with 
\begin{equation} 
\dot{m}_{0}=4\pi\alpha\chi_{1}{\Bigl(}\frac{\gamma \tilde{P}(0) \theta_{0}}{a_{\rho} \chi_{1}}{\Bigr)}^{-3/2}\,
\frac{G^{2}M_{\rm BH}\rho_{0}\,\tS}{v_{s}(R_{\rm BH})^{3}} \, ,
\label{eqn:mdot.Rbh}
\end{equation}
and $\eta_{L}$ is as in Equations~(\ref{eqn:eta.L.decoupled})
and (\ref{eqn:eta.L.coupled}). Therefore 
$\mdot=1$ at 
\begin{equation}
t_{\dot{m}}=\tacc\,\mdot_{0}^{1/\eta_{L}}.
\label{eqn:tq.of.tacc}
\end{equation}

To solve for $\mdot_{0}$, we consider the properties of the central 
regions of the merger, for which the stellar bulge has, in general, nearly finished 
forming by the time the ``blowout'' occurs \citep{SDH05a,H05g}. 
We define a characteristic 
bulge radius $a$ by 
\begin{equation}
a\equiv\frac{G\,\Mbul}{\sigma^{2}}, 
\label{eqn:define.a}
\end{equation}
which is similar to the half-mass or half-light bulge radius (with the 
exact proportionality constant relating the two depending on the 
shape of the bulge profile). 
The definition for $\Raccr$ gives the relation $\Raccr=\mu\,a$ (where
$\mu$ is the ratio of black hole to bulge stellar mass).
We can relate the spheroid and black hole 
properties via the black hole-bulge mass 
relation of \citet{MH03} and $M_{\rm BH}-\sigma$ relation of \citet{Tremaine02} 
(see Equation~[\ref{eqn:Mbh.scalings}]), which 
then imply 
\begin{equation}
a=a_{0}\bigfrac{\sigma}{\sigma_{0}}^{2}, 
\label{eqn:a.of.sigma}
\end{equation}
\begin{equation}
\tdyn\equiv\frac{a}{\sigma}=t_{\rm dyn,\,0}\,\bigfrac{\sigma}{\sigma_{0}},
\label{eqn:tdyn}
\end{equation}
where we have defined the bulge dynamical time above and 
$a_{0}=14.5\,$kpc, $t_{\rm dyn,\,0}=7.13\times10^{7}\,$yr 
for $\sigma_{0}=200\,{\rm km\,s^{-1}}$. It is also useful to define 
an effective gas mass fraction (just before the ``blowout'')
\begin{equation}
\rho_{0}(\Raccr) \equiv \fgas \,{\rho}^{\ast}\,
\frac{3\,\Mbul}{4\pi\,a^{3}} \bigfrac{\Raccr}{a}^{-k_{\rho}},
\end{equation}
where $\fgas$ is the total bulge mass fraction in gas ($\fgas\equiv M_{\rm gas}/\mbul$), 
evaluated at the time of the ``blowout'' (usually $\sim1\%$) and 
${\rho}^{\ast}$ is a correction for the shape of the galaxy profile (i.e.\ for the fact that 
our definition of $a$ is not equivalent to the half mass radius), with e.g.\ 
${\rho}^{\ast}=[2\,(2+\sqrt{2})^{2}]/[3\,(3+\sqrt{2})^{3}]=0.07$ and 
${\rho}^{\ast}=1/(3\,\pi)=0.11$ for \citet{Hernquist90} and isothermal sphere profiles, 
respectively. This gives
\begin{equation}
G\,\rho_{0}(\Raccr)=\frac{\fgas\,{\rho}^{\ast}}{\tdyn^{2}}\frac{3}{4\pi}\bigfrac{\Raccr}{a}^{-k_{\rho}}.
\end{equation}
Combining these equations and scaling by $\cs/\sigma$ gives
\begin{equation}
\mdot_{0}=3\alpha\fgas\,{\rho}^{\ast}\,\mu \bigfrac{\ts}{\tdyn}
\bigfrac{c_{s,\,0}(\Raccr)}{\sigma}^{-3}
\bigfrac{\Raccr}{a}^{-k_{\rho}}.
\end{equation}

The energy condition 
for blowout, namely that sufficient energy be 
released to rapidly unbind the surrounding gas, 
is given by $v_{s}(\Raccr)=v_{\rm esc}$, 
with a well-defined $v_{\rm esc}=\sqrt{2}\sigma$ at $R_{\rm BH}$.
Therefore, with these choices, 
\begin{equation}
\tacc=\eta\frac{\Racc}{v_{s}(\Raccr)}=\eta\,\mu\,\tdyn\bigfrac{v_{s}(\Raccr)}{\sigma}^{-1}.
\end{equation}
We can then use Equation~(\ref{eqn:tq.of.tacc}) to determine $t_{\dot{m}}$:
\begin{eqnarray}
t_{\dot{m}}&=&\eta\,(3\alpha\fgas\,{\rho}^{\ast})^{1/\eta_{L}}\bigfrac{\tdyn}{\ts}^{1-1/\eta_{L}}
\bigfrac{v_{s}(\Raccr)}{\sigma}^{-1}\nonumber\\
& &\mu^{1+(1-k_{\rho})/\eta_{L}}\bigfrac{c_{s,\,0}(\Raccr)}{\sigma}^{-3/\eta_{L}}\,\ts \, .
\label{eqn:tq.solved}
\end{eqnarray}

For the simplest, 
self-similar example of Sedov-Taylor expansion under conditions 
resembling our simulations, we have $\eta_{L}=2$, $\eta=2/3$, which 
corresponds to the solution in e.g.\ an isothermal sphere, wind, or 
any $v\propto r$ velocity structure, and indeed this 
provides a good first-order approximation to the blowout phase 
over a wide range of conditions \citep{H05g}. In this case, we can 
simplify the above equation. We expect $v_{s}(\Raccr)\sim\sigma$, 
as is essentially guaranteed by the blowout condition that the sudden energy 
input be sufficient to unbind the gas locally. We also expect 
$c_{s,\,0}(\Raccr)\sim10\ {\rm km\,s^{-1}}$, since it is inflows
of cold 
gas from the merging disks which are being accreted. 
This gives
\begin{eqnarray}
t_{\dot{m}} &=& 4.08\times10^{6}\,{\rm yr}\ 
\bigfrac{\mbh}{10^{8}\,M_{\sun}}^{1/2}\\
& &\bigfrac{\fgas}{0.01}^{1/2} 
\bigfrac{\mu}{0.001}^{1/2}
\bigfrac{v_{s}(\Raccr)}{\sigma}^{-1}
\bigfrac{c_{s,\,0}(\Raccr)}{10\,{\rm km\,s^{-1}}}^{-3/2}.\nonumber
\end{eqnarray}
Here, we have again used the $\Mbh-\sigma$ relation to replace 
the $\sigma$ dependence with an $\mbh$ term.
Direct comparison with the best-fit to the blowout in our simulations 
shows such a trend, with a scatter of a 
factor $\sim2$ about the relation above. 
This scatter is smaller than that implied by the
various
factors above, but this may be because several are correlated. 
%For example, in calculating the threshold for blowout, 
%it is likely that $\mu\propto 1/\fgas$ \citep[see, e.g.,][]{SR98,WL03,Murray05}, 
%meaning that their respective scalings cancel in the final 
%blowout timescale. 

\subsection{Molecular Cloud Accretion}
\label{sec:cloud.timescale}

Next, consider the accretion of a molecular 
cloud. As we describe in \S~\ref{sec:rates}, we expect
clouds to be moving at a speed $\sim10\,{\rm km\,s^{-1}}$ 
near the black hole, with characteristic gas 
densities $n\sim100\,{\rm cm^{-3}}$. 
This gives an initial Bondi-Hoyle accretion rate $\dot{m}\gg1$ 
(limited to $\dot{m}=1$) for 
all black hole masses of interest. Provided that the black hole has grown 
large enough to lie on the $M_{\rm BH}-\sigma$ 
relation (see \S~\ref{sec:m.sigma}), it will release sufficient
feedback
to unbind the molecular cloud in a 
timescale $\lesssim10^{7}\,$yr. Because this is 
small compared to the Salpeter time, the cloud 
crossing time, the cloud dynamical time, and the 
timescale for the moderate and low-accretion rate phases 
of the ``blowout'', we can effectively treat such an 
event as if it enters the blowout phase immediately upon 
the ``collision'' with the cloud. 

The calculation of the relevant timescale for the accretion rate decay is identical 
to that in \S~\ref{sec:merger.timescale}, since the typical cloud radius 
is larger than the black hole radius of influence, especially at the relatively small masses of interest. 
However, the surrounding medium is not the same, and the 
condition for ``blowout'' is modified by the need to unbind a local cloud of 
cold gas as opposed to the entire gas content of central regions
of the galaxy.
For a molecular cloud, we expect $k_{\rho}=0$ (at least 
effectively, since the black hole does not necessarily lie at the cloud center), 
$\eta=2/5$, and $\rho_{0}(R_{\rm BH})=\rho_{0}=m_{p}\,n_{\rm cl}$ 
with $n_{\rm cl}\sim100\,{\rm cm^{-3}}$. The speed $v_{s}$ at some 
radius and time is given by 
$v_{s}=\eta R_{\rm BH}/t$. We again define $t_{\rm BH}$ 
as the time at which $R_{s}=R_{\rm BH}$.

The similarity condition for the blast wave gives 
\begin{equation}
R_{s}={\Bigl[}\frac{\xi E_{b}(R_{s})}{\bar{\rho}(R_{s})}{\Bigr]}^{1/5}\,t^{2/5}
\label{eqn:similarity}
\end{equation}
where $E_{b}$ is the blast wave energy, $\bar{\rho}=\rho_{0}$ is the 
mean density inside the blast wave (necessarily equal to the mean density 
outside by mass conservation and since $k_{\rho}=0$), and 
\begin{eqnarray}
& & \xi=\frac{3}{4\pi\eta^{2}\sigma_{b}}\\
\nonumber & & \sigma_{b}=E_{b}/M\,v_{s}^{2}
\end{eqnarray}
are generally constant for a self-similar blast wave (and 
$\sigma_{b}\approx1$ for the early stages of the 
blast wave where $v_{s}\gg v_{\rm esc}$). 

Consider first
the simple case of an energy-conserving blast wave, in which 
the accretion rate declines quickly enough that the energy 
input is dominated by the initial accretion (or applicable if 
e.g.\ the ionization or initial blowout of the inner regions 
de-couples the blast front from the black hole feedback). 
Simulations of particular wind mechanisms find that driving 
is efficient for $\dot{m}\gtrsim0.1$ \citep[e.g.,][]{BalsaraKrolik93}, 
which also suggests this is an accurate approximation. 
The blowout condition essentially defines the 
blast wave energy, as that required to unbind and expel the cloud
\begin{equation}
E_{b}\gtrsim E_{\rm binding}\sim M_{\rm cl}\,\sigma^{2}.
\end{equation}
This implies $\eta=2/5$, and plugging this in 
we obtain 
\begin{equation}
v_{s}=\eta\frac{R_{s}}{t}=\sigma\,f_{b}^{1/2}\,\bigfrac{R_{cl}}{R_{s}}^{3/2}
\end{equation}
where 
\begin{equation}
f_{b}\equiv \bigfrac{M\,v_{s}^{2}}{E_{\rm binding}}
\end{equation}
is approximately constant, at least until the late stages of the 
blast wave evolution. 

Using this result in Equation~(\ref{eqn:mdot.Rbh}), 
again normalizing to $R_{\rm BH}$, we obtain
an equation for $\dot{m}(t)$. We again re-normalize to 
$\dot{m}=(t/t_{\dot{m}})^{-\eta_{L}}$, 
giving $t_{\dot{m}}=t_{\rm BH}\,\dot{m}_{0}^{1/\eta_{L}}$.
Solving gives 
\begin{eqnarray}
\nonumber & & t_{\dot{m}}=\frac{R_{\rm BH}}{\sigma}\,\frac{\eta}{f_{b}^{1/2}}
\dot{m}_{0}^{1/\eta_{L}}
\bigfrac{R_{\rm BH}}{R_{\rm cl}}^{3\,(\eta_{L}^{-1}-1)/2}\\
& & \dot{m}_{0}\equiv
\frac{4\pi\alpha\chi_{1}}{f_{b}^{3/2}}
{\Bigl(}\frac{\gamma \tilde{P}(0) \theta_{0}}{a_{\rho} \chi_{1}}{\Bigr)}^{-3/2}\,
\frac{G^{2}\,M_{\rm BH}\,\rho_{0}\,\tS}{\sigma^{3}} \, .
\label{eqn:tmdot.full}
\end{eqnarray}
Finally, we use $k_{\rho}=0$, $\eta_{E}=0$, adopt the 
$M_{\rm BH}-\sigma$ relation as an initial condition, 
and take $\gamma=5/3$ ($\alpha=1/4$ for the 
Bondi solution) to obtain 
\begin{eqnarray}
& & t_{\dot{m}}=0.40\,\frac{R_{\rm cl}}{\sigma}\,\tau_{\ast}\\
\nonumber & & \dot{m}_{0}=1.67\,f_{b}^{-3/2}\,\bigfrac{\sigma}{100\,{\rm km\,s^{-1}}}\\
\nonumber & & \tau_{\ast}=f_{b}^{-19/18}\,\bigfrac{\sigma}{100\,{\rm km\,s^{-1}}}^{13/27}\,
\bigfrac{n_{\rm cl}}{100\,{\rm cm^{-3}}}^{10/27}\,\bigfrac{R_{\rm cl}}{100\,{\rm pc}}^{-1/18}
\label{eqn:tmdot.calc}
\end{eqnarray}
where the $\tau_{\ast}$ term is a small, order unity correction which 
depends weakly on the 
properties of the system. 

This more detailed analysis 
recovers the expected scaling, that the characteristic timescale for 
accretion decay is given by the cloud size divided by the local escape 
velocity;
i.e.\ approximately the velocity at which the cloud will 
be unbound and expelled. Although we have solved this 
in detail for the $\gamma=5/3$, $k_{\rho}=0$ case, 
a similar calculation for different values of $\gamma$, $k_{\rho}$, 
$\eta_{L}$, and $\eta_{E}$ gives a nearly identical equation 
for $t_{\dot{m}}$, with small, order unity numerical changes to the 
coefficient of $R_{\rm cl}/\sigma$ as expected from 
Equation~(\ref{eqn:tmdot.full})
and slightly different power-law coefficients for the terms in $\tau_{\ast}$, 
with similar weak dependencies on $\sigma$, $n_{0}$, and $R_{\rm cl}$ 
in $\tau_{\ast}$.

One important point to check is that our approximation of a 
constant $M_{\rm BH}$ is reasonable. 
For $t_{\dot{m}}$ 
we have 
\begin{equation}
t_{\dot{m}}\approx3.9\,\tau_{\ast}\times10^{5}\,{\rm yr}\,
\bigfrac{R_{\rm cl}}{100\,{\rm pc}}\,\bigfrac{\sigma}{100\,{\rm km\,s^{-1}}}^{-1}
\end{equation}
which is much less than the Salpeter time ($t_{\dot{m}}\sim10^{-2}\,\tS$), 
and so the black hole mass is roughly constant during blowout. 
This also means that once the system enters blowout,
the accretion rate 
drops below $\dot{m}=1$ immediately, and so a pure-power 
law decay is a good approximation when considering the 
probability of viewing the black hole at a given $\dot{m}$. 

For the luminosity-dependent Seyfert lifetime, a power-law  
accretion time history of the form in Equation~(\ref{eqn:mdot.pwrlaw})
gives 
\begin{equation}
\frac{{\rm d}t}{{\rm d}\log\dot{m}}=\eta_{L}\,t_{\dot{m}}\,\ln{10}\,
\dot{m}^{-1/\eta_{L}}
\label{eqn:lifetime}
\end{equation}
which we can use to infer the probability of 
observing a system at a given accretion rate. Since 
$M_{\rm BH}$ is roughly constant once the 
blowout begins, this can be directly 
translated to a distribution in luminosity using 
$L=\dot{m}\,L_{\rm Edd}=
\dot{m}\,\epsilon_{r}\,M_{\rm BH}\,\tS^{-1}\,c^{2}$.

\section{Accretion of Molecular Clouds}
\label{sec:mol.cloud.accr}

\subsection{Rate of Cloud-Collision Events}
\label{sec:rates}

If we consider a black hole in a system with a significant disk or other 
cold gaseous component which extends near the center, then there 
will be some probability for the black hole to cross paths with a dense molecular 
cloud. 
The timescale between collisions with clouds is
\begin{equation}
t_{\rm event}=(n\,\pi R_{\rm cl}^{2}v_{\rm cl})^{-1}
\end{equation}
where $n=N_{\rm cl}/V$ is the number density of cold 
clouds, $R=R_{\rm cl}$ is their typical size (giving 
a cloud cross-section $\pi R_{\rm cl}^{2}$), and $v_{\rm cl}$ is their 
characteristic random velocity. Although the gravitational interaction 
with the black hole is a long-range force, the effective cross section for 
interaction with a cloud will be $\approx \pi R_{\rm cl}^{2}$ if the 
black hole radius of influence, 
\begin{equation}
R_{\rm BH}\equiv\frac{G\,M_{\rm BH}}{\sigma^{2}}\approx 1\,{\rm pc}\,
\bigfrac{M_{\rm BH}}{10^{6}\,M_{\rm sun}}^{1/2}
\end{equation}
is much smaller than the cloud size, which for a typical large molecular cloud 
with $R_{\rm cl}\sim10-100\,$pc is satisfied for all black hole masses of interest, 
especially 
for the small $M_{\rm BH}$ characteristic of late-type systems.  

The typical random cloud velocity is essentially the vertical velocity dispersion 
of the disk, $v_{\rm cl}=c_{s}^{\rm disk}\sim10\,{\rm km\,s^{-1}}$, 
in order to maintain pressure support.
In any case, the value of $c_{s}^{\rm disk}$ is relatively 
unimportant as we demonstrate that it ultimately cancels below. The cloud 
number density is
\begin{equation}
n=\frac{N}{V}=\frac{1}{V_{\rm cl}}\frac{N\,V_{\rm cl}}{V}=\frac{v_{ff}}{(4\pi/3)\,R_{\rm cl}^{3}},
\end{equation}
where $v_{ff}$ is the volume filling factor of cold clouds, so 
\begin{equation}
t_{\rm event}\sim \frac{1}{v_{ff}}\frac{R_{\rm cl}}{c_{s}^{\rm disk}}
\approx 10^{10}\,{\rm yr}\,\bigfrac{v_{ff}}{0.001}^{-1}\bigfrac{R_{\rm cl}}{100\,{\rm pc}}
\bigfrac{c_{s}^{\rm disk}}{10\,{\rm km\,s^{-1}}}^{-1}.
\end{equation}
Therefore, a large
fraction of black holes in disk-dominated systems should 
have undergone such an event, although we determine the effective ``duty cycle'' 
more completely below. 

Next, we consider how $v_{ff}$ and $t_{\rm event}$ scale with 
host galaxy properties.
Because most of the gas in the ISM resides in cold clouds, the 
cloud filling factor is
\begin{equation}
v_{ff}=\frac{\rho_{\rm ISM}}{\rho_{\rm cl}}\approx\frac{n_{\rm ISM}}{n_{\rm cl}}
\end{equation}
(where $\rho_{\rm ISM}$ and $\rho_{\rm cl}$ are the average ISM and cloud gas densities, 
respectively)
independent of the size distribution or shapes of clouds. 
For an exponential disk with total gas mass $M_{\rm gas}=f_{\rm gas}\,M_{\rm d}$, 
the surface density profile is
\begin{equation}
\Sigma_{\rm gas}(r)=\frac{f_{\rm gas}\,M_{\rm d}}{2\pi\,r_{\rm d}^{2}}
\exp(-r/r_{\rm d}) \, .
\end{equation}
In the central regions of the disk (taking $r\rightarrow0$, $z\rightarrow0$), 
this resembles
an isothermal sheet with constant scale height $h$, 
\begin{equation}
\rho_{\rm ISM}=\frac{f_{\rm gas}\,M_{\rm d}}{4\pi\,r_{\rm d}^{2}\,h}.
\end{equation}
The scale height $h$ is in general given by $h\sim r_{\rm d}\,(c_{s}^{\rm disk}/v_{c})$, 
where $v_{c}$ is the circular velocity. As $r\rightarrow0$, the potential 
is dominated by the bulge, i.e.\ $v_{c}(\rm disk)\rightarrow0$ while 
the bulge $\sigma\rightarrow\,$constant, so
$h\sim r_{\rm d}\,(c_{s}^{\rm disk}/\sigma)$, giving 
\begin{equation}
\rho_{\rm ISM}=\frac{f_{\rm gas}\,M_{\rm d}}{4\pi\,r_{\rm d}^{3}}\,
\frac{\sigma}{c_{s}^{\rm disk}}.
\end{equation}
For a disk formed by collapse at fixed baryon fraction and 
spin parameter $\lambda$, $M_{\rm d}\propto r_{\rm d}^{3}$
and the quantity 
\begin{equation}
\frac{1}{m_{p}}\frac{M_{\rm d}}{4\pi\,r_{\rm d}^{3}}
=\frac{1}{m_{p}}\frac{M_{\rm d}}{4\pi\,r_{\rm d}^{3}}{\Bigr|}_{\rm MW}
\approx 0.4\,{\rm cm^{-3}}
\end{equation}
is constant across disks, 
where we take $M_{d}=10^{11}\,M_{\sun}$ and 
$R_{d}=9\,$kpc for the Milky Way in normalizing this; 
i.e.\ essentially a baryon fraction 
$m_{b}\approx0.05$ and spin parameter $\lambda=0.04$ for 
$V_{\rm vir}\approx160\,{\rm km\,s^{-1}}$. 
Note that for this $V_{\rm vir}$, calculating the 
gas density with $f_{\rm gas}=0.1$ at $r=8\,$kpc (including 
the scaling of circular velocity, scale height, and 
surface density to this radius) gives the 
standard $n_{\rm ISM}\approx0.3\,{\rm cm^{-3}}$ 
for the local ISM. 

For a standard cloud density $n_{\rm cl}=100\,{\rm cm^{-3}}$, 
this then gives for the volume filling fraction at the center of the disk 
\begin{equation}
v_{ff}=\frac{\rho_{\rm gas}}{\rho_{\rm cl}}
=f_{\rm gas}\,\tilde{n}\,\frac{\sigma}{c_{s}^{\rm disk}}
\end{equation}
where 
\begin{eqnarray}
\tilde{n}&\equiv& \frac{1}{m_{p}\,n_{\rm cl}}\,
\frac{M_{\rm d}}{4\pi\,r_{\rm d}^{3}}\nonumber \\
&\approx& 4.0\times10^{-3}
\bigfrac{n_{\rm cl}}{100\,{\rm cm^{-3}}}^{-1}
%\bigfrac{m_{b}}{0.05}
%\bigfrac{\lambda}{0.04}^{-3}
\end{eqnarray}
is approximately constant. 
The characteristic timescale for a collision 
with a molecular cloud is then 
\begin{equation}
t_{\rm event}\sim\frac{1}{f_{\rm gas}\,\tilde{n}}\,\frac{R_{\rm cl}}{\sigma}
\end{equation}
and the factors of $c_{s}^{\rm disk}$ cancel
(as a higher $c_{s}^{\rm disk}$ will increase the 
speed of clouds and thus number of collisions, but also 
``puff up'' the disk and reduce the density of cold clouds). 
In addition to producing an appropriate Eddington ratio distribution, 
there is observational support for stochastic cloud collisions 
with AGN on such a timescale, as e.g.\ dust clouds in the central regions 
of nearly low-luminosity AGN appear to trace accretion events with a
similar frequency \citep{Lauer05}.

The expected spectrum of
cloud sizes is $n(R\rightarrow R+dR )\propto R^{-4}\,dR$ 
so that the filling factor for clouds in some interval of 
$R$ goes as $dR/R\propto d\log{R}\propto d\log{M_{\rm cl}}$, 
and there are equal contributions to 
$v_{ff}$ from each logarithmic interval in cloud size from 
the smallest 
$\sim1\,$pc to the largest
$\gtrsim100\,$pc. Therefore, a 
more complete calculation of $t_{\rm event}$ considering
the probability of collision with clouds in each size interval 
gives only a logarithmic $\ln{(R_{\rm cl,\,max}/R_{\rm cl,\,min})}$
correction to the rate at which clouds collide with the black hole and ultimate 
the ``duty cycle'' of activity. We subsume this factor into $\tilde{n}$, as 
it is within the present uncertainties. However, this 
range of cloud masses can be important for e.g.\ the 
scatter in the expected $M_{\rm BH}-\sigma$ relation 
from events fueled in this fashion. 

Finally, the mass of a cloud needed for an accretion 
``event'' is quite
modest (see \S~\ref{sec:mass.blowout}), as low as 
$\sim10^{5}\,M_{\sun}$, and the ultimate duty cycle derived in 
\S~\ref{sec:duty} is effectively independent of the mass of cold gas inflow. 
Therefore, only a small fraction of the $\sim10^{8}-10^{9}\,M_{\sun}$ of 
cold molecular gas typically concentrated within the central regions of 
late-type galaxies \citep{Kaneko89,Heckman89,Meixner90,
GDF97,Galliano03,Mason05,Elitzur05} needs to pass randomly
near the black hole, and from our derivation above events 
are expected only of order once per Hubble time. With these considerations, 
there is no ``angular momentum problem,'' at least on the scales of interest. 

Unlike bright quasars, which require large gravitational torques to 
sustain high accretion rates $\dot{m}\sim1$ for $\sim10^{7}-10^{8}$\,yr 
with large $M_{\rm BH}\gtrsim10^{8}\,M_{\sun}$ masses, 
no disturbance to the galactic gas, even in the central disk, is required for 
these quiescent Seyferts. Of course, processes such as 
such as disk and bar instabilities \citep[e.g.,][]{NormanSilk83,NormanScoville88,
SBF89,Lin88,Lubow88}, minor mergers \citep[e.g.,][]{Roos81,Gaskell85,
Hernquist89,MH94,HM95,DeRobertis98,Tanaguchi99}, 
or magnetic instabilities \citep[e.g.,][]{KrolikMeiksin90} may 
nevertheless play 
a role, and it is still of interest to understand in detail
the transport of material from scales of order tens of pc considered 
here to the small $\sim$AU sizes of an accretion disk, but 
these processes, at least on the large scales we consider here, will 
ultimately serve to increase the effective random (non-rotational) 
velocity dispersion of clouds.

There are however two limits in which this activity will be suppressed, 
i.e.\ $t_{\rm event}$ can be become much larger than a Hubble time. 
The first, in which $f_{\rm gas}\rightarrow0$ and there is simply no 
gas supply to fuel this mode of accretion, is straightforward. The second, 
in which $\sigma\rightarrow0$, implies that the central gas in 
the AGN is not dynamically hot -- i.e.\ it has relatively little disordered motion 
and there is no disturbance which can bring gas to the black hole. Such 
systems will also have very small black holes, given the $M_{\rm BH}-\sigma$ 
relation, so their contribution to the observed Seyfert luminosity 
function (which we discuss in \S\ref{sec:morph.contrib}) will be small. 

\subsection{Light Curves and Duty Cycles}
\label{sec:duty}

When colliding with a molecular cloud, a 
black hole on the $M_{\rm BH}-\sigma$ relation
will immediately enter the ``blowout'' phase. 
Even if the black hole is slightly 
undermassive and accretes
before entering this phase, this time is short compared to 
the eventual time at low or moderate accretion rates
(see the discussion in \S~\ref{sec:m.sigma.slope} and \ref{sec:low.m.timescale}), and 
the distribution of duty cycles can be well approximated 
by neglecting these times. 
In \S~\ref{sec:cloud.timescale} we determined that the 
light curve in such an event is given by 
$\dot{m}=(t/t_{\dot{m}})^{-\eta_{L}}$ (Equation~[\ref{eqn:mdot.pwrlaw}]), 
where $t_{\dot{m}}\propto R_{\rm cl}/\sigma$ is given 
in Equation~(\ref{eqn:tmdot.calc}). 
This yields 
a differential time per logarithmic interval in luminosity, 
$dt/d\log{\dot{m}}\propto t_{\dot{m}}\,\dot{m}^{-1/\eta_{L}}$, given in Equation~(\ref{eqn:lifetime}). 
The effective ``duty cycle'' for an object as a function of 
the accretion rate $\dot{m}$ is then 
\begin{eqnarray}
\frac{{\rm d}f}{{\rm d}\log{\dot{m}}}&=&
\frac{{\rm d}n_{\rm event}}{{\rm d}t}\frac{{\rm d}f}{{\rm d}\log{\dot{m}}}\nonumber\\
&=&\eta_{L}\frac{t_{\dot{m}}}{t_{\rm event}}\,\ln10\ \dot{m}^{-1/\eta_{L}}
\equiv \delta_{0}\ \dot{m}^{-1/\eta_{L}} \, .
\label{eqn:dfdlogmdot}
\end{eqnarray}
Using our solutions for $t_{\rm event}$ (\S~\ref{sec:rates}) 
and $t_{\dot{m}}$ (\S~\ref{sec:cloud.timescale}), we obtain
\begin{equation}
\delta_{0}\approx10^{-3}\,\bigfrac{f_{\rm gas}}{0.1}\,\eta_{L}\,\tau_{\ast}\,\ln10 \, ,
\label{eqn:duty.calc}
\end{equation}
where $\tau_{\ast}$ is defined in Equation~(\ref{eqn:tmdot.calc}).

In general, $\eta_{L}\sim1$, so we can easily infer some
important properties of the duty cycle. First, this implies a 
duty cycle at large accretion rates $\dot{m}\gtrsim0.1$ of 
$\sim1\%$, similar to that estimated observationally from 
e.g.\ \citet{Kauffmann03,YLK05,Dong05}. The duty cycle becomes large 
at lower accretion rates, going to $\sim1$ at $\mdot\lesssim10^{-3}$, 
again similar to that measured observationally from 
e.g.\ \citet{Hao05,Best05} who find a large fraction of late-type galaxies 
hosting moderate/low accretion rate Seyferts. Note, however, that technically these are theoretical 
upper limits to the duty cycles, for if accretion proceeds intermittently (i.e.\ 
in short, potentially super-Eddington ``bursts'') the same average accretion 
rate on the timescales relevant for our calculations is maintained, although 
the timescale for such bursts is still constrained by the observed episodic 
quasar lifetime \citep[see e.g.,][]{Martini04}.

This also implies an effective minimum accretion rate 
$\dot{m}_{\rm min}$, at which 
the total time spent with $\dot{m}>\dot{m}_{\rm min}$ is 
equal to $t_{\rm event}$. This is not to 
say that this is a hard minimum to the Seyfert accretion rate, 
but rather that by the long timescales expected for decay to 
accretion rates much below this, a subsequent collision 
with a molecular cloud is expected, which will ``reset'' the system to a 
high accretion rate. In detail, for $\dot{m}<\dot{m}_{\rm min}$ with 
Poisson statistics for the excitation rates, 
an exponential cutoff is expected to introduce a term 
$\exp[-({\rm d}t/{\rm d}\log{\dot{m}})/t_{\rm event}]$. 
Because this is a rapid cutoff, we can temporarily
approximate it as an absolute cutoff and determine 
$\dot{m}_{\rm min}$ as where the duty 
cycle $f=\int {\rm d}f\rightarrow 1$. 
This gives 
\begin{equation}
\dot{m}_{\rm min}=\bigfrac{\eta_{L}\,\delta_{0}}{\ln 10}^{\eta_{L}}
={\Bigl(}\eta_{L}^{2}\,\frac{t_{\dot{m}}}{t_{\rm event}}{\Bigr)}^{\eta_{L}} \, ,
\label{eqn:mdot.min}
\end{equation}
which for $\eta_{L}\lesssim1$ gives $\dot{m}_{\rm min}\lesssim10^{-3}$.

Finally, note that the terms involving cloud sizes have completely 
canceled in this derivation. Thus, while the cloud size may be important for 
e.g.\ scatter in the $M_{\rm BH}-\sigma$ relation (see \S~\ref{sec:m.sigma}), 
it does not enter into our ultimate calculation of the Seyfert luminosity 
function and distribution of accretion rates. Therefore, the considerable 
uncertainties in the properties of clouds at the centers of galaxies, and the 
``typical'' giant cloud size related to Seyfert activity (and the immediate 
source of such clouds) do not affect our calculations. Rather it is 
the well-constrained quantities such as gas fraction and 
our theoretically determined $\eta_{L}$ which determine these predictions. 

\subsection{Seyfert Luminosity Function}
\label{sec:LF}

From the duty cycle as a function of accretion rate, 
we can determine the Seyfert luminosity function 
implied by this mode of fueling. 
We assume the 
black hole mass remains roughly constant 
during the blowout (see \S~\ref{sec:mass.blowout}), 
giving ${\rm d}f/{\rm d}\log{L}={\rm d}f/{\rm d}\log{\dot{m}}$ 
with $L=\dot{m}\,\epsilon_{r}\,M_{\rm BH}/\tS\,c^{2}$, 
and assume a constant $\epsilon_{r}=0.1$, 
as expected for a standard \citet{SS73} thin disk. 

To derive the expected Seyfert luminosity function, then, 
we require the distribution of black hole masses and corresponding host 
galaxy properties, in particular 
their masses, gas fractions, and velocity dispersions. 
We use the local galaxy luminosity functions 
from the CfA redshift survey in $B$-band \citep{Marzke94a,Marzke94b} 
separately determined for each morphological classification of 
E, S0, Sa/b, Sc/d, and Sm/Im. We convert these to $K_{20}$ 
luminosity functions with the conversions for each type from 
this survey as in \citet{Kochanek01} (see their Table 5), and 
then to mass functions with the typical $K$-band mass-to-light estimates 
based on mass and type from \citet{Bell03}, which incorporate 
corrections for galaxy evolution and detailed stellar population synthesis 
from the models of \citet{FRV97}. 

Grouping the E and S0 galaxies as ``early-type'' and Sa/b, Sc/d, and
Sm/Im galaxies as ``late-type,'' we compare these directly to the
$K$-band luminosity functions of \citet{Kochanek01}, and find
agreement (as the authors derive). We similarly compare to the mass
functions of early and late-type galaxies determined by \citet{Bell03}
in both $g$ and $K$ band, and find reasonable agreement.  Because the
more recent \citet{Bell03} mass functions are estimated from the much
larger combined SDSS-2MASS local galaxy sample, with more detailed
correction for stellar mass-to-light ratios, we re-normalize
our mass functions slightly (a small $\lesssim30\%$
correction) to reproduce the \citet{Bell03} late and early type mass
functions.

Next, we estimate the 
bulge-to-disk ratio of each morphological type following 
\citet{AllerRichstone02,Hunt04a} in the $B$-band and $H$-band, 
respectively (adopting a similar procedure to correct to a mass ratio). 
This gives a bulge-to-total mass ratio of approximately 
$B/T=(1.0,\,0.57,\,0.19,\,0.077)$ [$\pm(-,\,0.2,\,0.4,\,0.5)$ dex]
for galaxies of types (E, S0, Sa/b, Sb/c), respectively. 
The bulge-to-total mass ratio of Sm/Im galaxies is uncertain, and 
these galaxies may have no bulges whatsoever, but in any case 
we find below that their contribution is sufficiently small 
that they can be neglected even taking a maximal 
$B/T$ for such systems. 

From the stellar mass function and $B/T$ ratio for each morphological
category, we construct a cumulative bulge or disk mass function
and mean $B/T$ as a function of mass. We do so and compare with the
bulge and disk mass functions of \citet{TascaWhite05}, and the
estimated mean $B/T$ as a function of mass from the size and surface
brightness analysis in \citet{Shen03}, and find agreement in both
cases, suggesting that this decomposition is reasonable. Although more
detailed properties such as the mean disk gas fraction $f_{\rm gas}$
are generally unnecessary for our subsequent analysis, we determine
them from the compilations of
\citet{RobertsHaynes94} and \citet{Kauffmann03b}.

From the bulge mass function, the 
$M_{\rm BH}-M_{\rm bulge}$ relation \citep{Magorrian98,MH03,HaringRix04} 
then determines the black hole mass function. We adopt the relation of 
\citet{MH03}, which is essentially identical to that
measured in simulations of spheroid and black hole formation 
\citep{Robertson05b}, and is also equivalent to the 
observed $M_{\rm BH}-\sigma$ relation determined in 
\citet{Tremaine02}, given the relation between radius and 
stellar mass determined in \citet{Shen03}. 
This gives
\begin{eqnarray}
\nonumber & & M_{\rm BH}=0.001\,M_{\rm bulge}\\
& & M_{\rm BH}=10^{8.13}\,M_{\sun}\,\bigfrac{\sigma}{200\,{\rm km\,s^{-1}}}^{4.02}.
\label{eqn:Mbh.scalings}
\end{eqnarray}
The scatter in these relations is observed (and determined in simulations)
to be $\approx0.35-0.4$ and $0.3,\,$dex, 
respectively (see also \citet{Novak05}). 
We assume that the PDF for black hole mass at a given bulge mass is 
distributed lognormally, with a dispersion equal to the above dispersions. 
For each bulge mass, then, we convolve the bulge mass function with 
the PDF for black hole mass, and determine the resulting black hole 
mass function (BHMF). 

\begin{figure}
    \centering
    \plotone{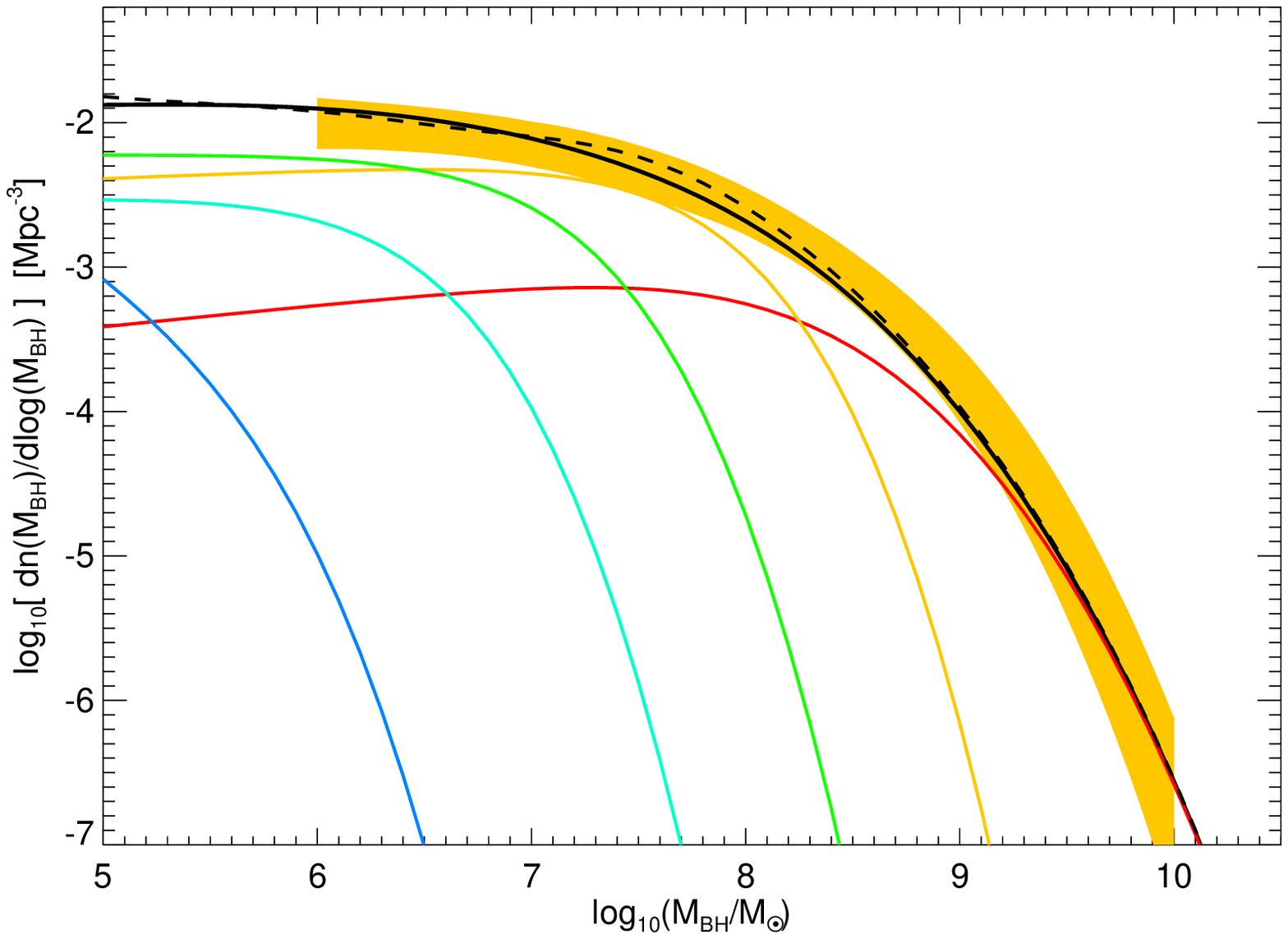}
    \caption{Inferred black hole mass function from the morphological 
    mass functions of \citet{Marzke94a,Marzke94b,Kochanek01,Bell03}, 
    with the typical bulk-disk decompositions of \citet{AllerRichstone02,
    Shen03,Hunt04a,TascaWhite05} and $M_{\rm BH}-M_{\rm bulge}$ relation of 
    \citet{MH03,Robertson05b} or, equivalently, 
    $M_{\rm BH}-\sigma$ relation from \citet{Tremaine02}. The mass function from 
    E (red), S0 (yellow), Sa/b (green), Sc/d (cyan), Sm/Im (blue; assumes a 
    maximal bulge B/T$=0.02$), and all types (black solid) are shown, and 
    compared to the $1\sigma$ range of the inferred mass function from 
    \citet{Marconi04}, based on the observations of 
    \citet{Marzke94a,Kochanek01,Nakamura03,Bernardi03,Sheth03} (yellow shaded 
    range). The results adopting the modified $M_{\rm BH}-\sigma$ relation 
    suggested in \S~\ref{sec:m.sigma} are also shown (black dashed).
    \label{fig:bhmf}}
\end{figure}

In Figure~\ref{fig:bhmf}, we show the resulting BHMF, calculated 
individually for each morphological category of 
E (red), S0 (yellow), Sa/b (green), Sc/d (cyan), and Sm/Im (blue). 
The cumulative BHMF obtained by summing these 
contributions is shown as the black line. For comparison, the 
BHMF determined in \citet{Marconi04}
is shown as the shaded yellow range, which
also agrees with other measurements
of the BHMF by e.g.\ 
\citet{Salucci99,MS02,YT02,Ferrarese02,AllerRichstone02,Shankar04}. 
That the agreement with our estimate is good suggests that our 
adopted conversions and decompositions are reasonable.
For the Sm/Im case, we have assumed a maximal 
$B/T=0.02$, only a factor of $\approx3$ below that 
of the Sc/d galaxies, and it is apparent that the contribution to the 
integrated BHMF and number density at any mass of interest is negligible. 
Therefore, we ignore these galaxies in our 
subsequent analysis, as their $B/T$ is in detail quite 
uncertain (if it is non-zero at all). 

We now have the observed BHMF, with the corresponding gas fraction,
bulge size and velocity dispersion, and disk mass and size for each
system.  Our calculations above then
allow us to estimate the rates, duty cycles, and lifetimes of Seyfert
activity fueled by the quiescent accretion of cold gas, and the
corresponding Seyfert luminosity function.

For each black hole (i.e.\ each point in the joint 
PDF of $M_{\rm BH},\ \sigma,\ f_{\rm gas},\ M_{\rm d}$, 
and host galaxy type), we convolve the black hole distribution with the 
duty cycle as a function of these properties. In other words, the 
observed luminosity function is given by 
\begin{equation}
\phi(L)\equiv\frac{{\rm d}\Phi(L)}{{\rm d}\log{L}}
=\int \frac{{\rm d}f}{{\rm d}\log{L}}(\bar{x})\,{\rm d}n(\bar{x})
\end{equation}
where $\bar{x}=(M_{\rm BH},\,f_{\rm gas},\,\sigma,\,M_{\rm d})$, 
for example. Again, we have $L\propto\dot{m}$ 
so ${\rm d}\log{L}={\rm d}\log{\dot{m}}$. 
Given our blast wave solution for ${\rm d}f/{\rm d}\log{\dot{m}}$, 
the distribution of host galaxy properties then completely 
determines the predicted Seyfert luminosity function 
(insofar as it is attributable to this fueling mechanism).

\begin{figure*}
    \centering
    \plotone{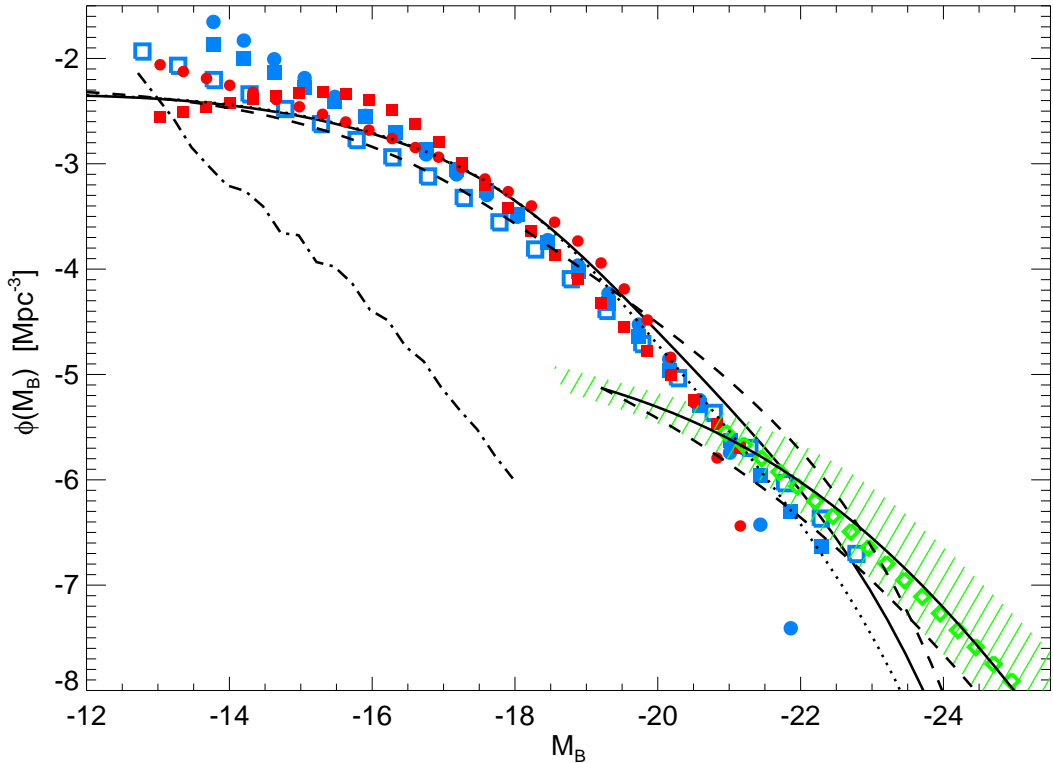}
    \caption{Predicted intrinsic 
    Seyfert luminosity function from our blast wave model.
    Solid black line extending to the upper left
    shows the prediction assuming $\eta_{L}=18/31$ as calculated 
    in \S~\ref{sec:deriv}, dashed and dotted lines show the predictions varying $\eta_{L}$ 
    to $\eta_{L}=1/2$ and $\eta_{L}=1$, respectively. 
    The observed SDSS Seyfert luminosity function from narrow-line H$\alpha$
    (i.e.\ isotropic, intrinsic luminosity function) from \citet{Hao05} is shown 
    (blue), with both a double-power law (squares) and Schechter function (circles) 
    fit to the observations, converted to intrinsic $B$-band using the 
    corrections derived by the authors (filled) and a constant conversion following 
    \citet{Mulchaey94,Elvis94} (open), each over the observed range.
    The luminosity function from [O{\small II}] measurements
    is shown in the same style (red). Green diamonds and shaded region show
    the quasar luminosity function extrapolated to $z=0$ from \citet{Ueda03} and 
    \citet{Richards05}, respectively, with predicted contribution of 
    merger-driven, 
    spheroid-forming quasar activity from \citet{H05e} (solid black line 
    in lower right for bolometric corrections of \citet{Marconi04}, 
    dashed a constant $L=17\,L_{B}$). Contribution from stellar winds 
    is shown (dot-dashed) as estimated in \S~\ref{sec:winds}. The expected 
    Seyfert luminosity function from accretion of cold gas and our model for 
    feedback-driven, self-regulated evolution dominates and agrees well with the observed 
    luminosity function 
    from $-15\lesssim M_{B}\lesssim-23$, with relatively weak dependence 
    on the theoretical uncertainties for all but the brightest systems. 
    \label{fig:LF}}
\end{figure*}

In Figure~\ref{fig:LF} we show the 
Seyfert luminosity function predicted by 
our blast wave model and the assumption of fueling by the
accretion of molecular clouds. Our prediction,
in terms of the bolometric luminosity, is 
converted to a $B$-band luminosity function using
bolometric corrections following \citet{Marconi04}, 
based on optical through hard X-ray observations
\citep[e.g.,][]{Elvis94,George98,VB01,Perola02,Telfer02,Ueda03,VBS03},
with an X-ray reflection component generated by the PEXRAV model
\citep{MZ95}, and the appropriate Jacobian factors inserted. 
We have not calculated the effects of obscuration, 
and thus the luminosity function plotted is in terms of 
the {\em intrinsic} $B$-band luminosity, not 
necessarily that observed. 

For comparison, we plot the local ($0<z<0.15$) AGN luminosity
functions from the SDSS, determined in \citet{Hao05}. The luminosity function estimated
from H$\alpha$ emission-line measurements is shown as blue points,
over the luminosity range for which measurements exist. We show both
the best-fit Schechter function (circles) and double power-law
(squares) to the data.  We convert to the
$B$-band following \citet{Hao05}, based on the color corrections
determined therein and from
\citet[e.g.,][]{Schneider02,Schneider03}. To show the effects of
different bolometric corrections, the open squares adopt the constant
bolometric corrections of \citet{Mulchaey94,Elvis94}.  In either case,
the authors note that this gives agreement with previous, but much
shallower and more poorly constrained $B$-band AGN luminosity function
measurements from \citet{HuchraBurg92} and \citet{UlvestadHo01}. The
luminosity function shown is determined from the H$\alpha$ narrow-line
component, and is thus expected to be isotropic to Seyfert 1 and
Seyfert 2 galaxies, i.e.\ tracing the intrinsic luminosity which our
prediction shows.

We also show (red points) the luminosity function determined from
[O\,{\small II}] emission-line measurements, which is also
expected to trace the intrinsic luminosity, with circles and squares
again showing the best-fit Schechter and double power-law functions
over the observed range.  The correction to the $B$-band
follows \citet{Mulchaey94,Elvis94}, and gives agreement with
the H$\alpha$ determination over the range where the observations are
most well-constrained.  However, this does illustrate the potential
importance of systematic effects, especially at low luminosities.

There is some ambiguity about the proper value of 
$\eta_{L}$ (recall $L\propto t^{-\eta_{L}}$) 
to use in blast wave solution for the evolution of 
the quasar light curve, depending on the exact derivation from 
\S~\ref{sec:deriv}. The solid line in the figure shows our prediction 
for $\eta_{L}=18/31$, the exact solution to the perturbative 
accretion flow in a medium with $\gamma=5/3$ and $k_{\rho}=0$ 
(i.e.\ assuming no strong density gradients in the cloud). Allowing 
for variation in $\gamma$ and $k_{\rho}$ in a reasonable range, 
assuming the black hole feedback immediately decouples from the 
blast wave once the blowout begins, or 
using instead e.g.\ the estimate of the Bondi rate at $R_{\rm BH}$ 
in the interior of the blast wave yield $\eta_{L}$ in the 
range $1/2\lesssim \eta_{L}\lesssim 3/2$. Therefore, we show
in the figure our prediction for 
$\eta_{L}=1/2$ (dashed) and $\eta_{L}=1$ (dotted).

The primary change is at high accretion rates -- a lower $\eta_{L}$
means a slower decay in the accretion rate, giving more time at high
accretion rates and high luminosities (see \S~\ref{sec:edd.ratios}).
However, the difference in
the predictions is small even at the highest luminosities,
and completely negligible at $M_{B}\gtrsim-20$, where the
observations are most well-constrained. In fact, the systematic
uncertainty from different bolometric corrections (compare e.g.\ the
open and closed observational points) is a larger source of
error at all luminosities than this theoretical uncertainty in
the exact blast wave evolution.

We also show for comparison the extrapolation of the 
SDSS-2dF quasar luminosity function from \citet{Richards05}
to $z=0$ \citep[see also][]{Boyle00}, with the $1\sigma$ range 
indicated as the 
green shaded range at high luminosities. Similarly, 
we show the extrapolation of the \citet{Ueda03} 
hard X-ray luminosity function to $z=0$ as open 
green diamonds over the same luminosity interval, 
converted to a $B$-band luminosity function again using the 
bolometric corrections of \citet{Marconi04}. We 
show the corresponding (lower right) predictions from the model for 
merger-induced quasar activity of \citet{H05e,H05g}, which reproduces the 
observed bright quasar luminosity functions over a wide range of 
redshifts and luminosities, as the solid black line 
\citep[adopting the bolometric corrections of][]{Marconi04} and 
dashed black line \citep[adopting the corrections of][]{Elvis94}.  

The difference in shape between the quasar and Seyfert luminosity
functions is related to several effects.  First, the mass function of
black holes and, correspondingly, the systems ``driving'' accretion is
different (the merger mass function for quasars, and the blue galaxy
mass function for Seyferts).  Second, our feedback-driven model gives
slightly different decay solutions for both. We explicitly calculate
different timescales for decay in the two regimes in
\S~\ref{sec:timescales}, and owing to changes in the external mass
profile and initial energy injection, the profile for the decay
(i.e.\ faint-end slope) will be different. Essentially, in more violent
quasar systems, the feedback drives the system to lower luminosities
more quickly, resulting in a shallower faint-end slope of the
luminosity function \citep{H05g}.

The agreement between the predicted and observed luminosity functions
is good over a wide range of luminosities, from $M_{B}\gtrsim-14$ to
$M_{B}\lesssim-22$, and further if we consider the predicted quasar
contribution from other fueling mechanisms. At the lowest luminosities
$M_{B}\gtrsim-14$, our prediction agrees with the [O\,{\small II}]
determinations, but underpredicts the luminosity function estimated
from H$\alpha$ measurements. However, there are several sources of
uncertainty at low luminosities. A detailed comparison of the Seyfert
1 and Seyfert 2 H$\alpha$, [O\,{\small II}], and [O\,{\small III}]
luminosity functions in \citet{Hao05}, including a comparison between
different selection criteria from \citet{Kewley01} and
\citet{Kauffmann03}, suggests that there may be significant
contamination by star formation at these luminosities in H$\alpha$,
resulting in a significantly higher estimate (note that we compare
with the stricter AGN cut the authors adopt, although contamination
is still possible).

Additionally, measurement errors and bin-to-bin 
variation are significant at these luminosities, $\sim0.3\,$dex. 
Moreover, the luminosity function must turn over strongly 
near these luminosities. Although the Seyfert number density 
implied by integrating the observed luminosity function
over the observed range implies that approximately $20\%$ 
of galaxies host a Seyfert in this luminosity interval 
(an observation necessarily reproduced by our modeling since 
our predicted luminosity and mass functions agree with those observed), 
extrapolating this luminosity function only $1.5$ magnitudes fainter 
would imply that there are more Seyferts than there are galaxies. 
Finally, alternative fueling mechanisms such as stellar winds 
may become important at these lowest luminosities, 
a point discussed in \S~\ref{sec:winds}.

\subsection{Contribution of Different Morphological Types}
\label{sec:morph.contrib}

Our analysis enables us to decompose the contributions to the AGN
luminosity function from different galaxy types.  At the brightest
end, above the break in the extrapolated quasar luminosity function,
the systems are traditional ``bright quasars'', which may be entering
the blowout phase in the final stages of a galaxy merger, expelling
the remaining gas in a newly-formed elliptical galaxy, a model
described in detail by Hopkins et al.\ (2005a-e; 2006a-e).  The volume density
of these objects is very low, $\ll10^{-7}\,{\rm Mpc^{-3}}$, and
thus while they may dominate the bright AGN population at high
redshifts they will not be observed locally even in large surveys such
as the SDSS.

At luminosities below the break in the extrapolated quasar luminosity
function and at the brightest end of the AGN luminosity function,
there is a substantial contribution from black holes relaxing after
the blowout stage in their host and black hole-forming mergers. These
follow a similar decay to the blast wave solution described above and
in \citet{H05f} (both analytically and from simulations of mergers)
for the specific case of post-merger blowout ($\dot{m}\propto
t^{-2}$), a steeper typical decay than for Seyferts.  These host
galaxies rapidly redden and evidence of disturbance fades quickly, and
they will be seen as relatively normal ellipticals with large black
holes at moderate to low accretion rates, with possible evidence for
recent ($\lesssim\,$Gyr) merger or star formation activity.  This
population, with precisely these properties and a similar fractional
contribution to that we predict at the bright end of the local AGN
luminosity function, is well known observationally
\citep[e.g.,][]{Kauffmann03,Sanchez03,Sanchez04}, 
even in cases where the AGN dominates the observed spectrum 
\citep{VandenBerk05}.

The bulk of the luminosity range from $-15>M_{B}>-20$ is dominated by
late-type systems at moderate to high accretion rates, fueled by the
accretion of cold gas. At the brightest luminosities, there is some
contribution to the AGN luminosity function from relaxing ellipticals,
but at lower luminosities (as is evident from the predicted
post-merger AGN luminosity function prediction in Figure~\ref{fig:LF})
this contribution becomes small. Ellipticals do not significantly contribute to the
AGN luminosity function determined by this fueling mechanism, as they
do not have a supply of cold gas. The luminosity function from this
mode of fueling is mainly determined by systems of intermediate mass
($M_{\rm BH}\sim10^{7}\,M_{\sun}$) in Sa/b systems and to a slightly
lesser extent (owing to their lower gas content) by S0s. Sc/d galaxies
may not be an insignificant contribution to the Seyfert luminosity
function, but they do not dominate owing to both their lower
characteristic black hole masses (by a factor $\sim10-30$) and lower
number density (by a factor $\sim2$) compared to both Sa/b and S0
systems. In general, systems with small bulges and black holes 
$\lesssim10^{6}\,M_{\sun}$ (see \S~\ref{sec:rates}) may only 
contribute significantly at the lowest luminosities $M_{B}\gtrsim-15$.
As discussed in \S~\ref{sec:LF} above, Sm/Im galaxies have
such low black hole masses (if any) that they contribute negligibly.

This distribution of AGN hosts for this luminosity interval is
consistent with observations \citep{Kauffmann03,Sanchez04,Best05},
specifically those of e.g.\ \citet{Dong05} who find that Sa/b and S0
systems make up most of the low-moderate luminosity contribution to
the Seyfert luminosity function, with a relatively small contribution
from Sc/d systems. Similarly, the range of masses and host galaxy
types for which large $\dot{m}\gtrsim0.1$ accretion rates are
predicted agrees with observational estimates suggesting that
present-day black hole growth (in the sense of the population of
high-Eddington ratio objects) is dominated by late-type, seemingly
normal systems with black hole masses in the range given above
\citep{Cowie96,Steffen03,Barger03,Ueda03,Heckman04}, as discussed in
more detail in \S~\ref{sec:edd.ratios}.

At the lowest luminosities, there is a substantial contribution from
low-Eddington ratio accretion in late-type systems, but also from
relaxed ellipticals at low accretion rates. Fueling by stellar winds,
either from young star clusters in systems which still have cold gas
or smaller contributions from aging stellar populations in old bulges
has long been recognized as a significant fuel source for accretion
\citep[e.g.,][]{Shull83,Matthews83,David87}, however the high
velocities of these winds yield relatively low Bondi rates, and a wide
variety of observations further show that such systems tend to be
accreting at rates significantly below the Bondi estimate
\citep{FC88,BB99,DiMatteo00,NIA00,QG00,DCF01,Loewenstein01,Bower03,Pellegrini05}.

Nevertheless, these can provide a significant contribution to the
lowest-luminosity systems, and many ``dead'' ellipticals with
accretion rates $\dot{m}\sim10^{-6}-10^{-4}$ are expected and observed
\citep[e.g.,][]{Ho02,Heckman04,Marchesini04,Jester05,Pellegrini05}. These
may also be fueled by mechanisms other than stellar winds or the
accretion of hot (virialized) gas, but this seems to account for most
of the objects here, as calculated in \S~\ref{sec:winds}, especially
if one accounts for feedback removing some of the accreted mass in a
steady-state solution \citep{Soria05b}.  We show the prediction for
the contribution from stellar wind and hot gas fueling in
Figure~\ref{fig:LF} as the dot-dashed line, which demonstrates the
lower accretion rates of such systems, but the narrower luminosity
range results in a steeper stellar-wind induced luminosity
function which becomes important at the lowest luminosities.

\subsection{Distribution of Eddington Ratios}
\label{sec:edd.ratios}

\begin{figure*}
    \centering
    \plotone{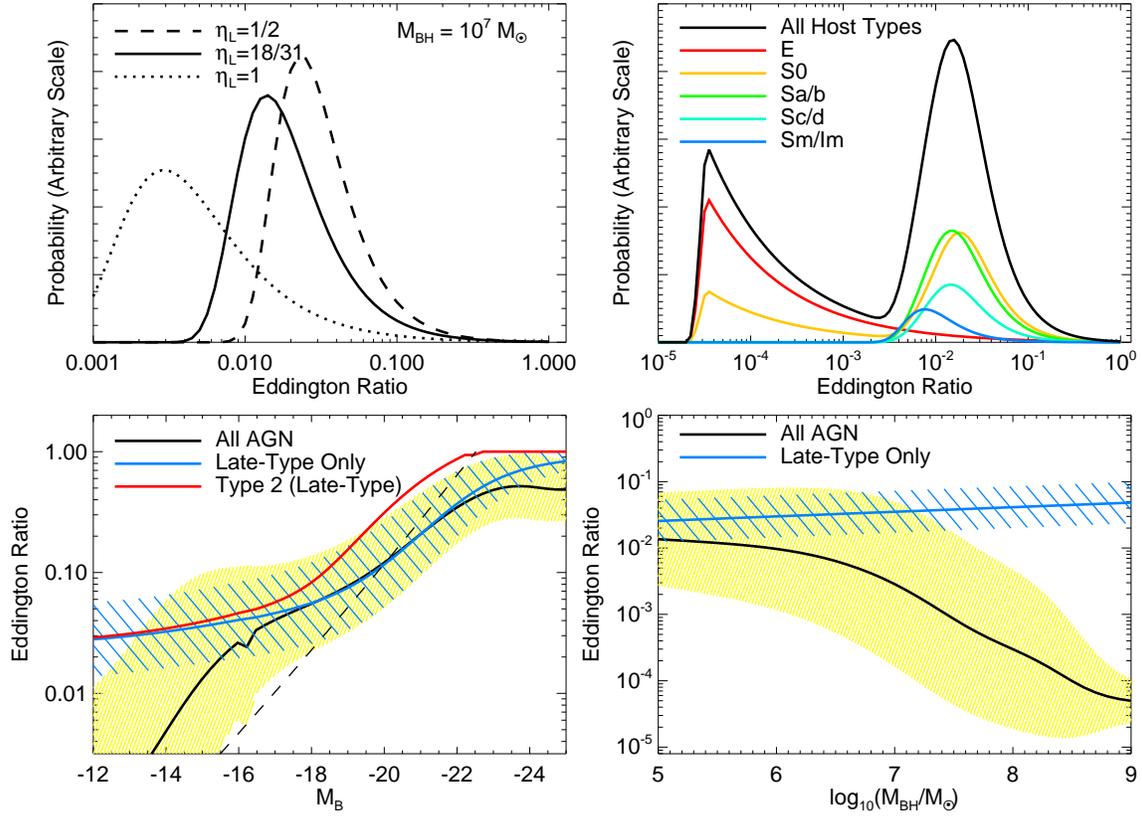}
    \caption{Upper left: Predicted distribution of Eddington ratios
    $\dot{m}\equiv \dot{M}/\dot{M}_{\rm Edd}$, for late-type 
    active (i.e.\ in some stage of ``blowout'') systems with 
    $M_{\rm BH}=10^{7}\,M_{\sun}$, predicted by our blowout model 
    in Equation~(\ref{eqn:dfdlogmdot}). Results for different 
    rates of accretion rate decay $\eta_{L}$ are shown, as labeled (note that the 
    vertical axis is a linear scale). 
    Upper right: Cumulative Eddington ratio distribution in active systems 
    (systems having a high-accretion rate event in a Hubble time)
    with $M_{B}\lesssim-12$, as a function of morphological type, 
    with late type Eddington ratios estimated as above and the E/S0 contribution 
    from post-spheroid forming merger decay estimated following 
    \citet{H05e,H05f,H05g}. Lower left: Mean Eddington ratio and 
    rms dispersion (shaded range) as a function of $M_{B}$ from 
    the Eddington ratio PDFs and predicted Seyfert luminosity function in 
    Figure~\ref{fig:LF}, for all AGN (black, with yellow dispersion), AGN in late-type hosts 
    (blue; modeled herein), and Type 2 AGN in late-type hosts (red; 
    dispersion not shown for clarity but similar to that of all late-type AGN). 
    Note that this does not imply higher Eddington ratios for Type 2 AGN as a whole, 
    as the luminosity-dependence of the Type 2 fraction is important for such a 
    statistic (see Figure~\ref{fig:obscuration}).
    Dashed line shows the expectation if 
    $M_{\rm BH}=3\times 10^{7}\,M_{\sun}$=\,constant (i.e.\ 
    implying the luminosity function is dominated by differences in Eddington ratio).
    Lower right: Mean Eddington ratio and rms dispersion, in the same style, 
    as a function of $M_{\rm BH}$. 
    \label{fig:edd}}
\end{figure*}

From the evolution of the accretion rate in our blast wave solution,
we can predict the accretion rate distribution as a function of e.g.\
black hole mass, luminosity, and host galaxy properties.
Figure~\ref{fig:edd} shows (upper left) the predicted Eddington ratio
distribution ${\rm d}f/{\rm d}\log{\dot{m}}$ for ``active'' (i.e.\ in
some stage of blowout) late-type systems determined in
Equation~(\ref{eqn:dfdlogmdot}), with the appropriate exponential
cutoff at low $\dot{m}$ (such that $\int{\rm d}f=1$).  We show this
for $M_{\rm BH}=10^{7}\,M_{\sun}$, where $\tau_{\ast}\approx1$ (as
determined in Equation~[\ref{eqn:tmdot.calc}]) and typical of the
black holes which dominate the Seyfert luminosity function,
but our calculation in \S~\ref{sec:duty} demonstrates that this
distribution depends only weakly on $M_{\rm BH}$ in late-type systems.

We show results for three values of $\eta_{L}$: $\eta_{L}=1/2$
(dotted), $\eta_{L}=18/31$ (solid), and $\eta_{L}=1$ (dashed).
Because $L\propto t^{-\eta_{L}}$, larger values of $\eta_{L}$
correspond to a more rapid falloff in the accretion rate and therefore
broader Eddington ratio distributions extending to lower accretion
rates.  In what follows, we adopt $\eta_{L}=18/31$, and a different
$\eta_{L}$ will not change the trends in Eddington ratio which we find
but will systematically shift the typical Eddington ratios as shown in
the figure. As is clear from comparison with Figure~\ref{fig:LF}, the
exact choice of $\eta_{L}$ has little effect on our predicted Seyfert
luminosity function, within the reasonable range of $\eta_{L}$
predicted by our blast wave model, $1/2\leq \eta_{L}\leq 3/2$.
However, $\eta_{L}$ dominates the systematic uncertainty in the
estimated Eddington ratio distribution.

Figure~\ref{fig:edd} also shows (upper right) the predicted cumulative
Eddington ratio distribution as a function of host galaxy morphology.
Here, the small differences in the Eddington ratio distribution among
late-type galaxies are caused by the weak dependence of $\tau_{\ast}$
and the duty cycle on host galaxy properties ($\sigma$ and $f_{\rm
gas}$).  The Eddington ratio distribution of ellipticals and inactive
S0s is estimated from the predicted formation times and blast wave
decay of merger-induced quasar activity \citep{H05e,H05f,H05g}.  The
``cumulative'' Eddington ratio is, in general, ill-defined, and here
we plot the distribution in active (i.e.\ with an event in $t_{H}$)
systems with $M_{B}\lesssim-12$. 

The lower left panel of Figure~\ref{fig:edd} shows the (logarithmic)
mean Eddington ratio and rms dispersion (shaded ranges) as a function
of $M_{B}$ from the Eddington ratio PDFs above and predicted Seyfert
luminosity function in Figure~\ref{fig:LF}. We show this for all AGN
(black, with yellow shaded range), AGN in late-type hosts as modeled
herein (blue), and Type 2 AGN in late-type hosts as calculated in
\S~\ref{sec:obscuration} below (red; dispersion not shown for clarity
but similar to that of all late-type AGN). The $B$-band magnitude
$M_{B}$ represents the intrinsic magnitude as in Figure~\ref{fig:LF},
and does not account for extinction. In the lower right panel of the
Figure, we show the same quantity as a function of black hole mass.

\begin{figure*}
    \centering
    \plotone{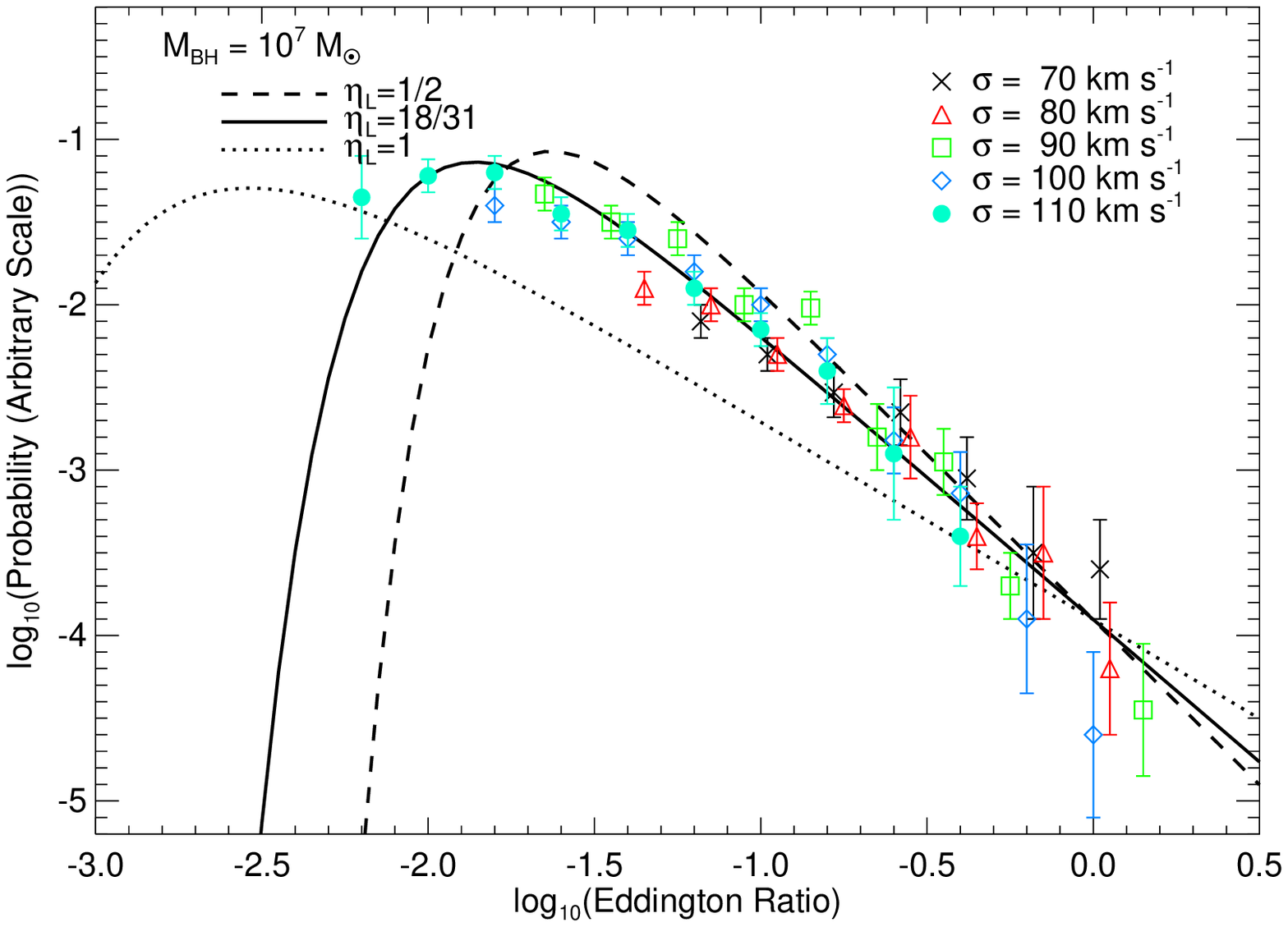}
    \caption{Predicted Eddington ratio distribution of active Seyferts 
    from our blast wave model in \S~\ref{sec:deriv}
    and for estimated duty cycles in \S~\ref{sec:duty}, as in Figure~\ref{fig:edd}
    (upper left), for our exact solution 
    $\eta_{L}=18/31$ and the range $\eta_{L}=1/2,\,1$ as labeled, for 
    black holes with $M_{\rm BH}=10^{7}\,M_{\sun}$ (although this depends only 
    weakly on $M_{\rm BH}$). The observed accretion rate distribution from 
    \citet{YLK05} is plotted for each of several values of $\sigma$ as labeled, 
    corresponding to 
    $M_{\rm BH}\approx2\times10^{6}-2\times10^{7}\,M_{\sun}$.
    Both the power law-trend at high $\dot{m}$ (dependent on the 
    feedback-regulated light curve decay) and flattening/turnover 
    at low $\dot{m}$ (dependent on excitation rates and duty cycles) 
    in the observed distribution agree very well 
    with the predictions of our model ($\chi^{2}/\nu=0.6$), whereas models 
    without feedback-driven self regulation of the accretion rate produce 
    much flatter distributions. 
    \label{fig:edd.YLK}}
\end{figure*}

Figure~\ref{fig:edd.YLK} shows our predicted Eddington ratio
distribution, as in the upper left panel of Figure~\ref{fig:edd} (but
in log-log).  Recall that this shows our prediction of ${\rm d}f/{\rm
d}\log{\dot{m}}$, the fraction of objects per logarithmic interval in
$\dot{m}$.  We compare our results with those measured in
\citet{YLK05} from a sample of $\gtrsim20,000$ local SDSS AGN
(points), from estimates of [O{\small III}] luminosities and black
hole masses inferred from measurements of $\sigma$.  We show the
observations for five values of $\sigma$, from $\sigma=70-110\,{\rm
km\,s^{-1}}$ (binned by $0.05$\,dex in $\log{\sigma}$).  We show this
range of $\sigma$ because it corresponds approximately to $M_{\rm
BH}=2\times10^{6}-2\times10^{7}\,M_{\sun}$, appropriate for comparison
with our predictions for $M_{\rm BH}=10^{7}\,M_{\sun}$.  Furthermore,
at larger $\sigma$, there is a significant contribution from ``dead''
ellipticals (as seen in e.g.\ the black hole mass function of
Figure~\ref{fig:bhmf} or Eddington ratio as a function of $M_{\rm BH}$
in Figure~\ref{fig:edd}), and while we can predict the combined
Eddington ratio distribution at these $\sigma$ by combining our
predictions with those of \citet{H05e,H05f}, the relation to our blast
wave model and physical interpretation is less direct.

The agreement between our predicted Eddington ratio distribution and
the observations is good ($\chi^{2}/\nu=0.6$ for the comparison with
the $M_{\rm BH}=10^{7}\,M_{\sun}$, i.e.\ $\sigma\approx100\,{\rm
km\,s^{-1}}$ data), over the observed range.  The agreement for
$1/2\leq\eta_{L}\leq5/6$ is less good, but still acceptable
($\chi^{2}/\nu\sim1.5-2.0$).  Note that we follow \citet{YLK05} and
exclude the lowest Eddington ratio bin for each $\sigma$ owing to
significant incompleteness. In order to compare with their
observations, we have convolved our prediction with the distribution
of [O{\small III}] luminosities (i.e.\ allowed for a small dispersion
in bolometric correction), which is why our prediction is non-zero
(although small) above an Eddington ratio of unity. The predicted
Eddington ratio distribution in this mass range depends only weakly on
$\sigma$, evident in e.g.\ our Equation~(\ref{eqn:tmdot.calc}) and in
Figure~\ref{fig:edd}, and therefore the observations over this entire
range in $\sigma$ all agree with our prediction in a nearly
self-similar manner.

That the observed \citet{YLK05} $\dot{m}$ distributions agree with our
exact blast wave solution with $\eta_{L}=18/31$ and reasonably well
with e.g.\ the range of solutions $1/2\leq\eta_{L}\leq 5/6$ is
somewhat surprising.  The slope this implies (i.e.\ logarithmic slope
of ${\rm d}n/{\rm d}\log{\dot{m}}$) of $-1/\eta_{L}=-31/18$ is
substantially different from the slope of $\approx-0.8$ fitted to the
low-$\dot{m}$ behavior of the $\dot{m}$ distributions by the authors,
which would imply $\eta_{L}\approx 1.26\pm0.1$. The authors use this
fit to argue for a self-similar quiescent disk evolution model to
explain the late-time $L\propto t^{-\eta_{L}}$ behavior.  However,
this fit from \citet{YLK05} is derived primarily from the
low-$\dot{m}$ behavior of their $\sigma\approx200\,{\rm km\,s^{-1}}$
objects, which we have noted (following numerous observational
studies) are a different population, dominated by ellipticals decaying
from larger accretion rates or fueled by e.g.\ stellar winds and
diffuse hot gas
\citep[e.g.,][]{Kauffmann03,Pellegrini05}. Furthermore, the authors
acknowledge that this does not provide a good fit to objects with
larger $\sigma$, which is surprising since these objects are otherwise
observed to be a similar AGN population to those with
$\sigma\approx200\,{\rm km\,s^{-1}}$ (i.e.\ both are dominated by
early-type systems).  The authors also acknowledge that this model for
the activity predicts that the active black holes will be observed at
masses substantially {\em below} their final ($M_{\rm BH}-\sigma$
relation) masses, by about $1\,$dex, whereas the observations
\citep[e.g.,][]{Barth05} suggest the opposite trend.  In
\S~\ref{sec:m.sigma} we demonstrate that our model predicts this
observed tendency rather than smaller black hole masses.

Additionally, the model fitted by \citet{YLK05} does not provide a
good match to the high-$\dot{m}$ distributions, even for those values
of $\sigma$ for which the overall fit is most acceptable, giving e.g.\
typical $\chi^{2}/\nu\sim4-6$ for $\sigma\sim100-200\,{\rm
km\,s^{-1}}$ (at $\dot{m}\gtrsim10^{-2}$).  Over the entire observed
range, our model provides an improved fit, $\chi^{2}/\nu\approx0.6$
compared to $\chi^{2}/\nu\approx1.6$ from the fits of \citet{YLK05};
alternatively, $\Delta\chi^{2}\approx11.0$ for e.g.\ $\sigma=110\,{\rm
km\,s^{-1}}$.  These arguments suggest that while the self-similar
quiescent disk evolution discussed in \citet{YLK05} may be important
for some range of early-type systems at low accretion rates, the
observations in the range of black hole masses shown in
Figure~\ref{fig:edd.YLK} are better explained by our model for
feedback-driven blast wave evolution.

The turnover at low-$\dot{m}$ in the Eddington ratio distribution is,
for any power-law distribution of accretion rates, set entirely by the
duty cycle at high $\dot{m}$. In our language, for a given slope
$\eta_{L}$, this turnover occurs at a value $\dot{m}_{\rm min}$
determined by Equation~(\ref{eqn:mdot.min}) as a function of the duty
cycle (i.e.\ lifetime and rate of activation) at high $\dot{m}$. There
is some uncertainty in the exact normalization of this turnover, as
our theoretical prediction depends on scaling arguments and the
detailed numerical prefactors are sensitive to the local kinematics
near the black hole.  Therefore, our prediction for the location of
the turnover is more uncertain than our prediction of $\eta_{L}$.

Currently, the observations do not resolve the turnover.  Doing so
would constrain the duty cycle as a function of $\dot{m}$; for
example, the difference between a sharp turnover as we have assumed
and a shallow (e.g.\ symmetric power-law) turnover for the
$\dot{m}_{\rm min}$ of our $\eta_{L}=18/31$ prediction amounts to a
factor of $\sim2$ change in the duty cycle $\delta_{0}$.  Therefore,
observations of the {\em shape} of the accretion rate distribution at
a given $\sigma$, probing only moderately fainter luminosities than
\citet{YLK05}, can constrain this quantity without the large
systematic uncertainties inherent in determining the normalization of
the distribution when the turnover is not resolved.  We note, however,
that at least the flattening of this distribution at low $\dot{m}$ is resolved
by the observations of \citet{YLK05} -- if our prediction were
simplified to a pure power-law, it would overpredict by
$\sim3-6\sigma$ each of the lowest four $\dot{m}$ observations
plotted.

\begin{figure*}
    \centering
    \plotone{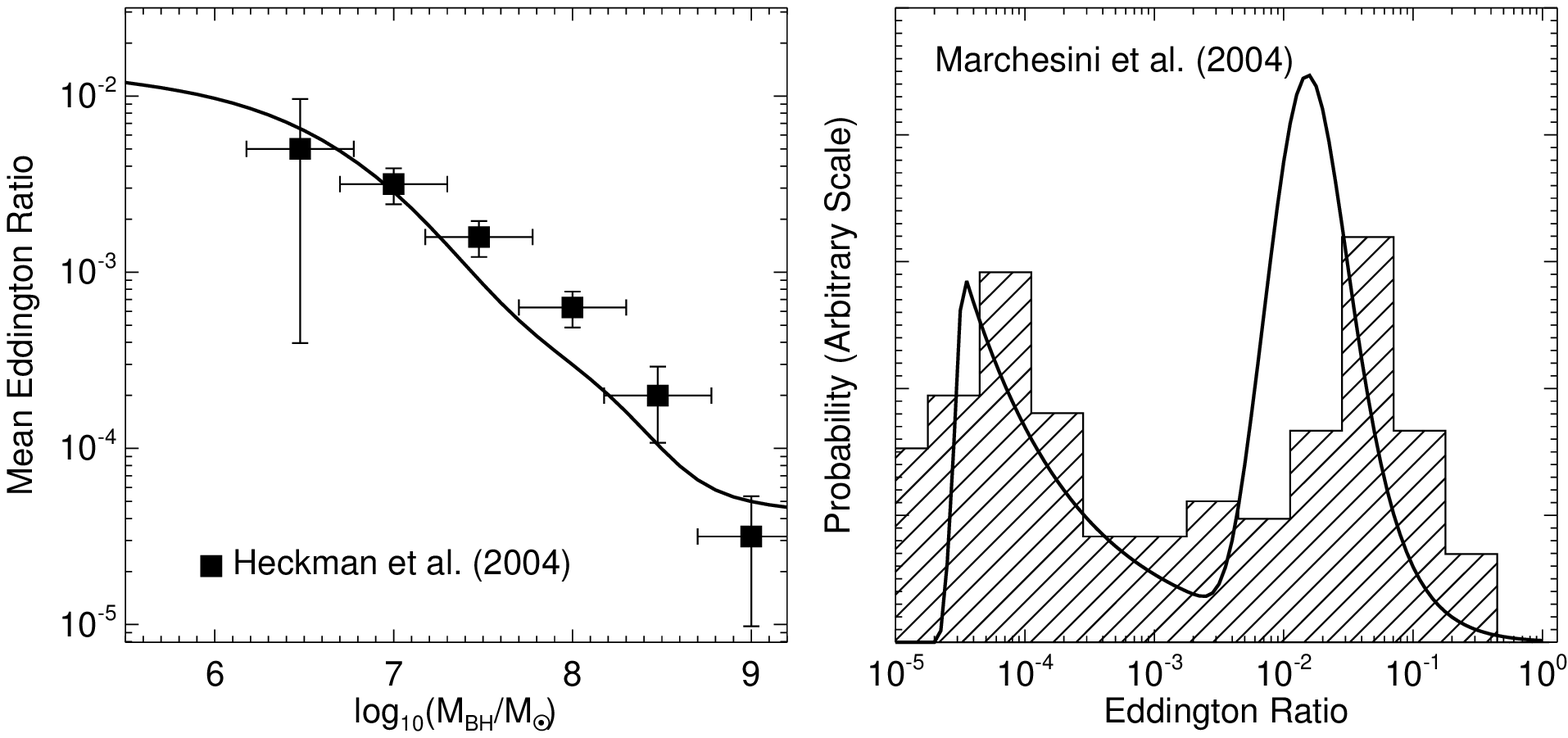}
    \caption{Left: Mean Eddington ratio as a function of black hole mass from 
    Figure~\ref{fig:edd}, compared to estimates from the observations of 
    \citet{Heckman04} (squares). The point at $M_{\rm BH}\approx3\times10^{6}\,M_{\rm sun}$
    shows the absolute systematic uncertainty in the mean Eddington ratio distribution from 
    fitting to the (high Eddington ratio only) distributions of \citet{Heckman04}, 
    subsequent points show the relative uncertainty in the trend of 
    Eddington ratio vs.\ $M_{\rm BH}$. The observed trend of decreasing 
    Eddington ratios with $M_{\rm BH}$ is reproduced, as larger-$M_{\rm BH}$ 
    populations are increasingly dominated by large spheroids without cold 
    gas supplies for stochastic accretion.     
    Right: Tentative Eddington ratio distribution 
    from Figure~\ref{fig:edd} (upper left), summed over 
    all morphological types, compared to the observationally 
    estimated Eddington ratio distribution from \citet{Marchesini04} (histogram). Note, however, 
    that the vertical axis here is linearly scaled, and the predicted bimodality can easily be erased 
    by e.g.\ different values of $\eta_{L}$. 
    \label{fig:edd.HM}}
\end{figure*}

Although incompleteness in the lowest-$\dot{m}$ bin at each $\sigma$
may be significant, the observations appear to favor the turnover in
the Eddington ratio distribution moving to larger $\dot{m}$ at lower
$\sigma$, an expectation broadly consistent with our result in the
lower-left panel of Figure~\ref{fig:edd}.  Figure~\ref{fig:edd.HM}
shows our prediction for the mean Eddington ratio as a function of
black hole mass from Figure~\ref{fig:edd}, compared to the estimates
of \citet{Heckman04} (squares).  Neither the observations of
\citet{Heckman04} nor those of \citet{YLK05} resolve the turnover in
accretion rate distributions at low-$\dot{m}$, and therefore cannot
determine a ``mean'' accretion rate in a proper sense.  Therefore, we
instead consider the cumulative distribution of accretion rates from
the observations for $M_{\rm BH}\approx3\times10^{6}\,M_{\rm sun}$,
and fit this to a distribution of Eddington ratios similar to our
prediction in Figure~\ref{fig:edd.YLK} above, but allowing both the
slope and duty cycle to vary freely.  From these fits, then, we can
constrain the peak and mean of the Eddington ratio distribution.  The
vertical error bars of the point at $M_{\rm BH}=3\times10^{6}\,M_{\rm
sun}$ in Figure~\ref{fig:edd.HM} show the $1\sigma$ allowed range from
these fits, and reflect the systematic uncertainty in the absolute
value of the mean accretion rates at a given $M_{\rm BH}$ or $\sigma$
determined observationally.

While the mean Eddington ratios are not well-determined
observationally, the trend of Eddington ratio with mass is more
well-constrained.  Using the best-fit absolute normalization from the
fit to the $M_{\rm BH}=3\times10^{6}\,M_{\rm sun}$ data, we show the
expected mean Eddington ratio as a function of $M_{\rm BH}$ at $M_{\rm
BH}=10^{7},\ 3\times10^{7},\ 10^{8},\ 3\times10^{8},\ {\rm and}\
10^{9}\,M_{\sun}$ from the observed trends in \citet{Heckman04} (see
their Figure~3).  Here, the vertical error bars represent the {\em
relative} error in Eddington ratio as a function of mass. Thus, the
typical mean Eddington ratios do agree with our predictions, but the
uncertainties are large because the downturn of the $\dot{m}$
distribution is not measured observationally. However, the trend of
mean Eddington ratio with black hole mass is reasonably
well-constrained, and agrees with our predictions for black hole
masses $M_{\rm BH}=3\times10^{6}-10^{9}\,M_{\sun}$.

The trend arises because for higher black hole masses, the population
is increasingly dominated by ellipticals which do not have a supply of
cold gas. These systems are presumably either fading from earlier
bright quasar activity or accreting in a quasi-steady state from
virialized hot gas or stellar mass loss, at accretion rates well below
those of late-type Seyferts.  Furthermore, the mean age of ellipticals
may increase with mass, here calculated following the determination as
a function of bulge and black hole mass in \citet{H05f}, further
contributing to the trend of decreasing $\dot{m}$ with increasing
$M_{\rm BH}$ even in pure elliptical systems.

From Figure~\ref{fig:edd.HM}, our modeling predicts that most of the
black hole mass growth in the local Universe occurs in relatively
low-$M_{\rm BH}$ systems with $M_{\rm BH}\lesssim10^{7}\,M_{\sun}$,
consistent with observations \citep{Heckman04}.  Of course,
observations of quasars, compared to our predictions for the evolution
of cloud-fueled activity in late-type galaxies in
\S~\ref{sec:redshift}, suggest that this is not true at higher
redshifts, where more massive black holes are formed.  Although this
constitutes the low-redshift end of ``cosmic downsizing''
\citep[e.g.,][]{Cowie96} seen observationally in AGN and quasar
evolution
\citep{Page97,Miyaji00,Miyaji01,LaFranca02,LaFranca05,Cowie03,
Ueda03,Fiore03,Hunt04b,Cirasuolo05,HMS05}, our modeling does not imply
that the local activity is fundamentally cosmological.  The transition
to lower black hole masses growing at lower redshifts through black
hole growth via merger-driven quasar fueling is cosmological and
traces galaxy downsizing \citep{H05f}.  But, the relatively high rates
of accretion in late-type galaxies compared to higher-$M_{\rm BH}$
systems predicted here is more a function of the high-$M_{\rm BH}$
ellipticals fading with time, rather than any sudden ``turning on'' of
the Seyfert population.

In Figure~\ref{fig:edd.HM} we also compare our result for the total
distribution of accretion rates across all morphological types from
the upper right panel of Figure~\ref{fig:edd} (solid line) to that
determined by \citet{Marchesini04} (shaded histogram), which agrees
with the estimates of e.g.\ \citet{Ho02,Jester05,H05i} and extends to
lower $\dot{m}$ than that of e.g.\ \citet{YLK05} and includes inactive
systems.  The predicted Eddington ratio distribution is bimodal, with
Seyferts comprising the majority of the high-$\dot{m}$ peak and
relaxed ellipticals dominating at low $\dot{m}\sim10^{-4}$ accretion
rates.  There is broad agreement, and the slight horizontal offset
between the theory and measurements is within the systematic
uncertainty of the observational estimate of $\dot{m}$.

While this agreement is suggestive, we caution that the predicted
bimodal distribution is not firm.  The vertical axis shows a linear
scale, and, from Figure~\ref{fig:edd.YLK}, a small change in our
choice of $\eta_{L}$ or the duty cycle at high $\dot{m}$ (which
determines the $\dot{m}_{\rm min}$ where the high-$\dot{m}$
distribution turns over) would broaden the distribution to
$\lesssim10^{-3}$, erasing the bimodality. Furthermore, our prediction
depends on the Eddington ratio distribution of quiescent ellipticals,
estimated from their accretion rate decay according to our blast wave
model following mergers and simulations of merger-driven black hole
growth \citep{H05e,H05g}, but once these low accretion rates are
attained, other processes may dominate the gas inflows and accretion,
and thus our estimate may not be appropriate for this end of the
distribution.

In the lower-left panel of Figure~\ref{fig:edd}, we show our
prediction (solid line) for the mean accretion rate as a function of
luminosity, compared (dashed line) to the corresponding Eddington
ratio for a black hole of mass $3\times10^{7}\,M_{\sun}$ at each
$M_{B}$ (i.e. approximately a logarithmic slope of $-1/2.5$).  This
slope is what would be expected if the observed luminosity function
were purely an Eddington ratio sequence.  Although the differences
between our predicted distribution and that shown are important, the
rough similarity between the two, at least for $M_{B}\gtrsim-20$,
implies that the range of the observed luminosity function is
dominated by differences in Eddington ratio, and not by different
black hole masses at similar relative accretion rates.  Since e.g.\
host galaxy mass and luminosity are correlated with black hole mass, we
expect that these quantities will not be strongly correlated with the
observed AGN luminosity, as a wide range of AGN luminosities can have a
similar distribution in constituent black hole masses.  This has been
seen in observations of local AGN and quasars and their host galaxies
\citep[e.g.,][]{Bahcall97,McLure99,Hamilton02,
WooUrry02,ODowd02,Jahnke03,Hao05,VandenBerk05}.

Note that the prediction for the Type 2 Eddington ratios in this
panel, while slightly above those of Type 1 objects at high
luminosities, does not imply that we predict typically higher
accretion rates in Type 2 Seyferts.  Rather, as shown in
\S~\ref{sec:obscuration}, our modeling indicates that Type 2 Seyferts
are a small fraction of the population at these high luminosities, but
a large (comparable) fraction at low luminosities. Therefore, our
modeling predicts that the characteristic accretion rates of Seyfert
2s should be lower than those of Seyfert 1s. In general, the
differences in Eddington ratio distributions between Type 1 and Type 2
systems is a more sensitive function of their characteristic
luminosities and determined by e.g.\ the dependence of the Type 2
fraction on luminosity, than it is a difference between the two
populations at a given luminosity.  (This refers to intrinsic
luminosity, as Type 2 systems will, by definition, be extinguished at
some frequencies.)

\subsection{Redshift Evolution of Seyfert Activity}
\label{sec:redshift}

Our estimate of the duty cycle as a function of $\dot{m}$ depends
primarily on the gas fraction and $\sigma$, and the convolution to give a
luminosity function depends on the black hole mass function in
late-type galaxies. From the evolution of these quantities it is
straightforward to determine the evolution in this luminosity
function.

Because the mass gained in a ``blowout'' event is small (at most a
factor $\sim2$; see \S~\ref{sec:mass.blowout}), we determine the black
hole mass function at each redshift of interest from the evolution of
the late-type galaxy mass function, using the method of
\S~\ref{sec:LF}.  From the spectral-type separated mass functions of
\citet{Fontana04} determined up to $z=2$ from the K20 survey, we
expect little evolution in the late-type mass function up to $z\sim1$,
and then a decrease in the number density of these galaxies.  In
detail, we trace this evolution up to $z=0.5$ using the $K$ and
$J$-band luminosity functions of e.g.\
\citet{Feulner03,Pozzetti03,Dahlen05}.  Normalizing the mass functions
to be the same in the overlapping redshift interval $0.2<z<0.55$, we
extend this with the morphologically classified spiral mass functions
from \citet{Bundy05} and \citet{Franceschini06} up to $z=1.4$, and
then repeat this with the \citet{Fontana04} observations to infer
the mass functions to $z=2$.

These estimates are roughly consistent with the $B$-band evolution in
late-type luminosity functions which is more well-constrained
\citep[e.g.,][]{deLapparent04,Giallongo05,Faber05}, measured evolution
in the cumulative mass function \citep{dePropris99,Drory04,Drory05},
and evolution in black hole mass functions from spheroid luminosity
functions and integration of the continuity equations for AGN
luminosity functions \citep{YuLu04,Tamura05}.  Given this, our approach
appears to be reasonable, but the observations have large
uncertainties.  However, we find that there is little evolution in the
Seyfert luminosity function in any case, and these uncertainties are
not important.

We estimate the evolution in the typical gas fractions of disks by
assuming that gas is consumed on a timescale related to star formation
$t_{\rm SF}\sim 4-8$\,Gyr, giving $f_{\rm gas}\sim
\exp{(-t_{H}(z)/t_{\rm SF})}$, where $t_{H}(z)$ is the Hubble time at
redshift $z$.  We take $t_{\rm SF}\sim6\,$Gyr, as this gives a
Milky-Way like $f_{\rm gas}\sim0.1$ at $z=0$.  This prescription and
characteristic timescale follow from observations
\citep[e.g.,][]{Kennicutt98,RowndYoung99,MartinKennicutt01},
cosmological simulations
\citep[e.g.,][]{SH03b,HS03,Nag04,Nag05a,Nag05b}, simulations of star
formation rate evolution in isolated disk galaxies
\citep{Li05a,Li05b}, and comparison of these simulations and the
predictions for quasar activity with merger rates and merger
luminosity functions \citep{H05h}. As gas fractions increase, the
event rate for cold gas accretion can increase, producing a higher
duty cycle at large $\dot{m}$ from Equation~(\ref{eqn:duty.calc}).
Although our choice for the evolution of the gas fraction is somewhat
arbitrary, we find a sufficiently weak dependence that observational
uncertainty is not significant.

\begin{figure*}
    \centering
    \plotone{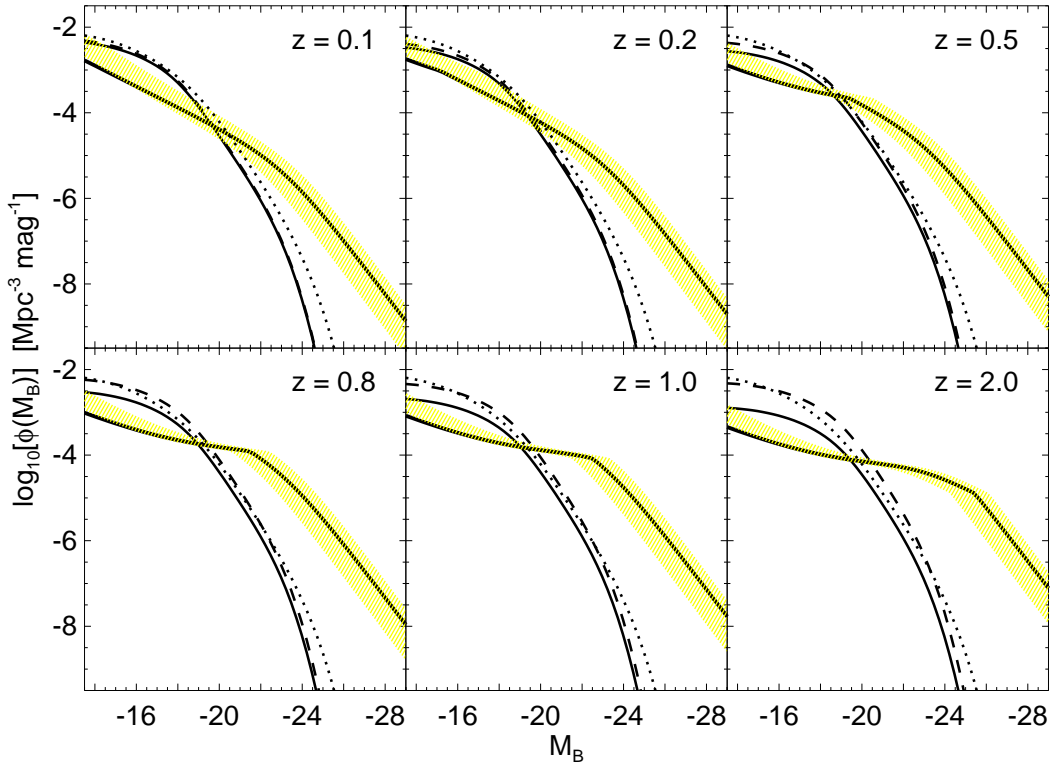}
    \caption{Predicted Seyfert luminosity function owing to the stochastic 
    fueling, at several redshifts (as labeled).  The 
    observed quasar luminosity function 
    from \citet{Ueda03} (black line with yellow shaded range) is shown,
    converted to $M_{B}$ as in Figure~\ref{fig:LF}.
    The solid line (no shaded range) shows our prediction for 
    the Seyfert luminosity function,
    determined from the evolution in late-type mass functions 
    \citep{dePropris99,Drory04,Drory05,Fontana04,deLapparent04,
    Bundy05,Dahlen05,Giallongo05, Franceschini06}
    and gas fractions 
    \citep{Kennicutt98,RowndYoung99,MartinKennicutt01,SH03b,Li05a,Li05b}. 
    Dashed lines shows the ``maximal evolution'' 
    (i.e.\ allowing gas fractions to increase with $z$ but not allowing 
    any mass function decrease). Dotted lines show the $z=0$ prediction of 
    Figure~\ref{fig:LF}, i.e.\ no evolution. Regardless of our assumptions, 
    the evolution in the luminosity function of systems fueled by quiescent, 
    non-merger driven cold gas accretion is weak, 
    and the quasar luminosity function at the redshifts of 
    peak quasar activity (and quasar luminosity density) 
    must be dominated by other fueling mechanisms. 
    \label{fig:LF.z}}
\end{figure*}

Figure~\ref{fig:LF.z} shows our predicted Seyfert luminosity function
(black lines with no shaded range) at various redshifts.  The solid
lines show the full prediction, accounting for both evolution in the
late-type galaxy mass functions and the evolution in typical gas
fractions of disk galaxies. The dashed line shows an estimate of
``maximal evolution,'' i.e.\ allowing the gas fraction to increase
with redshift, but not allowing any decrease in the mass functions.
The dotted line shows the $z=0$ prediction from Figure~\ref{fig:LF} at
each redshift -- i.e., no evolution.

For comparison, at each redshift, we show the fitted quasar luminosity
function of \citet{Ueda03} (black solid line with yellow range showing
the uncertainty owing to different bolometric corrections), converted
to $M_{B}$ as in Figure~\ref{fig:LF}. The ``low'' ($0.2\leq z \leq 1$)
redshift behavior of this luminosity function has also been compared
with the luminosity functions of e.g.\ \citet{Richards05,HMS05} which
have more objects at these redshifts, and they give consistent
results.  Alternatively, we could plot the expectation for
merger-driven quasar activity calculated from the models and
simulations of quasars in mergers from Hopkins et al.\ (2005a-e; 2006a-e),
which is essentially identical to the plotted quasar luminosity
functions.

From Figure~\ref{fig:LF.z} we see that the quasar luminosity function
evolves rapidly, while our predicted Seyfert luminosity function
hardly evolves even out to $z=2$.  While our estimates of the
evolution of e.g.\ gas fractions and late-type mass functions are
uncertain, the similarity (relative to the evolution in the quasar
luminosity function) of the expected evolution, no evolution, and
maximal evolution cases means that unless an extreme model for
late-type galaxy evolution were adopted (at odds with the
observations), there would be essentially no difference in the
predicted evolution of this luminosity function. Thus, these
observational errors are not a significant source of theoretical
uncertainty for the key conclusions from Figure~\ref{fig:LF.z}.

From the figure, it appears that our prediction may be systematically
high, by about a factor of $\sim2$, at the lowest luminosities, which
is possible, given the uncertainties in our estimate of the duty
cycle.  If the normalization at these luminosities is matched to the
extrapolated faint end of the quasar luminosity function, then it
implies an even smaller contribution from this activity at these
redshifts.  However, it is important to note that the faint end quasar
luminosity function plotted here is extrapolated below the
observations at the higher redshifts, and may be incomplete at the
lowest $M_{B}\gtrsim-18$ luminosities, corresponding to or slightly
below the faintest bins in even the deep luminosity functions of e.g.\
\citet{Ueda03,HMS05,LaFranca05} at low ($z\lesssim1$) redshifts.

For local, $z=0$ AGN, the contribution from stochastic cold gas
accretion dominates the luminosity function up to $M_{B}\approx-22$,
near the traditional division between Seyfert and quasar activity (see
Figure~\ref{fig:LF}).  However, above $z\sim0.1$, this shifts to lower
luminosity $M_{B}\approx-20$, and the relative contribution at higher
luminosities steadily decreases.  This luminosity function does still
represent a significant contribution to the faint end (i.e.\ below the
``break'') of the quasar luminosity function up to $z\sim0.5$.
However, by $z=1$, the ``Seyfert'' (i.e.\ our predicted) contribution
to the quasar luminosity function is only significant at about $4-5$
magnitudes (i.e.\ two orders of magnitude in luminosity) below the
break, and by $z=2$ only important about 7 magnitudes, or 3 orders of
magnitude in luminosity, below the break. Thus, at the epochs
generally associated with ``quasar activity,'' the contribution from
this accretion mechanism is negligible, even at luminosities far below
the deepest observations.

That the quasar luminosity function evolves more rapidly than our
prediction for the Seyfert luminosity function indicates that
different modes of fueling are likely responsible for the two
populations.  Indeed, merger-induced quasar activity can account for
the evolution in the quasar luminosity function, and merger rates of
gas-rich galaxies evolve more rapidly than the mass function of
gas-rich galaxies.  Note that at any time the number of gas-rich
galaxies undergoing mergers is small, so a relatively large number of
mergers providing the observed quasar activity does not demand strong
evolution in the disk galaxy mass function.  \citet{H05f,H05h} study
the evolution of quasar, red galaxy, and merger luminosity functions
and argue that a self-consistent mapping between the two implies that
a significant fraction of the brightest quasar activity must result
from mergers.

We can use our predictions to determine the contribution of quiescent
fueling to the buildup of the black hole mass density of the Universe
and cosmological backgrounds, such as the cosmic X-ray background.
From the analysis in \S~\ref{sec:m.sigma} or the mean Eddington ratio
plotted as a function of $M_{\rm BH}$ in Figure~\ref{fig:edd}, this
process will contribute negligibly to the growth of black hole masses
above $\sigma\sim100\,{\rm km\,s^{-1}}$.  Larger-mass black holes
($M_{\rm BH}\gtrsim10^{7}\,M_{\sun}$) dominate the present black hole
mass density
\citep{Salucci99,MS02,YT02,Ferrarese02,AllerRichstone02,Shankar04},
and thus the contribution of quiescent fueling to the total black hole
mass density is negligible.  Even at relatively low $M_{\rm
BH}=10^{6}-10^{7}\,M_{\sun}$, the quiescent mode of fueling will
increase black holes masses by only up to a factor $\sim1.2-2.0$ (see
\S~\ref{sec:mass.blowout}).  This can also be seen in
Figure~\ref{fig:bhmf}, where the dashed line shows the inferred BHMF
if we allow for the modified $M_{\rm BH}-\sigma$ relation predicted
from this fueling mechanism in \S~\ref{sec:m.sigma}; in other words
allowing for this fueling mechanism to grow small-mass black holes
substantially.  Because this also results in an increased scatter in
the relation, the effect is ``blurred out,'' and we see in
Figure~\ref{fig:bhmf} that the net effect on the black hole mass
function is small, within observational uncertainty. Therefore, while
the contribution to the mass growth of low-mass black holes is
non-negligible (although probably insufficient to grow them by orders
of magnitude from much smaller seeds), the contribution to the
integrated black hole mass buildup and cumulative black hole mass
density is small, satisfying e.g.\ the constraints from \citet{H05e}
that non-merger driven fueling mechanisms are relatively unimportant in 
these integrated quantities.

Given this small contribution to the integrated black hole mass
density, it follows from energetics that the contribution from this
fueling mechanism to the X-ray background is also small.  Although
observations \citep[e.g.,][]{Miyaji00,Ueda03,Barger05} have suggested
that much of the X-ray background is built up at $z<1$, where this
fueling mechanism is relatively more important, we find a significant
contribution only at faint luminosities, which only approach the
break in the quasar luminosity function for $z\lesssim0.5$.  This is
consistent with our previous estimates for the X-ray background
spectrum determined by merger-induced quasar activity in \citet{H05e},
which also suggests that other mechanisms should not dominate the
X-ray background, even at redshifts $z<1$.  This distinction cautions
against extrapolating models for obscuration based on local AGN (e.g.\
traditional toroidal models) in X-ray background synthesis
\citep[e.g.,][]{SW89,Madau94,Comastri95,Gilli99,Gilli01,TU05}, as
obscuration mechanisms in these different populations are distinct physically
\citep{H05b,H05d}, and indeed are observed to follow quantitatively
different, albeit qualitatively similar, trends in obscuration as a
function of luminosity (see \S~\ref{sec:obscuration}).

\section{Black Hole Growth and the $M_{\rm BH}-\sigma$ Relation}
\label{sec:m.sigma}

\subsection{Slope of the $M_{\rm BH}-\sigma$ Relation}
\label{sec:m.sigma.slope}

In our scenario, when a molecular cloud of mass $M_{\rm cl}$
encounters a black hole of mass $M_{\rm BH}$, there will be a short
period of high accretion.  For a cloud moving at a speed $\sim
c_{s}^{\rm disk}\sim10\,{\rm km\,s^{-1}}$, the Bondi accretion rate
will be
\begin{equation}
\dot{m}=\frac{4\pi\alpha G^{2} M_{\rm BH} \rho_{\rm cl} \tS}{(c_{s}^{\rm disk})^{3}}\gg1
\end{equation}
and accretion will be Eddington-limited with $\dot{m}=1$.
The accretion rate will drop rapidly once feedback energy
unbinds the cloud and heats or expels the surrounding gas.

If feedback impacts the gas energetically, then the ``blowout'' phase
begins roughly when the coupled radiant energy is comparable to the
cloud binding energy.  The system will follow the ``blast wave''
solution if it accretes at $\dot{m}\approx1$ for a time $\Delta t\sim
t_{\dot{m}}\sim R_{\rm cl}/\sigma$ (see \S~\ref{sec:cloud.timescale}),
and an energy sufficient to unbind the cloud (i.e.\ a fraction
$f_{b}\sim 1$ of the binding energy of the cloud) couples to the
gas in a shorter time.  This timescale is less than or comparable to
the dynamical time of the cloud $\sim1/\sqrt{G\rho}$ and the cloud
crossing time $R/c_{s}^{\rm disk}$ (i.e.\ timescale of interaction
between the black hole and bulk of the cloud), so it does not matter
which we use in our analysis. They are also $\ll \tS$, the Salpeter
time, so we are justified in assuming that the black hole mass is
approximately constant and the accretion rate is $\dot{M}_{\rm
Edd}=M_{\rm BH}/\tS$ over this interval.  The blowout criterion
then becomes
\begin{equation}
\eta\,L\,\Delta t=\eta\,\epsilon_{r}\,\frac{M_{\rm BH}}{\tS}\,t_{\dot m}\,c^{2}
\gtrsim f_{b}\,\ M_{\rm cl}\,\tilde{\phi}\,\sigma^{2}
\end{equation}
where $\eta$ is the feedback coupling efficiency and $\tilde{\phi}$ is
a numerical coefficient which depends on the bulge profile and gives
the binding energy at $r=0$ as a function of the observed $\sigma$
(i.e.\ line-of-sight averaged velocity dispersion within the effective
radius). For a \citet{Hernquist90} spheroid,
$\tilde{\phi}=10.1$, so this factor is significant.  This then determines a
minimum black hole mass for blowout of
\begin{eqnarray}
\nonumber & & M_{\rm BH}\approx 1.25\,\alpha\,\times 10^{7}\,M_{\sun}\,
\bigfrac{\sigma}{100\,{\rm km\,s^{-1}}}^{3}\\
& & \alpha=\bigfrac{n_{\rm cl}}{100\,{\rm cm^{-3}}}\,
\bigfrac{R_{\rm cl}}{100\,{\rm pc}}^{2}\,
\bigfrac{0.001}{\eta\,\tau_{\ast}/f_{b}}
\label{eqn:m.sigma.E}
\end{eqnarray}
where $\tau_{\ast}$ is defined in Equation~(\ref{eqn:tmdot.calc}).  If
the black hole is initially formed along with the bulge as is expected
from models of the $M_{\rm BH}-\sigma$ relation with $M_{\rm
BH}\propto \sigma^{4}$, it may therefore have to grow 
by a non-negligible factor at small-$\sigma$ before a blowout occurs, and 
this implies a slightly shallower $M_{\rm
BH}-\sigma$ relation at low masses with $M_{\rm BH}\propto
\sigma^{3}$. 

The above derivation is similar to that of \citet{SR98}, except that
the characteristic timescale of an event and total gas mass are set
externally and depend differently on $\sigma$, from e.g.\ the
dynamical time and gas mass in a merger.  However, it may not be the
case that an energy criterion determines the $M_{\rm BH}-\sigma$
relation.  For ``galaxy-scale'' co-formation of bulges and black holes
in mergers, the bulge mass goes as $M\propto \sigma^{4}$, and the
dynamical time goes as $\propto a/\sigma \propto \sigma$, suggesting
$M_{\rm BH}\propto \sigma^{5}$, so the actual relationship may be shallower.

A momentum-based coupling and ``blowout'' suggests $M_{\rm
BH}\propto \sigma^{4}$, in better agreement
with observations \citep{Murray05}.  For momentum coupling, the
injected momentum is $\eta L/c$, and the criterion becomes
\begin{equation}
\eta\frac{L}{c}\Delta t=\eta\epsilon_{r}\frac{M_{\rm BH}}{\tS}\,t_{\dot{m}}\,c
\gtrsim M_{\rm cl}\,\sigma \, ,
\end{equation}
which gives a minimum black hole mass
\begin{eqnarray}
\nonumber & & M_{\rm BH}=1.13\alpha\times10^{7}\,M_{\sun}\,\bigfrac{\sigma}{100\,{\rm km\,s^{-1}}}^{2}\\
& & \alpha=\bigfrac{n_{\rm cl}}{100\,{\rm cm^{-3}}}\,
\bigfrac{R_{\rm cl}}{100\,{\rm pc}}^{2}\,
\bigfrac{1.0}{\eta\,\tau_{\ast}/f_{b}} \, ,
\label{eqn:m.sigma.mom}
\end{eqnarray}
where the more efficient coupling ($\eta\sim1$) for momentum-driven 
feedback is adopted 
following \citet{Murray05}, when
the molecular gas is optically thick to the 
AGN radiation, at least outside the dust sublimation radius. 

If the typical spheroid $M-\sigma$ relation is 
given by 
\begin{equation}
M_{\rm BH}^{i}=1.35\times10^{8}\,M_{\sun}\,\bigfrac{\sigma}{200\,{\rm km\,s^{-1}}}^{4}
\end{equation}
then the minimum black hole mass required to expel a
molecular cloud is greater than $M_{\rm BH}^{i}$ at a mass of $\sim
10^{7}\,M_{\sun}$, or $\sigma\approx 100\,{\rm km\,s^{-1}}$.  In
general, the $M_{\rm BH}-\sigma$ relation at any $\sigma$ will be
given by the larger of either the ``bulge-formation'' $M_{\rm BH}\propto
\sigma^{4}$ or the relation determined above, $M_{\rm BH}\propto
\sigma^{3}$ (or $M_{\rm BH}\propto \sigma^{2}$ for momentum coupling).
Thus, at sufficiently low masses, the shallower relation
predicted above will dominate, yielding a shallower slope to the $M_{\rm
BH}-\sigma$ relationship and introducing a break or some curvature in
the relation.

The location of this break is only loosely constrained since the
coupling efficiency is not well-determined, and we use the
normalizations above as they give a good match to observations for the
merger-induced $M_{\rm BH}\propto \sigma^{4}$ relation
\citep[e.g.,][]{DSH05,Murray05}.  However, the slope and scatter (see
\S~\ref{sec:m.sigma.scatter} below) induced by this fueling mechanism 
are robust predictions independent
of these uncertainties, and the ``break'' location is reasonably
constrained to $\sim10^{6}-10^{7}\,M_{\sun}$ based on estimates of
the feedback efficiency from the normalization of the merger-driven
$M_{\rm BH}\propto\sigma^{4}$ relation \citep{DSH05}, although this
does not apply if e.g.\ the accretion mode and coupling change at
lower masses or luminosities.

\begin{figure*}
    \centering
    \plotone{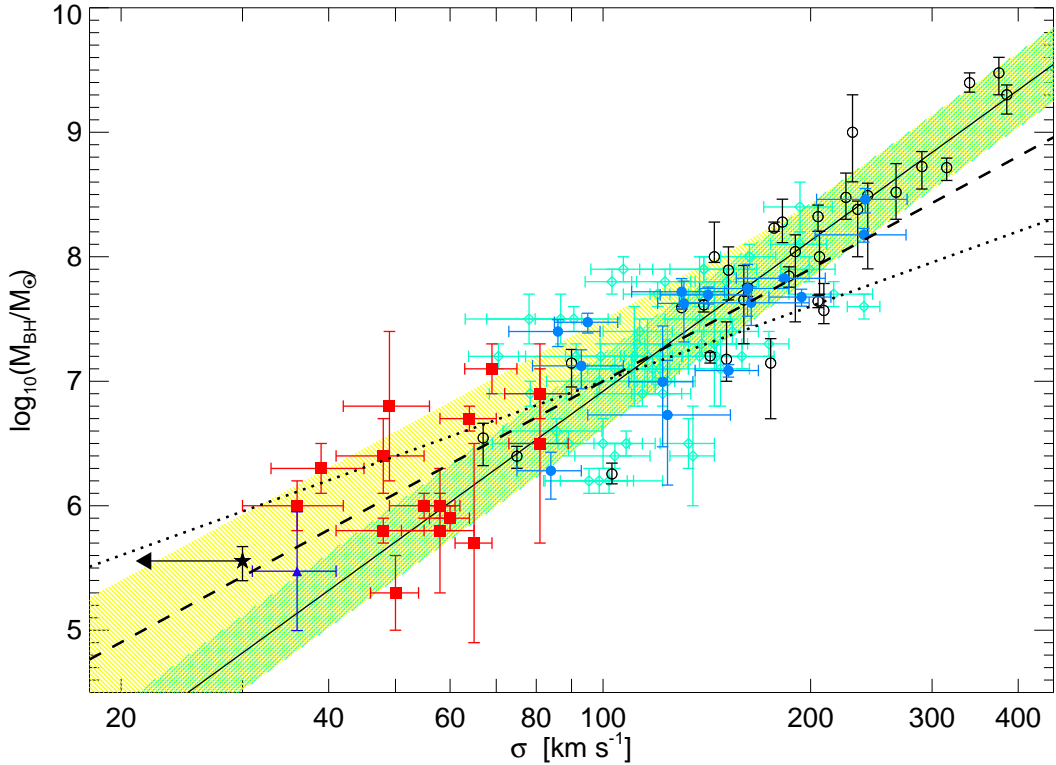}
    \caption{$M_{\rm BH}-\sigma$ relation measured by 
    \citet{Tremaine02} (black solid line; $M_{\rm BH}\propto\sigma^{4}$), 
    with estimated intrinsic 
    $0.27$\,dex scatter (green shaded range), and the predicted 
    relation at low-$\sigma$ from our model for Seyfert triggering 
    as calculated in \S~\ref{sec:m.sigma.slope}, given 
    an energy-based feedback coupling 
    (dashed line; $M_{\rm BH}\propto\sigma^{3}$) 
    or momentum-based coupling (dotted line; $M_{\rm BH}\propto\sigma^{2}$). 
    Note that these relations apply only 
    at low $\sigma$ where they lie above the solid line. 
    Yellow range shows the additional scatter expected (for the 
    energy-based coupling, but similar for the momentum-based coupling) 
    calculated from the size spectrum of molecular clouds in \S~\ref{sec:m.sigma.scatter}. 
    For comparison, observations are shown from the compilations of 
    \citet{Tremaine02} (black open circles), 
    \citet{Onken04,Nelson04,Peterson04} (blue filled circles), 
    \citet{Barth05} (red squares), 
    \citet{Greene05} (cyan diamonds), 
    and of POX 52 from \citet{Barth04} (purple triangle) and 
    of NGC4395 from \citet{Peterson05} (black star), for which only an upper limit to 
    $\sigma$ exists \citep{FilippenkoHo03}. The observations tentatively 
    favor the shallower predicted slope and larger scatter at low-$M_{\rm BH}$ from 
    our model, but this remains to be tested with larger samples and more accurate 
    mass determinations. 
    \label{fig:m.sigma}}
\end{figure*}

In Figure~\ref{fig:m.sigma}, we show (solid line) the predicted
relation for both merger-driven co-formation of bulges and black holes
with $M_{\rm BH}\propto \sigma^{4}$ \citep{DSH05}, matching the
observed relation \citep{FM00,Granato00,Tremaine02}, and for our
Seyfert molecular-cloud mode of accretion as predicted by
Equation~(\ref{eqn:m.sigma.E}) for energy-driven coupling (dashed
line) and Equation~(\ref{eqn:m.sigma.mom}) for a momentum coupling
(dotted line).  For comparison, observations are shown from various
compilations as indicated in the figure caption.  Note that the
(highly uncertain) reverberation region geometric normalization factor
adopted in \citet{Greene05} differs slightly (by $\sim0.2$\,dex) from
that of \citet{Onken04,Nelson04,Peterson04,Barth05}.  Since we cannot
predict the normalization of the $M_{\rm BH}-\sigma$ relation to this
accuracy and are, in any case, interested only in the slope and
scatter of the relation, we adopt the same normalization for all the
samples used.

The comparison between the predicted energy-based relation and the
low-mass observations is suggestive.  We consider the 37 observed
systems below $\sigma\approx100\,{\rm km\,s^{-1}}$, primarily from the
\citet{Barth05} and \citet{Greene05} samples but also including Pox
52, NGC4395 (for which we adopt the upper limit $\sigma=30\,{\rm
km\,s^{-1}}$, as a lower value will only further favor our estimated
$M_{\rm BH}-\sigma$ relation over the extrapolated $M_{\rm BH}\propto
\sigma^{4}$ relation), and four \citet{Tremaine02} objects and four
\citet{Onken04} objects.  For these objects, if we ignore intrinsic
scatter in the relationship, the reduced $\chi^{2}$ of the
\citet{Tremaine02} fitted relation is $\chi^{2}/\nu=16.8$, whereas our
prediction gives $\chi^{2}/\nu=12.1$ (an absolute $\Delta
\chi^{2}\approx 174$).  These are both unacceptable fits, however, so
allowing for a constant intrinsic scatter of $0.27$\,dex gives
$\chi^{2}/\nu=1.72$ (marginally acceptable) for the \citet{Tremaine02}
fit, compared to $\chi^{2}/\nu=1.02$ for our prediction (absolute
$\Delta\chi^{2}\approx26$, still highly significant).  Further allowing
for the increased intrinsic scatter at low $\sigma$ we predict in
\S~\ref{sec:m.sigma.scatter} below, this drops to
$\chi^{2}/\nu=0.21$.  Above this $\sigma$, the comparison is not
meaningful because we do not predict any substantial correction to the
$M_{\rm BH}-\sigma$ relation.

The observations at low black hole masses appear to favor our
prediction over a pure bulge and black hole co-formation (i.e.\
merger-driven) $M_{\rm BH}-\sigma$ relation, which yields a scaling
$M_{\rm BH}\propto \sigma^{4}$, roughly independent of mass \citep[see
e.g.,][]{DSH05,Robertson05b}.  However, there is only a marginal
($\sim1-2\sigma$) detection of this trend, and both the measurement
errors and systematic uncertainties
\citep[e.g.,][]{Peterson05,Barth05,GH05} are large at low masses.
Future observations of the $M_{\rm BH}-\sigma$ relation at low masses 
will provide a strong test of our model, even potentially
distinguishing momentum coupling vs.\ energy coupling models.

\subsection{Scatter in the $M_{\rm BH}-\sigma$ Relation}
\label{sec:m.sigma.scatter}

If molecular cloud accretion is important for the growth of low-mass
black holes, this mode can influence the scatter in the observed $M_{\rm
BH}-\sigma$ relation.  For example, the critical $M_{\rm BH}$ for
entering the ``blowout'' phase in Equation~(\ref{eqn:m.sigma.mom}) is
proportional to $R_{\rm cl}^{2}\propto M_{\rm cl}^{2/3}$, so a
spectrum of sizes of molecular clouds will introduce scatter in the
$M_{\rm BH}-\sigma$ relation.  This scatter will be significant if it
is larger than the $\sim0.3\,$dex scatter intrinsic to the $M_{\rm
BH}-\sigma$ relation from spheroid-black hole co-formation. The
observed scatter in the relationship is approximately constant at high
masses \citep[e.g.,][]{Tremaine02,Novak05}, and simulations of
merger-induced spheroid formation also predict the scatter to remain
roughly constant at low masses \citep{DSH05,Robertson05b}.

As discussed by \citet{MO77}, observations of column density
distributions in the galaxy \citep[e.g.,][]{Hobbs74,Payne83,Welty94},
LMC \citep[e.g.,][]{Oestreicher96} and AGN hosts
\citep[e.g.,][]{Tytler87,Hopkins04} indicate that the number of clouds
per unit column density varies as approximately $N_{H}^{-2}$, implying
(for a constant cloud density) that the number of clouds
with radii in the range $R\rightarrow R+dR$ per unit volume is
$\propto R^{-4}\,dR$. The rate at which clouds collide with the black
hole is
\begin{equation}
\frac{dn_{\rm event}}{dt}\sim n_{\rm cl}\,\pi\,R_{\rm cl}^{2}\,v_{\rm cl}
\propto \frac{1}{R}\,d\ln{R} \propto \frac{1}{M_{\rm cl}^{1/3}}\,d\ln{M_{\rm cl}},
\label{eqn:P.of.R}
\end{equation}
where $n_{\rm cl}\propto R^{-4}\,dR$ and the independence
of $v_{\rm cl}$ and $R$ are used in the second equality.  Therefore, the
$+1\sigma$ ($68\%$ $R<R'$) collision probability corresponds to a cloud
about four times larger in radius than our ``typical'' cloud,
i.e.\ $M_{\rm cl}\sim70$ times bigger than normal, or $M_{\rm
cl}^{2/3}\sim17.1$ times normal. The $+1\sigma$ dispersion in the
$M_{\rm BH}-\sigma$ relation should then be $\approx1.2$\,dex at the
lowest $M_{\rm BH}$, larger than the $\sim0.3$\,dex
intrinsic to the $M_{\rm BH}-\sigma$ relation determined by
co-formation of bulge and black hole.

The lower-$M_{\rm BH}$ limit to the molecular cloud accretion $M_{\rm
BH}-\sigma$ relation will be determined by extrapolating the $M_{\rm
BH}-\sigma$ relation for co-formation of spheroids and bulges.
Although technically the PDF for encountering a cloud of size $R$
(Equation~[\ref{eqn:P.of.R}]) can be estimated down to the minimum
cloud size $\sim0.5\,$pc, this is unimportant for the scatter in the
$M_{\rm BH}-\sigma$ relation.  For $R\ll 10-100\,$pc (i.e.\ less than
``typical'' large cloud sizes), the minimum black hole mass to
``blowout'' from Equation~(\ref{eqn:m.sigma.mom}) is small, e.g.\
$M_{\rm BH}\sim10^{4}\,M_{\sun}$ for $R\ll 10\,$pc (and recall that we
are considering intervals in $\log{R}$). Thus the contribution to the $M_{\rm BH}-\sigma$ 
relation from
these small clouds (with masses $\lesssim10^{4}\,M_{\sun}$ for
$R\lesssim10\,$pc) would be larger than the $M_{\rm BH}$ expected from
the co-formation of spheroid and bulge only for very small systems
$\sigma\lesssim25\,{\rm km\,s^{-1}}$, below the range of interest
(although these terms do contribute a Coulomb logarithm term to the
rate of black hole activation, see \S~\ref{sec:rates}).  However,
since the typical timescale for an encounter with a sufficiently large
cloud is $t_{\rm event}\sim10^{10}\,$yr, up to $\exp(-t_{H}/t_{\rm
event})\sim21\%$ of systems may not have had an encounter with a
molecular cloud of substantial size at any point, and will have an
$M_{\rm BH}$ consistent with the $M_{\rm BH}-\sigma$ relation for
black hole-spheroid formation ($M_{\rm BH}\propto \sigma^{4}$), making
this the approximate $-1\sigma$ lower limit in the $M_{\rm BH}-\sigma$
relation.

In Figure~\ref{fig:m.sigma} we show, in addition to the predicted
$M_{\rm BH}-\sigma$ relation from molecular cloud accretion, the
predicted scatter at each mass (combined yellow and green shaded
areas).  We assume a constant $0.27\,$dex scatter for the
spheroid-black hole co-formation $M_{\rm BH}-\sigma$ mechanism with
$M_{\rm BH}\propto\sigma^{4}$ \citep{Tremaine02}, shown as the shaded
green range.  We could equivalently show the scatter as a function of
mass from simulations of merger-driven spheroid formation
\citep{Robertson05b}, which span a wide range in final bulge/spheroid
masses and have approximately constant $\sim0.3$\,dex scatter.

The observations indicate that the scatter increases at small black
hole mass.  \citet{Greene05} estimate a scatter of $\sim0.4$\,dex
primarily in the range $\sigma\sim100-200\,{\rm km\,s^{-1}}$, and this
appears to increase marginally to $\sim0.6-0.8$\,dex if we consider
the objects below $\sigma\sim75\,{\rm km\,s^{-1}}$.  The increased
scatter our model predicts at low black hole masses is
evident. Quantitatively, however, this change in scatter in the observations
is only marginally ($\sim1\sigma$) significant if we consider points
at $\sigma<200\,{\rm km\,s^{-1}}$, where the predicted (yellow)
scatter begins to increase over the constant (green) scatter, owing to
both the relatively small number of points at the lowest masses and
the relatively large measurement errors.

\subsection{Accretion of Molecular Clouds at Low $M_{\rm BH}$}
\label{sec:low.m.timescale}

The condition for blowout, assuming the
black hole energy or momentum couples to the surrounding medium,
implies a shallower $M_{\rm BH}-\sigma$ relation, with increased
scatter, at low black hole masses.  However, a precondition for this
is that the black hole be able to efficiently accrete a molecular
cloud. If the cloud is moving at a velocity $c_{s}^{\rm disk}$, then
the crossing timescale during which the cloud interacts with the black
hole is
\begin{equation}
t_{\rm cross}=\frac{R}{c_{s}^{\rm disk}}=1.0\times10^{7}\,{\rm yr}\,
\bigfrac{R}{100\,{\rm pc}}\,\bigfrac{c_{s}^{\rm disk}}{10\,{\rm km\,s^{-1}}}^{-1} \, ,
\end{equation}
significantly shorter than the Salpeter time, $\tS=4.2\times10^{7}\,$yr, 
so the black hole cannot grow by more than a factor of $\sim3$ in mass 
in this time interval. Any remaining accretion must come 
from mass captured in the passage of the cloud. 

For a cloud moving at a bulk velocity $c_{s}^{\rm disk}$, 
the portion of the cloud which passes within a radius 
\begin{eqnarray}
R&\lesssim& \frac{2\,G\,M_{\rm BH}}{(c_{s}^{\rm disk})^{2}}\nonumber\\
&\approx& 89\,{\rm pc}\,\bigfrac{M_{\rm BH}}{10^{6}\,M_{\sun}}
\end{eqnarray}
can be ``captured'' (i.e.\ bound to the black hole).  On average, a
fraction $\sim (R/R_{\rm cl})^{2}$ (given by the cross section of a
cloud passing over a black hole which does not successfully capture
it) of the cloud volume/mass is then available for accretion. For
large black holes, this is more than sufficient to power
the blowout, and indeed much will not be accreted once the blowout
begins. For small enough black holes, however,
relatively little mass can be captured.  The captured mass is
\begin{equation}
\frac{M_{\rm capt}}{M_{\rm BH}}
\sim 10\,\bigfrac{M_{\rm BH}}{10^{6}\,M_{\sun}}\,
\bigfrac{R_{\rm cl}}{100\,{\rm pc}}
\end{equation}
(with a maximum at $M_{\rm capt}=M_{\rm cl}$) and so a low-mass black
hole can grow only by at most about one order of magnitude from
an individual interaction with a typical cloud. If the black hole
grows rapidly from the beginning of the interaction, this will still
only increase this by a further factor $\sim2-3$, based on the
comparison of the crossing time to the Salpeter time above.  For the
black hole masses plotted in Figure~\ref{fig:m.sigma},
this is sufficient to account for the movement of black
holes from the spheroid-black hole co-formation $M_{\rm BH}-\sigma$
relation ($M_{\rm BH}\propto\sigma^{4}$) to the molecular
cloud-dominated $M_{\rm BH}-\sigma$ relation ($M_{\rm
BH}\propto\sigma^{3}$), and to provide the mass to power the
``blowout''. Of course, a significant ($+1\sigma$) fraction of black holes can
interact with larger clouds or torii/disk inflows of sizes up to
$\sim400\,$pc (as calculated in \S~\ref{sec:m.sigma.scatter} above),
allowing a factor $\sim200$ growth in the black hole mass.  

However, for low initial black hole masses
$\sim10^{2}-10^{3}\,M_{\sun}$, as for seed black holes from Population
III stars, this is only marginally sufficient to grow black holes to
substantial masses (to subsequently act as seeds for further Seyfert
accretion or growth in mergers).  These seeds likely formed at high
redshifts $z\sim10-20$, when gas densities were large and gas
fractions were of order unity, implying a higher effective ``collision
rate'' with dense gas. The Eddington ratio in these phases for
undermassive black holes is $\dot{m}\approx1$, so the timescale for
this growth from the captured mass is $\approx1-2\times10^{8}$\,yr,
short enough compared to the age of the Universe for the buildup
of seeds which could then grow larger in later mergers.

\subsection{Mass Gain in the Blowout Phase}
\label{sec:mass.blowout}

Black holes can also grow during the ``blowout''
phase.  For a power-law decay of the accretion rate with
$\mdot=(t/t_{\dot{m}})^{-\eta_{L}}$
(Equation~[\ref{eqn:mdot.pwrlaw}]), the mass gained can be
estimated simply if it is small
and we can approximate $\dot{M}=\dot{m}\,\dot{M}_{\rm Edd}
=\dot{m}\,M_{\rm BH}/\tS$, where $M_{\rm BH}$ is the initial black
hole mass.  This yields
\begin{eqnarray}
\frac{\Delta M_{\rm BH}}{M_{\rm BH}}&=&\int \frac{\dot{M}}{M_{\rm BH}} {\rm d}t=
\int \frac{{\rm d}t}{{\rm d}\log{\dot{m}}}\,\frac{\dot{m}}{\tS}\,{\rm d}\log{\dot{m}}\nonumber\\
&=&\eta_{L}\,\frac{t_{\dot{m}}}{t_{S}}\int^{1}_{\dot{m}_{\rm min}}\dot{m}^{-1/\eta_{L}}\,{\rm d}\dot{m}\nonumber\\
&=&\frac{t_{\dot{m}}}{t_{S}}\ \frac{\eta_{L}}{1-1/\eta_{L}}[1-\dot{m}_{\rm min}^{1-1/\eta_{L}}].
\end{eqnarray}
Because $\dot{m}_{\rm min}\ll1$, if $\eta_{L}\geq1$ then 
the above becomes 
\begin{eqnarray}
\frac{\Delta{M_{\rm BH}}}{M_{\rm BH}}&\approx&
\frac{t_{\dot{m}}}{t_{S}}\ \frac{\eta_{L}}{1-1/\eta_{L}}\
\ \ \ \eta_{L}>1\nonumber \\
&=&\frac{t_{\dot{m}}}{t_{S}}\ \ln(\dot{m}_{\rm min}^{-1})\ 
\ \ \ \eta_{L}=1
\end{eqnarray}
$\sim t_{\dot{m}}/\tS\ll 1$, and indeed the mass gained in the blowout
is small. Similarly, $\eta_{L}=1$ contributes an additional
logarithmic term $\ln(1/\dot{m}_{\rm min})$, which still leaves
$\Delta{M_{\rm BH}}/M_{\rm BH}\ll1$.
  
For $\eta_{L}<1$, this becomes slightly more complicated, 
but ultimately gives a similar result. Because $\dot{m}_{\rm min}\ll1$, 
this implies 
$\Delta{M_{\rm BH}}/M_{\rm BH}\propto \dot{m}_{\rm min}^{1-1/\eta_{L}}$. 
Using the determination of $\dot{m}_{\rm min}$ from 
Equation~(\ref{eqn:mdot.min}), this simplifies to 
\begin{equation}
\frac{\Delta{M_{\rm BH}}}{M_{\rm BH}}
=f{(\eta_{L})}\,\bigfrac{\sigma}{100\,{\rm km\,s^{-1}}} \, ,
\end{equation}
where $f{(\eta_{L})}$ is a function of $\eta_{L}$ which 
varies smoothly from e.g.\ $f(\eta_{L}=1/2)=0.4$ 
to $f(\eta_{L}\rightarrow0)\rightarrow1.8$.
Therefore, for sufficiently low values of $\eta_{L}$, 
the accretion rate decays slowly enough 
that some mass gain is possible in the 
blowout.

However, the growth is still relatively small, a factor $\lesssim2$,
and less for lower-mass disks.  A full numerical calculation
allowing for the increase in mass with low-$\eta_{L}$ changing the
Eddington luminosity as the accretion rate decays also gives a small
correction of a factor $\lesssim2$ to the duty cycle at the lowest
luminosities. This is consistent with detailed simulations of X-ray
driven winds, which suggest efficient wind driving for $L/L_{\rm
Edd}>0.08$ \citep{BalsaraKrolik93}, implying (given the large
observed energetics of AGN outflows) that most of the energy of the
system should be generated in this early phase.  This is also a
potential source of increased scatter in the $M_{\rm BH}-\sigma$
relation, but is at most comparable to the observed scatter
\citep[e.g.,][]{Tremaine02,Novak05} and that expected from
merger-driven co-formation of spheroids and black holes, and is much
smaller than the scatter expected from the accretion of different
mass molecular clouds given the blowout conditions as discussed in
\S~\ref{sec:m.sigma.scatter}.

\section{Obscuration and the Classic Molecular Torus}
\label{sec:torus}

\subsection{Obscuration as a Function of Luminosity}
\label{sec:obscuration}

\begin{figure}
    \centering
    \epsscale{0.7}
    \plotone{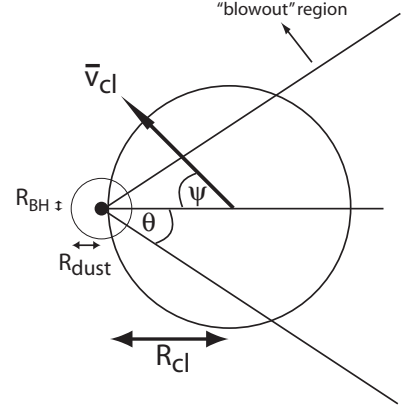}
    \caption{Illustration of
     the geometry at first interaction of a cloud and black hole, 
    with cloud radius $R_{\rm cl}$, black hole radius of influence $R_{\rm BH}$, dust sublimation 
    radius $R_{\rm dust}$, and cloud velocity $\bar{v}_{\rm cl}\sim c_{s}^{\rm disk}$. 
    \label{fig:obsc.diagram}}
    \epsscale{1}
\end{figure}

If a molecular cloud is accreted and not all is subsequently disrupted
by feedback, then some fraction of sightlines to the black hole will
pass through the surviving portion of the cloud and will be highly
obscured.  From our momentum coupling considerations of blowout, we
can estimate this fraction as a function of luminosity or black hole
mass.

Consider a cloud approaching the black hole at the time where the
black hole is about $\sim R_{\rm cl}$ from cloud center (i.e.\ first
``touching'' the cloud), as illustrated in
Figure~\ref{fig:obsc.diagram}.  Since $R_{\rm BH}\ll R_{\rm cl}$ this
indeed defines their ``initial'' interaction.  The cloud is moving
with a local velocity $c_{s}^{\rm disk}$, at an angle $\psi$ relative
to the axis connecting the black hole and cloud center.  Define
$\theta$ as the angle from this axis to a given point in the cloud,
with the black hole at the origin (defining a ray $\bar{r}$),
and let $\phi$ be the angle about this axis of symmetry.  Thus the
cloud velocity relative to the black hole is
$\bar{v}=c_{s}^{\rm disk}\,\hat{v}$ with
$\hat{v}=\sin{\psi}\,\hat{x}-\cos{\psi}\hat{z}$.  The linear momentum
of a given parcel of gas along such a ray originating at the black
hole, against the direction of the ray, is then
\begin{eqnarray}
{\rm d}{|{p}|}&=&{\rm d}M\,c_{s}^{\rm disk}\,(-\hat{r}\cdot\hat{v})\nonumber\\
&=&\rho_{\rm cl}\,c_{s}^{\rm disk}\,r^{2}\,\sin{\theta}\,(-\sin{\theta}\,\cos{\phi}\,\sin{\psi}+
\cos{\theta}\,\cos{\psi})\,{\rm d}r\,{\rm d}\theta\,{\rm d}\phi\nonumber\\
%&=2\pi\,\rho_{\rm cl}\,c_{s}^{\rm disk}\,r^{2}\,\sin{\theta}\,
%\cos{\theta}\,\cos{\psi}\,{\rm d}r\,{\rm d}\theta\\
&=&\frac{8\pi}{3}\,\rho_{\rm cl}\,R_{\rm cl}^{3}\,c_{s}^{\rm disk}\,\cos{\psi}\,\sin{\theta}\cos^{4}{\theta}\,
{\rm d}\theta\nonumber\\
&=&2\,M_{\rm cl}\,c_{s}^{\rm disk}\,\cos{\psi}\,\cos^{4}{\theta}\,{\rm d}{\cos{\theta}} \, ,
\end{eqnarray}
where the third equality comes from integrating over $\phi$ and $r$ (from 
$r\sim R_{\rm BH}\ll R_{\rm cl}$ to $r=2\,R_{\rm cl}\,\cos{\theta}$).

Consider the portion of the cloud with $\theta>\theta_{0}$, then the
total momentum input to ``clear'' a sightline in this fraction of the
cloud is
\begin{equation}
|p|=\frac{2}{5}\,\cos{\psi}\,M_{\rm cl}\,c_{s}^{\rm disk}\,\cos^{5}\theta_{0} \, .
\end{equation}
Since by our definitions $0\leq\psi\leq\pi/2$, 
$\EV{\cos{\psi}}=2/\pi$, so 
\begin{equation}
\EV{p}=\frac{4\pi}{5}M_{\rm cl}\,c_{s}^{\rm disk}\,\cos^{5}\theta_{0} \, .
\end{equation}

If the radiation from the black hole is directed isotropically and
impacts the cloud, at least initially, via momentum coupling, the
momentum imparted in some time interval $\Delta t$ is then ${\rm
d}p=\eta\,(L\,\Delta t/c)\,({\rm d}\Omega/4\pi)$, where ${\rm
d}\Omega$ is the opening angle considered and $\eta$ is the coupling
efficiency.  Considering the range above, with
$\theta_{0}\leq\theta\leq\pi/2$, this gives $\Delta\Omega=\int{\rm
d}\Omega=2\pi\,\cos{\theta_{0}}$, so the total momentum input from the
black hole in this region is
\begin{equation}
p_{\rm BH}=\frac{1}{2}\,\eta\,\frac{L\,\Delta t}{c}\,\cos{\theta_{0}} \, .
\end{equation}
This will be sufficient to overwhelm the 
momentum of the cloud and blow out material with $\theta>\theta_{0}$ 
for 
\begin{eqnarray}
\label{eqn:obsc.frac}
\cos{\theta_{0}}&>&\ {\Bigl[}\frac{5}{8\pi}\eta\,\frac{L\,\Delta t}{M_{\rm cl}\,c_{s}^{\rm disk}\,c}{\Bigr]}^{1/4}\\
&=&{\Bigl[}\frac{1}{4\pi}\eta\,\epsilon_{r}\,\dot{m}\,\frac{M_{\rm BH}}{M_{\rm cl}}\,
\frac{R_{\rm cl}\,c\,\tau_{\ast}}{\tS\,c_{s}^{\rm disk}\,\sigma}{\Bigr]}^{1/4}\nonumber\\
&=& 0.8\,(\eta\,\tau_{\ast}\,\dot{m})^{1/4}\,
\bigfrac{R_{\rm cl}}{100\,{\rm pc}}^{-1/2}\,\bigfrac{M_{\rm BH}}{10^{7}\,M_{\sun}}^{3/16}\nonumber\\
&\approx& 0.6\,(\eta\,\tau_{\ast})^{1/4}\,
\bigfrac{R_{\rm cl}}{100\,{\rm pc}}^{-1/2}\,
\bigfrac{L}{10^{11}\,L_{\sun}}^{1/4} \, ,\nonumber
\end{eqnarray}
where we have used $\Delta t=t_{\dot{m}}\sim R_{\rm cl}/\sigma$, the
characteristic timescale for the blast wave impact.

Thus, for sufficiently low luminosities or black hole masses, there is
some solid angle in which feedback cannot ``punch through''
the cloud, and the cloud will remain as an obscurer.  The covering
angle for this is
\begin{equation}
f=\frac{1}{2}(1-\cos{\theta_{0}}) \, .
\label{eqn:f.obsc}
\end{equation}
This is similar to our derivation of the blowout criterion as a function 
of mass in \S~\ref{sec:m.sigma.slope}, and here we incorporate the 
geometric effects of ``partial'' blowout.  These partially
obscured objects will 
be visible as Seyfert 2's, with properties consistent with the canonical 
molecular torus unification model.

\begin{figure}
    \centering
    \plotone{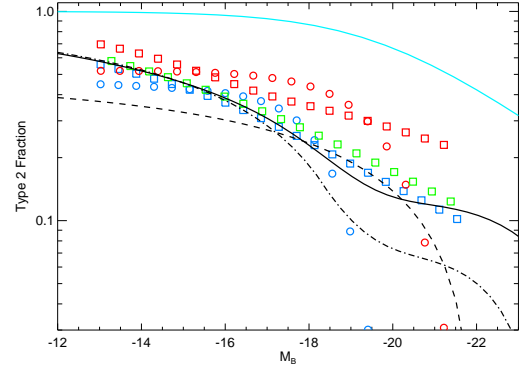}
    \caption{Predicted Seyfert 2 fraction as a function of luminosity, 
    calculated from our modeling of a pressure-driven blast wave (black lines). 
    Dashed line uses the 
    simplified luminosity criterion of Equation~(\ref{eqn:obsc.frac}), 
    dot-dashed line includes the more detailed dependence on 
    $M_{\rm BH}$ (integrating over the high-accretion rate period), 
    solid line further includes integration over a spectrum of 
    cloud sizes. Points show the observed Type 2 fraction from 
    the local AGN luminosity functions of \citet{Hao05}, using 
    Schechter (circles) and double power-law (squares) function fits to 
    H$\alpha$ (blue), [O{II}] (red), and [O{III}] (green) 
    observations, converted as in Figure~\ref{fig:LF}. 
    For comparison, the fitted Type 2 fraction from \citet{Simpson05}, 
    determined from the samples therein and from \citet{Ueda03,Hasinger04,GRW04}, 
    and dominated by quasars with $0\lesssim z\lesssim 3$ (i.e.\ merger-driven 
    quasar activity and obscuration in our modeling; see Figure~\ref{fig:LF.z}
    and Hopkins et al.\ 2006a) 
    is shown (cyan line).
    \label{fig:obscuration}}
\end{figure}

In Figure~\ref{fig:obscuration}, we show the predicted fraction of
Seyfert 2's as a function of luminosity from the luminosity-dependent
form of Equation~(\ref{eqn:obsc.frac}).  For comparison, we plot the
estimated Seyfert 2 fraction from observations as indicated, in the
manner of Figure~\ref{fig:LF}.  The predicted scaling agrees with the
observed trend, and also defines a cutoff at $L\sim
10^{11}-10^{12}\,L_{\sun}$, i.e.\ $M_{B}\sim -19$ to $-21$, above
which the requirement for obscuration is $\cos{\theta_{0}}>1$; i.e.\
the black hole is sufficiently luminous to destroy the entire cloud
regardless of angle.

There are, however, two detailed effects which will smooth this
cutoff.  First, the pure luminosity form of
Equation~(\ref{eqn:obsc.frac}), while giving the appropriate scaling,
is not strictly appropriate since the blowout in a given direction may
not depend on the {\em instantaneous} Seyfert luminosity, which may
have decayed to quite low luminosities at later times, but on the
momentum injected in the initial, high-luminosity ``blowout''
phase. This is given by re-considering Equation~(\ref{eqn:obsc.frac}),
as a function of black hole mass, with $\dot{m}=1$ for a period
$t_{\dot{m}}$ as determined in \S~\ref{sec:cloud.timescale}
(technically we integrate from $\dot{m}=1$ to $\dot{m}\lesssim0.1$,
since we define a differential time per logarithmic interval in
accretion rate).  This then provides a more precise estimate of the
fraction which will have been ``blown out'' by the blast
wave.  Accounting for this yields the dot-dashed line in
Figure~\ref{fig:obscuration}. At a given luminosity, there is a
distribution of black hole masses at various accretion rates, and this
effect smooths the obscured fraction as a function of luminosity.
However, this still defines a characteristic $M_{\rm
BH}\sim10^{8}\,M_{\rm sun} \gg M_{\rm cl}$ above which the blowout
will completely disrupt the cloud, with a corresponding Eddington
luminosity $L\sim10^{12}\,L_{\sun}$.

The second effect is the spectrum of cloud sizes.  A more or less
massive cloud also implies a correspondingly smaller or larger
covering angle of material which cannot be expelled.  We use the
spectrum of cloud sizes in \S~\ref{sec:m.sigma.scatter} to calculate
the dispersion in the $M_{\rm BH}-\sigma$ relation expected from this
mode of accretion, and an identical calculation gives the dispersion
in $M_{\rm cl}$ and correspondingly, via
Equation~(\ref{eqn:obsc.frac}), the variation in obscured fractions as
a function of $M_{\rm BH}$.  Averaging over this distribution
yields the total obscured fraction as a function of $M_{\rm BH}$ and
$L$, shown as the solid black line in Figure~\ref{fig:obscuration}.

These more detailed effects produce significant differences at
intermediate masses, where the transition between obscured and
unobscured populations is smoothed out.  The agreement with the
observations is improved, suggesting that this momentum balance
argument provides a plausible estimate of the obscured fraction as a
function of luminosity.

Note also that our argument does not necessarily depend on the black
hole luminosity affecting {\em directly} the surrounding medium via
radiation pressure or other momentum couplings. Rather, because the
blast wave impacting the medium will be pressure-driven, $\eta$ as
used above represents an effective efficiency of the Seyfert energy
and momentum ``ending up'' in the initially driven blast wave by the
time it impacts on the surrounding medium.

Our prediction is qualitatively similar to those of various modified
``luminosity-dependent'' torus models, and the fundamental qualitative
point, that black hole feedback which is stronger in higher-luminosity
systems is able to clear a larger opening angle, still obtains.
However, our result is distinct from some of these models in several
respects.  For example, the ``receding torus'' model
\citep[e.g.,][]{Lawrence91} assumes that the opening angle increases as the
inner torus radius increases, owing to a larger dust sublimation
radius $r_{\rm dust}\propto L^{1/2}$ with increasing dissociating
luminosity.  However, while this gives a reasonable scaling of the
obscured fraction for a geometrically flat (i.e.\ rectangular torus
cross-section) torus, it is clear from Figure~\ref{fig:obsc.diagram}
that for an increasing torus radius with $r$ (i.e.\ a circular torus
cross-section or elliptical/spherical cloud, and conical opening angle
models), removing a spherical region around the origin at the black
hole has almost no effect on the covering angle of the cloud.  This is
true until $r_{\rm dust}\gg R_{\rm cl}$ (for example, even at $r_{\rm
dust}=R_{\rm cl}$, for a spherical cloud or circular torus cross
section, this only decreases the Type 2 fraction by a factor of two from
$r_{\rm dust}=0$).  In other words, the dust sublimation radius would
have to reach $\sim10-100\,$pc before this mechanism would explain a
scaling of obscured fraction with luminosity in this fueling scenario,
into (or even beyond) the traditional narrow line regions.

Qualitatively, our mechanism is more similar to the disk-wind model
of \citet{KoniglKartje94,Elvis00,Proga04}, in which radiation pressure
flattens the initially vertical outflow and dust distribution in
objects with higher bolometric luminosities.  However, it can operate
on much larger scales and depends on momentum balance from the blast
wave (not necessarily direct radiation pressure) ejecting material
within some opening angle, rather than merely flattening the
distribution. It is also worth noting the striking similarity of the
obscured fraction calculated from Equations~(\ref{eqn:obsc.frac}) and
(\ref{eqn:f.obsc}) and the functional form of empirical calculations
which allow for e.g.\ evolution of the inner torus radius and torus
height as a function of luminosity \citep[e.g.,][]{Simpson05},
suggesting that feedback balance does capture the critical dependence
on black hole and fueling properties.

For comparison, the fitted Type 2 fraction as a function of luminosity
in higher-redshift quasars from \citet{Simpson05}, determined from the
SDSS and from \citet{Ueda03,Hasinger04,GRW04}, is shown as the cyan
line in Figure~\ref{fig:obscuration}.  While a similar qualitative
trend is apparent, the Type 2 fraction is much higher, completely
dominating at low-luminosities and contributing a substantial
$\sim30\%$ portion of the population even at $M_{B}\lesssim-23$,
whereas in the local samples the Type 2 fraction is at most comparable
to the Type 1 fraction at low luminosities and negligible ($\ll 10\%$)
at high luminosities.

The cause of this difference is suggested by Figure~\ref{fig:LF.z}, in
which the Seyfert (i.e.\ quiescent, cold gas driven) and quasar (i.e.\
merger-driven) luminosity functions are each plotted at several
redshifts.  Even by $z\sim0.1-0.2$, the quasar population is larger at
these luminosities, and indeed the \citet{Simpson05} sample is drawn
from surveys which are dominated by quasars at high redshifts spanning
the wide range $0.15\lesssim z\lesssim 3$. These objects, in our
modeling, are triggered by different processes, and therefore should
not necessarily follow the same trend -- indeed, the discrepancy
between the two observations cautions against extrapolation of
obscuration models (e.g.\ geometrical models) calibrated for local,
non-interacting populations. In \citet{H05a,H05b,H05e} we discuss this
obscuration in a merger-driven context in detail, and show that the
trends of \citet{Ueda03,Hasinger04,GRW04,Simpson05} can be reproduced
when the primary source of obscuration is produced by larger scale
($\gtrsim50$\,pc) gas inflows from the mergers which themselves power
accretion, and the distribution of column densities derives primarily
from time-dependent rather than line-of-sight dependent differences. A
similar qualitative trend is found because bright systems are more
likely to be near their peak luminosity, when they are able to expel
obscuring gas and dust, which is similar in principle to the model we
have described here for obscuration of cloud-fueled AGN.  However, in
our model for non-interacting Seyferts, the obscuration is removed
along certain sightlines at the beginning, not at the end of black
hole growth, and the distribution of $N_{H}$ does, presumably, derive
from geometric effects, as in canonical torus or disk-wind models.

\subsection{Building a Torus}
\label{sec:torus.properties}
 
Although it is not a requirement of this model, our picture for AGN
fueling can both derive from and seed the traditional molecular torus
of standard AGN unification models.  Assuming that a significant
fraction of an accreted cloud is captured on passage near the black
hole, essentially requiring $R_{\rm cl}\lesssim G\,M_{\rm
BH}\,(c_{s}^{\rm disk})^{-2}$, or $M_{\rm BH}\gtrsim 10^{5}\,M_{\sun}$
(ensured for the black hole masses of interest and for the predicted
$M_{\rm BH}-\sigma$ relation for this fueling mechanism, see
\S~\ref{sec:m.sigma}), then it will presumably develop an orbit,
circularize, and some fraction of the cloud will be tidally disrupted.

Given an initial impact parameter $b$ and velocity $c_{s}^{\rm disk}$,
the material can settle into a Keplerian orbit determined by
conservation of angular momentum. For the bulk of the cloud, $b\sim
R_{\rm cl}\gg R_{\rm BH}$, so the circular velocity is determined by
the spheroid potential,
\begin{equation}
v_{c}=\sigma\,\sqrt{\frac{r}{a}}
\end{equation}
where $a$ is the bulge scale-length. Using $a=10\,{\rm kpc}\,(\sigma/{\rm 200\,km\,s^{-1}})^{2}$ from 
\S~\ref{sec:merger.timescale} with the appropriate factor of $\tilde{\phi}$ to convert from 
observed $\sigma$ to a physical $v_{c}$ and $a$ assuming a Hernquist (1990) spheroid profile 
yields $v_{c}\approx200\,{\rm km\,s^{-1}}\,\sqrt{r/{\rm kpc}}$. 
Since this will include most of the mass, demanding the final angular
momentum $\sim M\,r\,v_{c}$ (in detail $=(3
\tilde{h}^{2}/4+1)\,M\,r\,v_{c}$ for a torus with $\tilde{h}=h/r$)
conserve the initial $\sim M\,R_{\rm cl}\,c_{s}^{\rm disk}$ (again, in
detail $M\,R_{\rm cl}\,c_{s}^{\rm disk}\,\sin{\psi}$ for the geometry
in \S~\ref{sec:obscuration}) yields a torus radius
\begin{eqnarray}
R_{\rm t}&=&R_{\rm cl}\,\bigfrac{a}{R_{\rm cl}}^{1/3}\,\bigfrac{c_{s}^{\rm disk}}{\sigma}^{2/3}\,
\bigfrac{\EV{\sin{\psi}}}{\frac{3}{4}\tilde{h}^{2}+1}^{2/3}\nonumber\\
&\approx& 20\,{\rm pc}\,\bigfrac{R_{\rm cl}}{100\,{\rm pc}}^{2/3}\,\bigfrac{c_{s}^{\rm disk}}{10\,{\rm km\,s^{-1}}} \, ,
\label{eqn:torus.size}
\end{eqnarray}
independent of $\sigma$, and assuming $\tilde{h}\sim1$ as we show
below.  Thus the characteristic torus size $\lesssim10$'s of pc (up to
a maximum $\sim100\,$pc) inferred from a wide variety of observations
\citep{KB88,GDF97,Bock00,Schinnerer00,Galliano03,Radomski03,
Weigelt04,Jaffe04,Prieto04,Elitzur05} and simulations of torus
structure and stability with radiative transfer models
\citep[e.g.,][]{PK93,GD94} is a natural consequence of the scales of
molecular clouds, and is ``built into'' this picture of Seyfert
fueling.

Given this radius, whether the molecular cloud is circularized into a
strict torus with volume $2\pi^{2}\,R_{\rm t}^{3}$ (assuming
$\tilde{h}\sim1$) or remains spherical/ellipsoidal, the change in
e.g.\ density and other gas properties is well-defined (varying by
e.g.\ $20\%$ using the exact solutions above and assuming it goes from
a sphere to a torus).  Therefore, the characteristic density is also
given by that of the original cloud, enhanced by the compression
expected from Equation~(\ref{eqn:torus.size}) above, explaining the
molecular and dust densities and masses inferred for molecular torii
around Seyferts, with $n\sim10^{2}-10^{4}\,{\rm cm}^{-3}$.  This gives
a line-of-sight column density through the torus plane of
\begin{equation}
N_{\rm H{\small I}}=2\,R_{\rm t}\,n_{\rm t}\approx 2\,R_{\rm cl}\,n_{\rm cl}\,
\bigfrac{R_{\rm cl}}{R_{\rm t}}^{2}
\approx1.6\times10^{24}\,{\rm cm^{-2}}.
\end{equation}
This is similar to the column densities of torii inferred from
phenomenological fits to the complete hard X-ray column density
distributions of Seyferts \citep{UP95,Treister04,Mainieri05} and from
synthesis models fitted to the X-ray background
\citep{SW89,Madau94,Comastri95,Gilli99,Gilli01,TU05}.

We can also calculate the expected height or flattening of this torus,
as a function of radius and Seyfert properties.  If the molecular
cloud, with density $n_{\rm cl}\sim100\,{\rm cm^{-3}}$ collapses from
the ISM with mean density $n_{\rm ISM}\sim1\,{\rm cm^{-3}}$, and
magnetic flux ($\equiv B\,{\rm d}A\sim B\,R^{2}$) is conserved, the
magnetic field will be enhanced by $B_{\rm cl}/B_{\rm ISM}\sim (n_{\rm
cl}/n_{\rm ISM})^{2/3}$.  For typical magnetic fields providing
pressure support in molecular clouds
\begin{equation}
\frac{\partial P}{\partial r}\sim\frac{1}{R_{\rm cl}}\,\frac{B_{\rm cl}^{2}}{4\pi}
=\rho\,\frac{\partial \phi}{\partial r}\sim \rho\,\frac{G\,M_{\rm cl}}{R_{\rm cl}^{2}} \, ,
\end{equation}
with $B_{\rm cl}=B_{\rm ISM}\,(n_{\rm cl}/n_{\rm ISM})^{2/3}$ and 
$M_{\rm cl}=(4\pi/3)\,R_{\rm cl}^{3}\,\rho$. 
For a diffuse ISM $B_{\rm ISM}\sim$ a few $\mu G$, 
this yields
\begin{equation}
R_{\rm cl}\sim 100\,{\rm pc}\,\bigfrac{B_{\rm ISM}}{4\,\mu G} \, ,
\end{equation}
giving the standard self-consistent picture for the magnetic support
(in addition to support from turbulent motions) for molecular clouds,
with these field strengths characteristic of the disordered fields of
the Milky Way ISM (e.g.\ Troland \& Heiles 1986; Rand \& Kulkarni
1989; for a review see e.g.\ Ferriere 2001).  Increasing magnetic
field strengths are also observed towards the center of the Galaxy, up
to a factor of $\sim10$ larger, supporting higher densities and even
larger giant molecular cloud complexes with $M_{\rm cl}$ up to
$\sim10^{8}\,M_{\sun}$
\citep[e.g.,][]{RandLyne94,Myers95a,Myers95b,Crutcher99}.

Because capture and circularization does not completely erase the
properties of the molecular cloud, the magnetic fields on cloud scales
can be preserved, continuing to provide pressure support.
This pressure defines an effective sound speed
$c_{B}^{2}=(B_{\rm cl}^{2}/4\pi)\,\rho^{-1}$, which for the above is
$c_{B}\approx 20\,{\rm km\,s^{-1}}\,(B_{\rm ISM}/4\,\mu G)$.  Note
that any other source of a similar pressure which is non-thermal
(i.e.\ does not scale adiabatically), such as e.g.\ turbulent motions,
will have a similar effect.  Since this dominates the pressure support
of the torus, and using the circular velocity at $R_{\rm t}\gg R_{\rm
BH}$ (i.e.\ dominated by the spheroid potential) the scale height is
\begin{eqnarray}
\tilde{h}&\equiv& \frac{H_{\rm t}}{R_{\rm t}}\sim \bigfrac{c_{B}}{v_{c}}
=\sqrt{\frac{a}{R_{\rm t}}}\,\bigfrac{c_{B}}{\sigma}\nonumber\\
&\approx& 0.70\,\bigfrac{R_{\rm cl}}{100\,{\rm pc}}^{-1/3} \, ,
\end{eqnarray}
with a maximum of $H_{\rm t}\sim R_{\rm cl}$.  Thus, the torus is not
expected to be strongly compressed, at the outer radii, with $H_{\rm
t}\sim R_{\rm t}$, again similar to traditional phenomenological
models and observational estimates of the torus structure and covering
angle \citep{KB88,Antonucci93,UP95,RMS99,Schmitt01}.

%Some fraction of the initial cloud will have an impact 
%parameter $b\lesssim R_{\rm BH}$ and reach a circular 
%orbit dominated by the black hole potential, $v_{c}^{2}=G\,M_{\rm BH}/r$. 
%If we define $b=\beta\,R_{\rm BH}$ with $R_{\rm BH}=G\,M_{\rm BH}/\sigma^{2}$ 
%and consider the 
%conservation of specific angular momentum ($b\,c_{s}^{\rm disk}=r\,v_{c}$), 
%\begin{equation}
%r=\beta^{2}\,\frac{G\,M_{\rm BH}\,(c_{s}^{\rm disk})^{2}}{\sigma^{4}}
%=0.04\,{\rm pc}\,\beta^{2}\,\bigfrac{c_{s}^{\rm disk}}{10\,{\rm km\,s^{-1}}}^{2}
%\end{equation}
%where we have substituted the $M_{\rm BH}-\sigma$ relation in the second equality. 
%This can also be written as a ``compression factor''
%\begin{equation}
%\frac{r}{b}=0.01\,\beta\,\bigfrac{c_{s}^{\rm disk}}{10\,{\rm km\,s^{-1}}}^{2}
%\,\bigfrac{\sigma}{100\,{\rm km\,s^{-1}}}^{-2}
%\end{equation}
%Thus this small fraction $\sim (R_{\rm BH}/R_{\rm cl})^{3}\lesssim 10^{-4}$ 
%of the initial cloud mass can immediately 
%collapse to small scales as it is low angular momentum 
%and reaches radii where the black hole dominates the potential. This extreme collapse 
%can also greatly enhance the magnetic fields, giving initial seed fields 
%internal to $r(\beta)$ of 
%\begin{equation}
%B_{\rm seed}\sim B_{\rm cl}\,\bigfrac{b}{r}^{2}
%\sim 1\,G\,\frac{1}{\beta^{2}}\,\bigfrac{B_{\rm ISM}}{4\,\mu G}\,
%\bigfrac{M_{\rm BH}}{10^{7}\,M_{\sun}}
%\end{equation}
%Thus this also provides a substantial source of seed magnetic fields 
%to the accretion disk which may develop. 

The timescale for all of this to occur will be of order the orbital
timescale, $t_{\rm orb}\sim r/v_{c}$, or $\sim R_{\rm cl}/c_{s}^{\rm
disk} \sim 10^{7}\,{\rm yr}$ for the bulk of the cloud at $R_{\rm
t}\sim R_{\rm cl}$.  Thus, if the cloud is captured, it will begin to
reach an equilibrium in the blast wave and orbit with the circular
velocity in roughly a crossing time. It is therefore expected that it
will be seen as such over the majority of the Seyfert lifetime. There
is a period of time $\sim10^{7}\,$yr, comparable to the timescale at
highest accretion rates $\dot{m}\gtrsim 0.1$, during which the system
may be settling into equilibrium.  We have shown the transition from
cloud to torus implies little change in the cloud properties, so
observational signatures of this transition phase, if it occurs, may
not be obvious.

\section{Global Consequences for the Host Galaxy}
\label{sec:host.fx}

Given the effects of the black hole feedback we have modeled, and the
fact that the blast wave can propagate to large ($\sim\,$kpc) scales
on reasonable ($\sim\,$Gyr) timescales, in principle there could be a
significant impact on the host galaxy ISM or gas structure.  The black
hole feedback expected from growth in mergers can sweep a large
fraction of the remaining gas in the galaxy into an unbound blast
wave, heating the remaining gas to the virial temperature and
terminating star formation \citep{H05e,H05g,SDH05a}.  However, the
case of fueling by molecular cloud accretion is quite different, and
we find that the blast wave is much weaker than required to
significantly impact the host galaxy.

If the blowout is defined by the criterion that $M_{\rm cl}$ be
accelerated to $v_{\rm esc}\sim \sigma$ (i.e.\ that the local gas be
unbound), then because $M_{\rm cl}\ll M_{\rm ISM}$, a collision
between the blast wave and the ISM will at most accelerate the ISM to
$\sim (M_{\rm cl}/M_{\rm ISM})\,\sigma$, if momentum considerations
are most important. For typical Galactic scales this gives a bulk
speed $\ll 1\,{\rm km\,s^{-1}}$, which is completely negligible on the
scales of the ISM.  Even under the extreme assumption that all the
blast wave energy can be converted into bulk motion gives a velocity
$\sim \sqrt{M_{\rm cl}/M_{\rm ISM}}\,\sigma \lesssim1\,{\rm
km\,s^{-1}}$. Thus, the blast wave itself is not expected to
significantly impact the host galaxy ISM.

If, instead of the momentum or energy in the blast wave shell, we make
an even more extreme assumption that all of the feedback energy or
momentum flux from black hole growth impacts the galaxy
ISM, the energy input could be much larger. We further make the
maximal assumptions that $\eta_{L}$ is small, so the black hole grows
substantially (by a factor $\sim 2$, see \S~\ref{sec:mass.blowout}),
and that the affected ISM is unable to radiate.  Let us also consider
the case of direct energy coupling (i.e.\ conversion of all feedback
energy into bulk motion), which gives an increased impact of
feedback by a factor $\sim c/\sigma \sim 10^{3}$ over e.g.\
momentum-based coupling.  Then, the feedback energy from the black
hole is $\sim \eta\,\epsilon_{r}\,M_{\rm BH}\,c^{2}\,\Delta\Omega$,
where $\Delta\Omega$ is the covering angle of the disk to the black
hole, since the radiation is isotropic.  Note that collimation will
tend to be perpendicular to the disk and along the angular momentum
vector, meaning it will only moderate the impact on the host galaxy.

The binding energy of the disk gas is $\sim f_{\rm gas}\,M_{\rm
d}\,v_{c}^{2}$. Using the fact that $M_{\rm BH}\sim \mu M_{\rm b}$,
where $M_{\rm b}$ is the bulge mass, this implies that the black hole
feedback will be significant (i.e.\ comparable to the binding energy
of the disk ISM gas) when
\begin{equation}
v_{c}^{2}\sim \frac{\epsilon_{r}\,\eta\,\mu}{f_{\rm gas}}\,\frac{M_{\rm b}}{M_{\rm d}}
\,\Delta\Omega\,c^{2} \, .
\end{equation}
Since the disk covering angle (weighted over the mass/volume of the disk) is 
$\theta\sim h_{\rm d}/r_{\rm d}\sim c_{s}^{\rm d}/v_{c}$, and 
the covering factor is $(1-\cos{\theta})\rightarrow \theta^{2}/2$, and 
we expect $\mu\sim0.001$, $\eta\sim0.01$, $c_{s}^{\rm disk}\sim 10\,{\rm km\,s^{-1}}$, 
we can solve for the maximum $v_{c}$ for which black hole feedback will 
have a large impact. This gives
\begin{eqnarray}
v_{c}\sim 36\,{\rm km\,s^{-1}}\,{\Bigl[}& &
\bigfrac{\epsilon_{r}}{0.1}\,\bigfrac{f_{\rm gas}}{0.1}^{-1}\,
\bigfrac{\eta}{0.01}\,\nonumber\\
& &\times\bigfrac{\mu}{0.001}\,
\bigfrac{{\rm B/T}}{0.1}{\Bigr]}^{1/4}
\end{eqnarray}
i.e.\ $M_{\rm d}\lesssim 10^{9}\,M_{\sun}$ with 
$M_{\rm BH}\lesssim 10^{5}\,M_{\sun}$. 
Again, this assumes the entire energy output of the black hole has a chance to 
couple to the ISM with no allowance for that ISM to dynamically relax or cool, 
and thus represents the most extreme estimate of which galaxies could 
be affected.  Such low-mass galaxies will 
typically be later types than the Sa/b's with typical 
B/T$\sim0.1$ above, more likely $\lesssim 0.02$ 
for Sc/d and Sm/Im galaxies, and gas fractions potentially quite a bit higher than $\sim0.1$, 
making the host even more robust against black hole feedback.
If we consider a momentum coupling ($\sim \eta\epsilon_{r}\,M_{\rm BH}\,c$ 
compared to $f_{\rm gas}\,M_{\rm d}\,v_{c}$) criteria instead, then we obtain 
\begin{eqnarray}
v_{c}\sim 0.2\,{\rm km\,s^{-1}}\,{\Bigl[}& &
\bigfrac{\epsilon_{r}}{0.1}\,\bigfrac{f_{\rm gas}}{0.1}^{-1}\,
\bigfrac{\eta}{1.0}\,\nonumber\\
& &\times\bigfrac{\mu}{0.001}\,
\bigfrac{{\rm B/T}}{0.1}{\Bigr]}^{1/3} \, ,
\end{eqnarray}
a negligible effect.

\section{The Contribution of Stellar Wind Fueling}
\label{sec:winds}

For comparison, we briefly consider the contribution of
stellar winds and hot (virialized) gas fueling to AGN activity.
First, consider the contribution of stellar winds from the galaxy as a
whole -- these will, in general, shock against one another and
virialize, contributing to the hot gas reservoir which can be accreted
through the Bondi radius.  Assuming the gas is virialized gives
$c_{s}^{2}=3\tilde{\phi}\,\sigma^{2}$, where the dimensionless
factor $\tilde{\phi}$ depends on the density profile and converts
between the observed mean projected velocity dispersion within the
effective radius ($\sigma$) and the potential at $z=0$
($\tilde{\phi}\approx10.1$ for a Hernquist [1990] spheroid
profile). We further use the radial hydrostatic equilibrium condition
to solve for $\rho$, which gives a profile having a core with a characteristic
$n_{\rm ISM}\sim1\,{\rm cm^{-3}}$ at the center.  Using this and the
$M_{\rm BH}-\sigma$ relation, we obtain
\begin{eqnarray}
\dot{m}&=& 4\pi\alpha\,\frac{G^{2}\,M_{\rm BH}\,\tS\,\rho}{\sigma^{3}}\,
\bigfrac{\sigma^{2}}{c_{s}^{2}+v^{2}}^{3/2}\nonumber\\
&\approx& 6.0\times10^{-5}\,
\bigfrac{n_{\rm ISM}}{1\,{\rm cm^{-3}}}\,
\bigfrac{\sigma}{200\,{\rm km\,s^{-1}}}
\label{eqn:wind.total}
\end{eqnarray} 
where we have also taken 
$\gamma=5/3$ (giving $\alpha=1/4$ for the Bondi-Hoyle solution) 
in the second equality. 

\citet{Ciotti91} consider a detailed numerical calculation of
injection of gas into the hot ISM from steady stellar mass loss,
supernova events, and galaxy-scale inflows and outflows. For each of
their range of conditions, we consider the temperature, gas velocity,
and density at the trans-sonic radius for inflow (in the solutions
they consider with a central mass concentration) and compare the
implied Bondi rate with that estimated above, and arrive at a similar
result (to within a factor $\sim2-3$). This also agrees with the Bondi
rates estimated from systematic measurements of hot gas temperature
and density profiles in the inner regions of a number of quiescent
ellipticals \citep[e.g.,][]{Pellegrini05,Soria05b}. The characteristic
$\mdot\sim10^{-5}-10^{-4}$ from galaxy scale hot virialized gas
appears to be a robust result. Note that while the total stellar mass
loss rate of a quiescent elliptical can be $\sim 1\,M_{\sun}\,{\rm
yr^{-1}}$ under typical conditions, the calculations of
\citet{Ciotti91} and others have shown that a far smaller fraction of
this gas actually reaches the center of the galaxy, and it does so
either at super-$\sigma$ bulk velocities (if un-shocked) or
$\sim\sigma$ sound speeds (if shocked and virialized), giving a much
lower actual Bondi rate for black hole accretion.

Following \citet{Soria05a,Soria05b}, we also estimate the total
stellar mass loss within the black hole radius of influence. Winds
within this radius will be captured by the black hole, and be
available for accretion. Note that, strictly speaking, this is
appropriate only for wind velocities $v_{w}<\sqrt{\tilde{\phi}}\sigma$
(the spheroid escape velocity at the center), as otherwise the capture
radius is smaller than the radius of influence.  For typical slow wind
speeds $\sim100-300\,{\rm km\,s^{-1}}$, however, this is at most
comparable to typical ellipticals $\sqrt{\tilde{\phi}}\sigma$, making
this a relatively small source of uncertainty. This will, however,
restrict the ability of e.g.\ a young stellar cluster near the nucleus
(with much higher $\dot{M}$ in winds than older stellar populations)
to contribute significantly to moderate or high-$\dot{m}$ populations,
as with typical $v_{w}$ up to $\sim1000\,{\rm km\,s^{-1}}$, the
effective volume in which such systems can have their winds captured
is dramatically reduced, and furthermore the large bulk velocity
reduces the estimated Bondi rate.

For steady mass loss from older stellar populations, we adopt 
the estimate of \citet{Ciotti91} from population synthesis for the mass loss 
rates as a function of the spheroid age
\begin{equation}
\dot{M}(t)=1.5\times10^{-11}\,\bigfrac{L_{B}}{L_{\sun,\,B}}\,
\bigfrac{t}{15\,{\rm Gyr}}^{-1.3}\,M_{\sun}\,{\rm yr^{-1}}. 
\end{equation}
Using a constant mass-to-light ratio
$\lambda=(M_{\sun}/L_{\sun,\,B})/(M_{\rm bulge}/L_{B})$, and modeling
the spheroid stars with a Hernquist (1990) profile to calculate the
mass fraction within $R_{\rm BH}\equiv G\,M_{\rm BH}/\sigma^{2}$, then
using the black hole-bulge mass correlation of \citet{MH03} to cancel
the dependence on $M_{\rm BH}/M_{\rm bulge}$, we obtain
\begin{equation}
\dot{m}\approx1.1\times10^{-4}\,\lambda\,\bigfrac{M_{\rm BH}/M_{\rm bulge}}{0.001}\,
\bigfrac{t}{10\,{\rm Gyr}}^{-1.3}
\label{eqn:wind.inner}
\end{equation}
which, for reasonable values of $\lambda$ (generally below unity 
since these stellar populations are old) from the stellar population synthesis 
models of \citet{BC03} gives $\dot{m}$ in the range 
$\sim10^{-4}$ (for relatively young ellipticals with ages $\sim2-4$\,Gyr) 
to $\sim10^{-5}$ (for older ellipticals with ages $\sim10$\,Gyr). 

In either case, fueling from stellar winds and quiescent Bondi
accretion of hot gas are roughly comparable, within the range
$\sim10^{-5}-10^{-4}$, without a strong dependence on host galaxy
properties.  Given our modeling of quasar formation in mergers with a
subsequent decay in the quasar light curve determined by the evolution
of the feedback-driven blast wave \citep{H05g}, we then expect that
ellipticals will, shortly after forming, be at moderate, declining
accretion rates for some time, contributing to the $-18\gtrsim
M_{B}\gtrsim-23$ end of the predicted elliptical Seyfert population
(i.e.\ the bright end of the local Seyfert luminosity function, but
faint end of the merger-driven quasar luminosity function).  However,
once a sufficiently low accretion rate given by
Equations~(\ref{eqn:wind.total}) \& (\ref{eqn:wind.inner}) is reached,
the hot gas in virialized equilibrium and stellar mass loss of aging
stellar populations near the black hole sustains a nearly constant
low-level accretion rate.

Accretion rates of black holes in early-type galaxies will then decay
from the redshift when the elliptical formed to these Eddington ratios
and ``pile up'' at $\mdot\sim10^{-4}$.  This is supported by
direct estimates of the Eddington ratio distribution in quiescent,
``dead'' ellipticals \citep[e.g.,][]{Ho02,Marchesini04,H05i}.
Furthermore, lower mass ellipticals which form later (e.g.\ Cowie et
al.\ 1996; see Hopkins et al.\ 2006b for details of our calculation)
may not decay quite to these lowest $\dot{m}$ values, ultimately
overlapping in luminosity. Therefore, a large fraction of the
elliptical population fueled by such mechanisms will be found in a
relatively narrow range in accretion rate and luminosity.

To estimate the contribution to the luminosity function, we must
account for changes in the radiative efficiency
\citep[e.g.,][]{NY95,EMN97,Quataert01,Narayan04} as a function of
luminosity and the fact that such low-$\mdot$ systems are observed to
be accreting substantially {\em below} the Bondi rate
\citep{FC88,BB99,DiMatteo00,NIA00,QG00,DCF01,Loewenstein01,Bower03,Pellegrini05}.
We do so by adopting a radiative efficiency
\begin{equation}
  \er = \left\{ \begin{array}{ll}
      0.1  & \mathrm{ if\ } \mdot > \mdotcrit \\
      0.1\,\Bigl( \frac{\mdot}{\mdotcrit}\Bigr) & \mathrm{ if\ } \mdot \leq \mdotcrit\ \, ,
\end{array}
    \right.
\label{eqn:rad.eff}
\end{equation}
for a transition between radiatively efficient and inefficient
accretion at $\dot{m}_{\rm crit}\sim 0.01$ as suggested by
observations of black hole binaries \citep{Maccarone03} and AGN
Eddington ratio distributions \citep{Marchesini04,Jester05}, and
theoretical extensions of accretion models
\citep[e.g.,][]{NY95,EMN97,MLMH00}.  This choice for the
efficiency follows from ADAF models \citep{NY95}, also allowing
for large mass loss through winds giving sub-Bondi accretion rates
\citep{BB99,Soria05b} and broadly accounts for observations of local
quiescent objects.

For a characteristic $\dot{m}\sim10^{-4}$ (i.e.\ $L/L_{\rm Edd}\sim
10^{-6}$) and early type black hole masses
$\sim10^{8}-10^{9}\,M_{\sun}$ with redshifts of formation $z\sim2$,
this implies a narrow, steep distribution in $M_{B}$ peaked around
$-11\lesssim M_{B}\lesssim -13.5$. In detail, we again use the
determination of the elliptical $\dot{m}$ distribution and luminosity
function in Figure~\ref{fig:LF} from our modeling of black hole and
spheroid co-formation in \citet{H05e,H05f} and the decays of those
light curves in our blast wave solution from \citet{H05g}, but impose
the larger of Equations~(\ref{eqn:wind.total}) \&
(\ref{eqn:wind.inner}) as a lower limit to the accretion rate (for
$t>1\,$Gyr) decay. Convolved over the early-type population, this
gives the dot-dashed line in Figure~\ref{fig:LF}.  Indeed, it is
suggestive that the estimated contribution from this mode of accretion
becomes important at about the luminosity where our prediction for
stochastic cold gas accretion flattens and may begin to fall below the
observations.  As discussed in \S~\ref{sec:morph.contrib}, this more
involved calculation ultimately implies a relatively basic point, that
there are a large number of ``dead'' quiescent early-type galaxies
with steady-state $\mdot\sim10^{-5}-10^{-4}$ accretion rates, which
make up a substantial portion of the AGN and LINER population at the lowest
luminosities.

\section{Discussion}
\label{sec:discuss}

We have developed a model for the light curve and accretion rate
evolution of accreting systems in feedback-driven, Sedov-Taylor type
blast waves. Our formalism is applicable to a wide variety of systems
with different equations of state, external gas profiles, physical
scales, and feedback coupling mechanisms. We have applied this general
model to the specific case of supermassive black hole accretion via
``stochastic'' (i.e.\ not cosmologically-induced) collisions with
molecular clouds (or the inflow of such clouds from disk, bar, or
torus processes), in quiescent systems with some supply of cold
(rotationally supported) gas. Feedback from accretion energy rapidly
unbinds nearby gas and the system is described by our feedback-dominated
decaying light curve solution.

Predictions from this model are consistent with many properties of
low-luminosity AGN, which are indeed observed to be in quiescent
systems with cold supplies of gas, and our
picture has many testable consequences.
Because the fueling mechanism is not cosmological in nature, these
predictions are essentially {\em a priori}, and do not require a
detailed cosmological modeling of the evolution in galaxy properties.
Moreover, this provides a context for considering such objects and
their large-scale fueling mechanisms, and for contrasting them and
their evolution with that of bright quasars driven by the cosmological
processes of interactions and galaxy mergers.

\subsection{Comparison with Observations}

Our model reproduces numerous observations of low-luminosity AGN,
including:

{$\bullet\ $}{\em Duty Cycles:} The duty cycle at high accretion rates
$\dot{m}\gtrsim0.1$ is expected to be $\sim1\%$ \citep[compare
e.g.,][]{Kauffmann03,YLK05,Dong05}. This rises to imply that a large
fraction of galaxies host a low-luminosity AGN, with a high active
fraction of $\sim20-50\%$ by $M_{B}\gtrsim-12$ (depending on the
location of the cutoff) \citep[e.g.,][]{Hao05,Best05}.  The
inflows of cold gas observed in \citet{Lauer05} in the central regions
of low-luminosity AGN also appear to follow a similar duty cycle and
periodicity to our predictions. Note, however, that technically these
are theoretical upper limits to the duty cycles, for if accretion
proceeds intermittently (i.e.\ in short, potentially super-Eddington
``bursts'') the same average accretion rate on the timescales relevant
for our calculations is maintained, although the timescale for such
bursts is still constrained by the observed episodic quasar lifetime
\citep[see e.g.,][]{Martini04}.

{$\bullet\ $}{\em Seyfert Luminosity Function:} We predict a local
Seyfert luminosity function which agrees with that observed
\citep[e.g.,][]{Hao05,UlvestadHo01,HuchraBurg92} over nearly ten
magnitudes, $-14\gtrsim M_{B}\gtrsim -23$. This result depends only
weakly on the theoretical uncertainties in our blast wave model and
more uncertain duty cycle calculation.  Although our prediction may
fall short of the observed luminosity function at low luminosities
$M_{B}\gtrsim-14$, this is where contamination from star formation
becomes a serious observational concern \citep[see the discussion
in][]{Hao05,Kauffmann03,Kewley01}, and also where the contribution from
quiescent, relaxed ellipticals fueled by hot gas accretion and stellar
winds may begin to dominate \citep[see \S~\ref{sec:winds}
above and][]{Ciotti97,Ciotti01,Pellegrini05,Soria05b}.  At high
luminosities, our predictions map neatly onto the quasar luminosity function 
observed \citep[e.g.,][]{Boyle00,Ueda03,Richards05}
and predicted from merger-induced activity \citep{H05e}, with an
different fueling mechanism but similar feedback processes
regulating the decay of the light curve.

{$\bullet\ $}{\em Morphological Properties:} From our analysis, we
predict the contributions of various morphological types to the
luminosity function.  At high luminosities, there will be a
significant contribution from relaxing ellipticals decaying from
recent ($\sim\,$Gyr) merger and starburst activity as seen in e.g.\
\citet{Kauffmann03,Sanchez03,Sanchez04,VandenBerk05}.  In the range $-20\gtrsim
M_{B}\gtrsim -23$ there will be a comparable contribution from S0 and
Sa/b systems at moderate to large accretion rates, with a
potentially non-negligible but relatively small contribution from Sc/d
systems (and smaller bulge/black hole systems). 
These Sa/b and S0 systems dominate the low-luminosity
activity over the range $-15\lesssim M_{B}\lesssim-20$.  Both of these
trends are seen observationally, e.g.\
\citet{Kauffmann03,Sanchez04,Best05} and specifically by \citet{Dong05}
who give a detailed breakdown of AGN activity in different late-type
morphological types. These objects are not necessarily associated with
mergers or interactions, unlike the highest-luminosity quasars (not seen at $z=0$), again
consistent with the observed morphologies
\citep[e.g.,][]{Kauffmann03}. Below these luminosities 
(i.e.\ luminosities characteristic of the lowest luminosity AGN and 
LINERs), ``dead''
early-type galaxies fueled by e.g.\ steady accretion of hot
(virialized) gas and stellar winds from old stellar populations make
up a large fraction of the active population, as seen by
\citet{Ho02,Heckman04,Marchesini04,
Jester05,Pellegrini05,Soria05b}. Independent evidence for our
hypothesis comes from observations of Seyfert clustering, which
suggest that Seyfert hosts have masses of typical gas-rich, late-type
(typically low-mass) systems, whereas large black holes with low
accretion rates show clustering typical of larger mass, gas-poor
ellipticals \citep{Constantin06}.

{$\bullet\ $}{\em Outflow Properties:} According to our model, even
low-luminosity AGN should be associated with some locally driven
Sedov-Taylor like wind or outflow, although this will not necessarily
globally impact the host galaxy. Many observations find such an
association, with characteristic entrained masses
$\sim10^{5}-10^{7}\,M_{\sun}$ and small-scale wind dynamical times
$\sim10^{4}-10^{6}\,$yr, driving outflows with mass outflow rates
several times the AGN accretion rate and implied energy coupling
efficiencies $\sim0.01-1$ \citep[see][for a review]{Veilleux05},
consistent with feedback-driven destruction of stochastically accreted
gas clouds. The velocity and temperature structures \citep[e.g.,][and
references therein]
{ShopbellBland98,Crenshaw00,KrolikKriss01,Kaspi02,Rice06}, small
energy injection zones \citep[e.g.,][]{SmithWilson01}, and detection 
rates \citep[e.g.,][]{Crenshaw99,Rupke05} of observed AGN winds, even in obscured
and starburst-dominated systems \citep[e.g.,][]{Rupke05}, support this
picture.  Of course, it has long been recognized that these
observations can be modeled by Sedov-Taylor blast waves driven from
near the black hole
\citep[e.g.,][]{Begelman83,SVS85,KoniglKartje94,Murray95,Elvis00}, but
our modeling self-consistently links the outflow and accretion rate
evolution.

{$\bullet\ $}{\em Eddington Ratio Distributions:} We predict the
distribution of Eddington ratios in low-luminosity AGN, as a function
of e.g.\ velocity dispersion, black hole mass, and luminosity, in
agreement with that observed by \citet{YLK05}, over three orders of
magnitude in $\dot{m}$, for a range of $\sigma\sim70-120\,{\rm
km\,s^{-1}}$. This includes both the power-law behavior of the
Eddington ratio distribution set by our blowout solution and the
turnover at the lowest $\dot{m}$ determined by the finite time between
new ``excitation'' events, resolved from comparison with \citet{YLK05}
at the $\sim4-6\,\sigma$ level. Predictions including the distribution of 
post-merger or hot gas-fueled ellipticals also agree well at larger $\sigma$, but 
this is less directly related to our modeling. The cumulative Eddington ratio
distribution, combined with that predicted for ``dead'' ellipticals
from \citet{H05e}, may be bimodal, as suggested by e.g.\
\citet{Marchesini04,Jester05}.  However, this bimodality is easily
washed out by relatively small changes in our estimated duty cycles,
and is therefore not as strong prediction of our model as is e.g.\ the
shape of the Eddington ratio distribution in active late-type systems
alone.

{$\bullet\ $}{\em Eddington Ratio vs.\ Luminosity and Mass:} We find
that the mean or median Eddington ratio is a strong function of the
observed (intrinsic) AGN luminosity. In other words, the observed
low-luminosity Seyfert luminosity function spans a wide range in
Eddington ratio and a relatively small range in black hole mass, so
the observed luminosity function is primarily an Eddington ratio
sequence determined by feedback-driven blowout. This agrees with e.g.\
\citet{Hao05,VandenBerk05}, who find little correlation between AGN
and host galaxy luminosities, suggesting that the observed luminosity
is primarily a function of Eddington ratio with characteristic 
black hole masses $\sim10^{7}\,M_{\sun}$.  The mean Eddington ratio
is also a function of black hole mass, declining at higher black hole
masses as the population is increasingly dominated by ``dead''
early-type systems which have decayed from their bright quasar epoch
and may be fueled by quiescent (e.g.\ hot gas or stellar wind
accretion) mechanisms. This trend agrees with that observed by
\citet{Heckman04}, and predicts that black hole growth at $z=0$ is
dominated by low mass systems, with $M_{\rm
BH}\lesssim10^{7}\,M_{\sun}$.

{$\bullet\ $}{\em The $M_{\rm BH}-\sigma$ Relation:} Because feedback
regulates accretion and causes massive systems to enter a
feedback-driven blowout phase in which the accretion rate declines and
relatively little mass is gained, our model reproduces and preserves
an $M_{\rm BH}-\sigma$ relation similar to that observed
\citep{Gebhardt00,FM00,Tremaine02}. At the high mass end, where the
observed $M_{\rm BH}-\sigma$ relation is most robust, black hole
masses are sufficiently large that any cold gas accretion described by
our modeling will enter the ``blowout'' phase effectively immediately,
yielding no change in the $M_{\rm BH}-\sigma$ relation.  At lower
luminosities, more detailed observations can test deviations from the
high-$M_{\rm BH}$ relation predicted by our modeling (see below), but
we note that these are subtle effects and are consistent with all
present observations of the relation down to quite low black hole
masses.

{$\bullet\ $}{\em Obscuration:} The same ``blowout'' criterion which
determines the applicability of our feedback-driven solution and the
$M_{\rm BH}-\sigma$ relation can be applied in a geometrical manner to
determine the opening angle in which feedback will be sufficient to
unbind gas and ``punch through'' unobscured sightlines to the black
hole.  Objects with larger black hole masses will, in their initial
high-$\dot{m}$ period, more efficiently unbind gas and blow out a
larger fraction of gas, giving (indirectly) an anticorrelation between
the obscured fraction and luminosity.  In detail, our prediction of
the obscured (Type 2) fraction as a function of observed (intrinsic)
luminosity agrees with that estimated observationally from the same
Seyfert luminosity functions we consider \citep{Hao05}.

Although qualitatively similar to ``receding torus'' models which 
attempt to explain
this effect \citep[e.g.,][]{Lawrence91} or models of obscuration
within a disk wind \citep[e.g.,][]{KoniglKartje94,
StoneNorman94,Murray95,Elvis00,Proga04}, our process operates over
larger scales and naturally explains the quantitative distinction (see
\S~\ref{sec:obscuration}) between obscured fractions in local objects
and (intrinsically) bright quasars
\citep[e.g.,][]{Ueda03,Steffen03,Hasinger04,GRW04,Zakamska04,Simpson05,Zakamska05}.
Our picture may, however, graft quite naturally onto such models
(especially feedback-driven disk wind models) on smaller scales, as
discussed below.

\subsection{Testable Predictions}

There are a number of additional implications of our model for which
observations either do not exist or are inconclusive, and
therefore can be used to test our theory.

{$\bullet\ $}{\em Evolution of Seyfert Luminosity Functions:} From the
evolution of the late-type galaxy luminosity function and gas
fractions, we determine the evolution with redshift of the luminosity
function determined by stochastic or steady-state cold gas
accretion. Although there is substantial uncertainty in the late-type
evolution, it is weak compared to the rapid evolution in the quasar
luminosity function. Probing even moderate redshifts (e.g.\
$z\sim0.1-0.5$) and separating the component of the quasar
luminosity function contributed by late-type, non-interacting galaxies
can test our prediction. We predict a relatively small and rapidly
declining relative contribution to the faint-end of the quasar
luminosity function at these redshifts, which has important
implications for the contribution to cosmic backgrounds and buildup of
black hole and spheroid mass, discussed below.

{$\bullet\ $}{\em Eddington Ratio Distributions at Low Luminosity:}
Although we have compared our predicted Eddington ratio distributions
to observations as a function of mass, velocity dispersion, and
luminosity, the observations are incomplete.  Both the AGN luminosity
function and Eddington ratio distribution must turn over and begin to
decline at some sufficiently low luminosity, and the large duty cycles
calculated at the lowest $M_{B}\gtrsim-14$ luminosities probed by
e.g.\ \citet{Hao05} imply that observations are rapidly approaching
this limit. Resolving the location and shape of this turnover, while
nearly independent of the blast wave structure we have modeled (which
determines the power-law slope of this distribution above the
turnover) determines the duty cycles and rate of excitation,
constraining and testing our estimates of the contribution
to luminosity functions from this fueling mechanism and its importance
in the buildup of low-mass black holes.

{$\bullet\ $}{\em The $M_{\rm BH}-\sigma$ Relation:} Because the mass
of individual gas clouds is not correlated with $\sigma$ as is the
total galaxy gas mass, and the timescale for ``blowout'' to occur is
not determined by the galaxy dynamical time in our scenario, we
predict a different slope for the $M_{\rm BH}-\sigma$ relation than
that indicated by models of spheroid and black hole co-formation (in
e.g.\ galaxy mergers) \citep[e.g.,][]{SR98,DSH05}. At high black hole
masses, black holes easily unbind a large gas mass rapidly and there
is no change in the $M_{\rm BH}-\sigma$ relation, but at low masses,
below an estimated break around $\sim10^{6}-10^{7}\,M_{\sun}$, the
slope of the $M_{\rm BH}-\sigma$ relation should become shallower,
$M_{\rm BH}\propto\sigma^{3}$ for energetically-determined feedback
(or $M_{\rm BH}\propto\sigma^{2}$ for direct momentum feedback). This
change in slope is a robust prediction, with a break set approximately
by the characteristic mass of a molecular cloud, but in detail the
``break'' location depends on uncertain factors such as the efficiency
of feedback coupling. Given this, it is possible that no change would
be observable down to $M_{\rm BH}\sim10^{5}\,M_{\sun}$, but even this
observation would set strong lower limits to the feedback efficiency
and means of feedback coupling. For reasonable coupling values
(similar to those which give the spheroid-formation induced relation),
this gives a break at $M_{\rm BH}\sim10^{7}\,M_{\sun}$, and this slope
change should be observable with larger samples of low-mass black
holes. In fact, existing observations by \citet{Barth04,Barth05} and
\citet{Greene05} favor a change in slope, but the large
observational errors make this only a $\sim1-2\sigma$ effect.

From the observed spectrum of AGN column density distributions and
molecular cloud sizes, we estimate their contribution to the scatter
in the $M_{\rm BH}-\sigma$ relation.  Again, at high $M_{\rm BH}$,
they are negligible, but at low luminosities we predict that the
intrinsic scatter in the $M_{\rm BH}-\sigma$ relation should increase
substantially, from $\sim0.27$\,dex to $\sim1$\,dex at $M_{\rm
BH}\lesssim10^{6}\,M_{\sun}$.  Present observations marginally
($\sim1\sigma$ level) favor this, but expanded observations of the
low-mass $M_{\rm BH}-\sigma$ relation can test this without any
dependence on the relatively large systematic uncertainties inherent
in the absolute $M_{\rm BH}$ normalization as measured by different
(e.g.\ reverberation-mapping vs.\ stellar dynamical) mass measurement
techniques.

{$\bullet\ $}{\em (Lack of) Global Effects on the Host Galaxy:} A
natural question following from our feedback-driven, blast wave
scenario is whether or not there will be a substantial impact on the
host galaxy. Indeed, for the case of growth-terminating ``blowouts''
in galaxy mergers \citep[e.g.,][]{H05g}, a powerful wind expels or
heats the remaining cold gas in the galaxy, enriching the X-ray halo,
reducing column densities, and shutting down star formation.  However,
the situation is quite different in our modeling of cold gas accretion
in quiescent systems.  In this case, black hole feedback is not forced
to overcome the entire galaxy supply of cold gas, which is driven to
the central regions and tightly bound by the entire galaxy potential
via gravitational torques in a merger.  Rather, only the relatively
weak feedback to unbind cold gas clouds is imparted before blowout is
entered, and most of the galaxy ISM remains in an organized disk
(which will not be impacted by the feedback). The feedback is
essentially negligible as far as the cold disk of the host is
concerned, although it may still enrich halo gas (as even a small
quantity of metal-rich gas expelled from the cloud ``blowout'' can be
significant) and will heat diffuse bulge gas. There is no necessary
causal connection between this AGN activity and star formation, in
tentative agreement with observations \citep[e.g.,][]{Laine06}, except
insofar as larger quantities of cold gas will both increase the duty
cycle for AGN excitation and enhance star formation. The hosts
will be relatively unperturbed, normal galaxies
which continue to be actively star forming.

{$\bullet\ $}{\em Testing Alternative Fueling Mechanisms:} 
In addition to the predictions above, there are a number of tests which 
distinguish our model of accretion of cold gas and subsequent feedback-driven 
light curve decay from other non-merger driven modes of accretion. For 
example, it has recently been suggested that stellar tidal disruptions can account 
for the entire low-luminosity AGN luminosity function \citep{Milosavljevic06}. 
However, stars cannot be disrupted outside the innermost stable circular orbit of 
the black hole for a black hole mass
$M_{\rm BH}\geq 2\times10^{7}\,M_{\sun}\,(M_{\ast}/M_{\sun})^{0.7}$ 
(where $M_{\ast}$ is the stellar mass). For the rates of stellar disruptions estimated 
therein and by e.g.\ \citet{WangMerritt04} to be sufficiently high to contribute 
the the observed luminosity functions, the typical stellar mass disrupted must be  
$\lesssim0.1\,M_{\sun}$, implying that the luminosity function is dominated 
by black holes with $M_{\rm BH}\lesssim2\times10^{6}\,M_{\sun}$ (typically 
in Sd or Sm/Im galaxies). If more recent black hole mass functions from 
\citet{Marconi04,Shankar04} are adopted in estimating these rates, the 
expected black hole masses go down even further. 

In contrast, our model predicts the luminosity function is 
dominated by $\sim 10^{7}\,M_{\sun}$ black holes, in Sa/b and S0 hosts. 
Furthermore, the bound mass in a disruption event is small, implying no correction 
to the slope or scatter of the $M_{\rm BH}-\sigma$ relation, and no association 
with cold gas supplies or outflows (the entrained mass and momentum of which 
are much larger than can be accounted for by individual disruptions). Finally, the 
timescales for such events are very short - typically $\sim1-10$\,yr. Although 
individual bursts of accretion may have shorter timescales than the time-averaged light curves 
we calculate, our characteristic timescales are consistent with both the 
integrated and ``episodic'' quasar lifetime constraints from a number of 
observations \citep[see e.g.][and references therein]{Martini04}, which estimate 
lower limits to the episodic lifetime of $\gtrsim10^{4}$\,yr, and direct attempts to 
measure the rates of short-duration X-ray flares \citep[which have traditionally 
been associated with stellar disruptions, e.g.][]{Hills75,MeszarosSilk77} have found 
rates $\sim2-3$ orders of magnitude below those needed to account for 
the low-luminosity AGN luminosity function \citep[e.g.,][]{Donley02}.

\subsection{Comparison With The Bright Quasar Population}

Our modeling, in providing a galaxy-level context for the evolution of
low-luminosity AGN in non-interacting galaxies, allows us to draw a
physical distinction between the fueling of AGN and bright quasars,
namely quiescent accretion vs.  galaxy mergers and interactions.  The
feedback mechanisms and explosive ``blowout'' decay modes we determine
here can be applied both to quiescent low-luminosity systems and to
more violent, merger-driven systems \citep{H05g}, and on small scales
(i.e.\ those relevant to detailed accretion and feedback mechanisms,
molecular torii, and disk winds) the systems may be similar. However,
there are fundamental qualitative distinctions that should be kept in
mind when making comparisons between the different classes of objects,
and which imply quite different properties and physical
interpretations of sometimes similar phenomena.

{$\bullet\ $}{\em Fueling Mechanisms:} The fueling mechanisms we have
described here and in merger-driven activity as in e.g.\ \citet{H05e}
are fundamentally different, the former being determined by a steady
cold gas supply in a quiescent system, the latter by the violent
torquing of cold gas throughout entire galaxies into the galaxy center
in major mergers.  Fueling by stellar winds or hot gas accretion may
represent yet a third qualitatively distinct mode of fueling, which
may not be be feedback-dominated but instead produce light curves
dictated by steady-state equilibrium with hot gas
\citep[e.g.,][]{Ciotti97,Ciotti01}, or slowly decaying injection of
gas from stellar winds inside the black hole radius of influence
\citep[e.g.,][]{Quataert04,Soria05b}.  Fueling by mergers is
principally a {\em cosmological} quasar fueling mechanism, determined
by galaxy-galaxy merger rates and the cosmological evolution in gas
fractions and morphologies, characteristic masses of merging objects,
and environmental dependencies of interactions.  The fueling we have
described here is non-cosmological, requiring only a local gas supply
in disk galaxies, and independent of e.g.\ environment, formation
mechanisms and times, and other cosmological influences
(although indirect effects or correlations of these properties with
e.g.\ morphology and gas fractions may be significant).

{$\bullet\ $}{\em Luminosities:} Objects fueled in these quiescent
systems span a different range in luminosities from the bright quasars
produced in gas-rich mergers.  Our mechanism can account
for local objects over the range $-15\gtrsim M_{B}\gtrsim-20$, with an
increasing contribution from post-merger decaying remnants above this,
and mergers driving the brightest activity (with too low a number
density to be seen at $z=0$, but rapidly increasing to dominate
populations down to $M_{B}\lesssim-18$ at moderate redshifts
$z\gtrsim1$).  Accretion of cold gas in quiescent systems cannot
explain the brightest quasars -- in our derivation of the $M_{\rm
BH}-\sigma$ relation, it is clear that black holes cannot grow to the
largest masses observed (compare the dashed and dot-dashed predictions
of maximum masses grown via this method to the observed solid line at
the highest black hole masses in Figure~\ref{fig:m.sigma}) through this
mode of accretion. Furthermore, even if this were possible, the high
Eddington ratios at these masses needed to explain bright quasars
cannot be maintained.  Even with an extremely low coupling efficiency
(further complicating attempts to match the $M_{\rm BH}-\sigma$
relation), a $M_{\rm BH}\sim10^{8}-10^{9}\,M_{\sun}$ black hole would
immediately unbind the gas around it and yield a bright quasar
lifetime of $\sim 10^{4}-10^{5}\,$yr, below observational lower limits
\citep[see e.g.,][]{Martini04}. The only way to force enough gas
accretion to ``overpower'' the blowout for a significant amount of
time (enough time to have a chance to see the observed quasar
population) would be to channel roughly the entire cold gas supply of
the galaxy to the central regions at once -- effectively requiring the
tidal torquing possible only though a major merger.  Major mergers
account for high-luminosity activity naturally, with simulations
producing typical bright quasar luminosities for the appropriate host
galaxy masses \citep{H05e,H05h}.

{$\bullet\ $}{\em Evolution:} As emphasized above, fueling AGN via
``stochastic'' cold gas accretion as we model is fundamentally
non-cosmological, whereas fueling via mergers is a cosmological
process. This results in little evolution in the luminosity function
of objects driven by quiescent fueling, even assuming maximal
evolution in disk galaxy populations and gas fractions up to
$z\sim2$. On the other hand, merger-driven activity evolves strongly
with redshift, dominating above even a moderate $M_{B}\sim-19$ by $z\sim1$. The
contribution from quiescent fueling is quickly relegated to the
faint-end of the luminosity function at higher redshifts, becoming
important only at luminosities one to two orders of magnitude ($\sim4$
magnitudes) below the break in the luminosity function at $z\sim1$,
and another order of magnitude fainter relative to the break at
$z\sim2$, the epoch of peak quasar activity. Thus, quiescent fueling
is essentially irrelevant at high redshifts, while merger rates are
high and merger-driven fueling dominates quasar activity.  At higher
redshifts, faint activity will be even more difficult to observe, but
the distinction is also somewhat less clear, as essentially all
systems are highly gas rich (and potentially unstable to gas
collapse), but merger rates are also high.

{$\bullet\ $}{\em Obscuration:} Although the qualitative trend of a
decreasing obscured fraction with increasing luminosity is similar in
both bright quasars and Seyferts fueled in the manner we model, there
are important distinctions. As discussed in \S~\ref{sec:obscuration},
observations of local AGN and moderate-redshift populations (i.e.\
merger-driven quasar dominated, see Figure~\ref{fig:obscuration}) give
a similar trend with luminosity but significantly different absolute
obscured fractions. The Seyfert (local, low-luminosity) Type 2
fraction is systematically lower, rising to comparable numbers of Type
1 and Type 2 objects at the lowest luminosities, whereas the quasar
samples rapidly become dominated by Type 2 objects at the lowest
luminosities \citep[compare,
e.g.,][]{Hao05,Ueda03,Hasinger04,GRW04,Simpson05}. Although both
trends with luminosity are caused by black hole feedback, they are
fundamentally distinct. In the local, non-interacting case the primary
source of obscuration is small-scale, geometrical obscuration
\citep[see e.g.,][]{RMS99}, and the feedback-based argument we use to
describe the obscured fraction (giving e.g.\ comparable obscured and
unobscured populations at low luminosity) is also a natural
consequence in many disk wind models and (potentially) torus models of
small-scale obscuration. However, many X-ray identified bright quasars
have been observed to be growing in starbursting or merging systems
\citep[e.g.,][]{Alexander05a,Alexander05b,Borys05}, obscured not by a
local geometrical structure but by the large-scale gas inflows
powering accretion, which can give a much larger obscured fraction at
low luminosities and, in detailed simulations, gives a trend of
obscuration with luminosity which agrees with that observed
\citep{H05b,H05e}. This distinction between quasi-static
geometrical and strongly time-dependent, interaction-driven
obscuration is important in understanding intrinsically bright Type 2
quasars \citep[which are observed to be interacting, with galaxy-scale
obscuring structures;][]{Zakamska05}, and in building any model of
obscuration for synthesis models of the cosmic X-ray or IR
backgrounds.

{$\bullet\ $}{\em Impact on the Host Galaxy:} Merger-driven quasars
are generally presumed to have a dramatic effect on their host
galaxies, driving a galaxy scale-outflow, expelling gas and heating it
to the virial temperature and terminating star formation
\citep{SDH05a,H05f}, and enriching the X-ray halo with metal-enriched
gas \citep{Cox05}. However, as discussed in \S~\ref{sec:host.fx}, we
do not expect the ``blowout'' from quiescent cold gas fueling to
result in any significant impact on the host galaxy, even for small
hosts. This can be understood by considering the different thresholds for
blowout. In the case of a merger, nearly the entire remaining gas mass
is torqued to the central regions of the galaxy, in a roughly
isotropic manner, and ``blowout'' is not entered until the energy is
sufficient to unbind this material. However, in the quiescent case,
the disk is largely unperturbed (and further has a limited covering
angle), so it does not need to be ``overcome'' to enter the blowout --
instead, only a small fraction of gas is unbound and it has little
impact on the host.  Recent observations support this picture, with
the outflow kinetic energy and entrained mass on large scales scaling
with the galaxy size (and therefore black hole mass) in a manner
similar to what we predict \citep{BaumMcCarthy00}, with Seyferts in
small hosts doing little ``damage'' to their environment.  This may
still be significant for metal enrichment of the halo and continued
heating of the small amount of diffuse bulge gas, or e.g.\ the
production of high-velocity clouds in the ejecta, but it does not
disturb the normal, star-forming host disk.

{$\bullet\ $}{\em Host Properties and the $M_{\rm BH}-\sigma$
Relation:} As discussed above, systems fueled by cold gas accretion
will be preferentially normal galaxies, S0s and Sa/b galaxies
dominating the high-accretion rate population, whereas at higher
luminosities the population will mainly be (at the faint end)
ellipticals decaying from previous merger activity and (at the bright
end) objects still affected by major interactions. The $M_{\rm
BH}-\sigma$ relationship as determined by these two fueling mechanisms
may also differ, with the fact that the immediate gas supply in
quiescent systems is not linked to the total gas supply of the galaxy
(whereas it is in a strongly torqued merger), ultimately producing a
somewhat shallower slope in the $M_{\rm BH}-\sigma$ relationship at
low black hole masses ($M_{\rm BH}\lesssim10^{6}-10^{7}\,M_{\sun}$),
and increased scatter in the relationship at these masses.

{$\bullet\ $}{\em Contribution to the Black Hole Mass Density and
Cosmic Backgrounds:} Given the relative evolution of the luminosity
functions determined by mergers and interactions as opposed to
quiescent or stochastic fueling, it is straightforward to calculate
the relative contribution to the black hole mass density and cosmic
backgrounds.  The rapid evolution in merger-driven quasar activity and
its dominance at bright luminosities, especially during the period of
peak quasar activity $z\gtrsim1$, implies that other fueling
mechanisms will not contribute significantly to these quantities.
Quiescent fueling in the manner we have modeled only adds
significantly to typical black hole masses at low mass ($M_{\rm
BH}\lesssim10^{6}\,M_{\sun}$), as can be seen from comparison of the
``increase'' above the merger-driven $M_{\rm BH}-\sigma$ relation in
Figure~\ref{fig:m.sigma}, below the $M_{\rm
BH}\gtrsim10^{8}\,M_{\sun}$ masses which dominate the black hole mass
density. From energetics, it is then clear that backgrounds such as
the cosmic X-ray background, which can be entirely accounted for by
merger-driven quasar activity \citep{H05e}, receive little
contribution from these alternative fueling mechanisms.

\subsection{Summary}

Our feedback-driven model for the evolution of accretion rates in
systems which are non-interacting and fueled by cold gas explains a
number of observations of AGN, and makes testable predictions for
future observations. The modeling allows us to distinguish
low-luminosity (stochastically cold gas accreting) ``Seyferts'' and
high-luminosity (merger-driven) ``quasars'' in a physically meaningful
manner based on their respective fueling mechanisms, even when both
have similar high accretion rates. This distinction is critical in
understanding the properties of local objects, and their contribution
(or lack thereof) to cosmological backgrounds and buildup of black
hole and spheroid mass. It is also important to understand such
distinctions when extrapolating properties of local, quiescent objects
to bright objects potentially fueled in mergers, which may be
appropriate on small scales (e.g.\ disk properties and detailed
accretion and feedback coupling mechanisms), but is inappropriate on
larger scales (fueling rates and mechanisms, obscuration mechanisms in
at least some fraction of cases, and impact on the host galaxy).

Our model for briefly fueling high accretion rate activity in
quiescent, non-interacting galaxies is intended to provide a
large-scale context, relating this fueling to the host galaxy and enabling a
consideration of the statistics of such activity.  We develop a
self-consistent model for the accretion rate evolution within a wind
or outflow, allowing the black hole light curve and accretion rate
statistics to be calculated {\em a priori}.  By no means does this
model attempt to describe the small-scale gas thermal and ionization
structure around the black hole.  The local phenomena around the
central engine, such as disk processes, the detailed coupling of
feedback mechanisms, ionization, broad and narrow-line region
evolution, and pc-scale molecular torii, which have been (and continue
to be) studied in detail both theoretically and observationally
\citep[see e.g.,][and references therein]{Krolik99}, are obviously
crucial to understanding AGN activity but are outside the scope of the
large-scale fueling and feedback mechanisms. Ultimately, as long as
feedback regulates AGN activity in a rapid, explosive manner, which we
have shown for the case of cold gas accretion in quiescent systems
requires only a small fraction ($\lesssim 10^{-3}$) of the accretion
energy to couple to the surrounding medium, the large scale evolution
of fueling and accretion rates should be described by our model.

However, our work emphasizes the need to understand the connection
between these small-scale phenomena, which are presumably universal in
AGN unification models \citep[e.g.,][]{Antonucci93}, and large-scale
models for fueling and feedback in both quiescent systems as presented
here or in interacting systems as in \citet{H05e}. Although 
care must be taken in extrapolating properties of local
quiescent objects to interacting systems, the broad agreement between
the predictions of feedback-driven blowout in both cases suggests that
feedback coupling may be quite similar in both. Our
model for feedback driven ``blowout'' and evolution of the Seyfert
light curve grafts naturally onto disk-wind models in which the broad
and narrow line regions are part of a high-velocity magnetically
\citep[e.g.,][]{StoneNorman94}, advectively \citep[e.g.,][]{NY94}, or
radiatively
\citep{SVS85,KoniglKartje94,Murray95,Elvis00,Proga00,Proga04} driven
outflow from the accretion disk, which immediately suggest mechanisms
for the coupling of feedback to the surrounding medium and a mapping
from local scales (i.e.\ the accretion disk region from which the wind
is driven) to large scales. Such a wind could directly drive an
outflow, with mass pile-up in a ``snowplow'' phase and the (initially
radiatively, energetically coupled outflow) powering a
momentum-driven wind. Alternatively, it could collide with gas on
cloud or central cold gas disk scales and shock, giving a pressure and
energy source which drives a pressure-driven Sedov-type
explosion.  It is suggestive that these mechanisms manifest similarly,
but there remain many alternative possibilities for the coupling of
feedback.  Studying this in the ``laboratory'' of local Seyferts,
where AGN-driven winds are well-known and studied in detail
\citep[see, e.g.][and references therein]{Veilleux05}, may potentially
represent the ultimate mapping between the well-studied properties of
AGN on small scales and the large-scale implications for the host
galaxy and AGN statistics as studied here.

\acknowledgments 
We thank Jeremy Goodman, Ramesh Narayan,
Chris McKee, J.P. Ostriker, Joel Primack, Eliot Quataert, and 
Michael Strauss for
helpful discussions and 
guidance in the development of this work, as well as
T.\ J.\ Cox, Brant Robertson, Sukanya Chakrabarti, and Yeuxing Li. 
This work was supported in part by NSF grants ACI
96-19019, AST 00-71019, AST 02-06299, AST 03-07433 and AST 03-07690,
and NASA ATP grants NAG5-12140, NAG5-13292, and NAG5-13381.

\end{document}